\documentclass[a4paper,11pt]{article}
\usepackage[utf8]{inputenc}

\usepackage{braket}
\usepackage{mathtools} 
\usepackage{amssymb} 
\usepackage{amsmath}
\usepackage{bbold}
\usepackage{amsfonts} 
\usepackage{stmaryrd} 
\usepackage{hyperref}
\usepackage{authblk}
\usepackage{xcolor}
\usepackage{graphicx}

\usepackage[
    backend=biber,
    style=ieee,
  ]{biblatex}
\usepackage{hyperref}

\newcommand{\B}{\mathcal{B}}
\newcommand{\hp}[1]{^{(#1)}}

\newcommand{\Sint}{\mathcal S_{\mathrm{int}}}
\newcommand{\Aut}{\mathrm{Aut}}

\newcommand{\Z}{\mathbb{Z}}

\title{On the large $N$ limit of 
Schwinger-Dyson equations
of a rank-$3$ tensor field theory}

\author[1,2]{R. Pascalie}
\author[2,5]{C.I. Pérez-S\'{a}nchez}
\author[1,3,4]{A. Tanasa}
\author[2]{R. Wulkenhaar}
\affil[1]{Universit\'{e} de Bordeaux, LaBRI, UMR 5800, 33400 Talence, France}
\affil[2]{Mathematisches Institut der Westf\"{a}lischen Wilhelms-Universit\"{a}t,
Einsteinstraße 62, D-48149 M\"{u}nster, Germany}
\affil[3]{H. Hulubei National Institute for Physics and Nuclear Engineering, P.O.B. MG-6, 077125 Magurele, Romania} 
\affil[4]{IUF,  1  rue  Descartes,  75231  Paris  Cedex  05, France}
\affil[5]{
 Faculty of Physics, University of Warsaw, 
 ul. Pasteura 5, 02-093 Warsaw, Poland}

\begin{document}

\maketitle

\begin{abstract}
 We analyse in this paper the large $N$ limit of the Schwinger-Dyson equations in a rank-$3$ tensor quantum field theory, which are derived with the help of Ward-Takahashi identities. 
 In order to have a well-defined large $N$ limit, appropriate scalings in powers of $N$ for the various terms present in the action are explicitly found. A perturbative check of our results is done, up to second order in the coupling constant. 
\end{abstract}

\section{Introduction}

Tensor models are witnessing a considerable regain of interest since the implementation of their large $N$ limit
(see \cite{Gurau:2009tw}, \cite{Dartois:2013he, Carrozza:2015adg, Benedetti:2017qxl, Carrozza:2018ewt, Klebanov:2017nlk} or the review articles \cite{Tanasa:2015uhr} and \cite{Bonzom:2016dwy}).
Recently, tensor models have been related to the celebrated Sachdev-Ye-Kitaev AdS/CFT toy-model \cite{sy, kitaev} in \cite{Witten} and \cite{Klebanov:2016xxf} - see also \cite{Gurau:2016lzk, Bonzom:2017pqs, Bonzom:2018jfo, Carrozza:2018psc}
and the lectures \cite{Klebanov:2018fzb}.

In \cite{Perez-Sanchez:2016zbh} and \cite{Sanchez:2017gxt}, the Ward-Takahashi identity (WTI) has been extensively used in order to derive the tower of exact Schwinger-Dyson equations (SDE) for  an $U(N)$-invariant tensor models whose kinetic part is modified to include a Laplacian-like operator (more exactly, this operator is a discrete Laplacian in the Fourier transformed space of the tensor index space). 

Let us emphasise here that this type of tensor model has been used as a test-bed for applying renormalisation techniques to tensor models - see \cite{BenGeloun:2011rc}, the thesis \cite{Carrozza:2013mna}
and references within. Recently, the functional renormalisation group has been used in \cite{Eichhorn:2018ylk} to investigate the existence of a universal continuum limit in tensor models, see also the review \cite{Eichhorn:2018phj}. The present paper provides a complementary non-perturbative tool to this approach. 

The  WTI has already been successfully used to study the SDE in the context of non-commutative quantum field theory - see \cite{Disertori:2006nq} and 
\cite{Grosse:2012uv}. In tensor models, a WTI appeared for the first time in 
\cite{DineWard}, whose consequences are still under investigation \cite{Lahoche:2018ggd}.

\medskip

This paper is a follow up of \cite{Perez-Sanchez:2016zbh} and \cite{Sanchez:2017gxt} in the sense that we study in detail the large $N$ limit of the SDE obtained {\it via} the use of  WTI. We thus find appropriate scalings in powers of $N$ for the various terms present in the action of a rank-$3$ model. Moreover, we analyse in detail a case where the boundary graph is disconnected (as explained in the following section, in tensor models, boundary graphs index the expansion of the free energy). This case has not been treated in \cite{Perez-Sanchez:2016zbh} and \cite{Sanchez:2017gxt}.

\medskip



Let us mention here that in \cite{Krajewski:2016svb}, scaling dimension for interactions in Abelian tensorial group
field theories with a closure constraint have been obtained. However, the mathematical physics techniques used in \cite{Krajewski:2016svb} (namely general  formulations  of  exact  renormalisation
group  equations  and  loop  equations  for  tensor  models  and  tensorial  group  field  theories) are different from the techniques used here.

\medskip

Our paper is organised as follows. In the following section we give the action of the tensor model we work with, and we recall tensor model tools used in the sequel, such as
the boundary graph expansion of the free energy and the
WTI.
Section \ref{constraints} is dedicated to the analysis of the scalings in powers of $N$ of the various terms present in the action. Having a well-defined large $N$ limit of the SDE imposes a series of constraints on these scalings. Section \ref{disco4pt} treats in detail the case 
of the $4$-point function with disconnected boundary graph. In section \ref{largeN} we find appropriate scalings in order to have a coherent large $N$ limit of the SDE. In the appendix a perturbative expansion check of these results is performed up to second order.

\section{The model and the tools}

Let us first consider a complex rank-$D$ bosonic tensor field theory with an action of the form
\begin{align}\label{action}
    \mathcal{S}[\varphi,\bar{\varphi}] &= \mathcal{S}_0[\varphi,\bar{\varphi}] +
    \Sint[\varphi,\bar{\varphi}] \\
    &= \sum \limits_{\mathbf{x}}
\bar{\varphi}^{\mathbf{x}}|\mathbf{x}|^2\varphi^{\mathbf{x}} + \lambda \sum \limits_{c=1}^D \sum \limits_{\mathbf{a},\mathbf{b}}\bar{\varphi}^{\mathbf{a}}\varphi^{\mathbf{b}_{\hat{c}}a_c}\bar{\varphi}^{\mathbf{a}_{\hat{c}}b_c}\varphi^{\mathbf{a}}, \nonumber
\end{align}
with $\mathbf{x}=(x_1,\hdots,x_D) \in \{ \frac{1}{N},\frac{2}{N}, \hdots, 1 \}^D$, $|\mathbf{x}|^2 = \sum\limits_{i=1}^D x_i^2 $, $\mathbf{a}_{\hat{c}}b_c =  (a_1, \hdots,a_{c-1},b_c,a_{c+1}, \hdots,a_D)$ for $D$-tuple. Note that the interaction terms in the action, called pillow interaction terms (sometimes referred, in the tensor model literature, as melonic bubbles), are invariant under the action of the group $\mathrm{U}(N)^D$. The tensor fields transform as
\begin{equation}
    \varphi^{\mathbf{x}} \rightarrow \varphi^{\mathbf{x}} = \sum\limits_{y_c}U^{(c)}_{x_cy_c} \varphi^{\mathbf{x}_{\hat{c}}y_c}, \quad
    \bar{\varphi}^{\mathbf{x}} \rightarrow \bar{\varphi}^{\mathbf{x}} = \sum\limits_{y_c}\bar{U}^{(c)}_{x_cy_c}\bar{\varphi}^{\mathbf{x}_{\hat{c}}y_c},
\end{equation}
for $U^{(c)} \in \mathrm{U}(N)$ and for each colour $c \in \{ 1,\hdots,D\}$. Each copy of the group $\mathrm{U}(N)$ acts on only one index of the tensor. Thus, the indices of the tensors have no symmetries and only indices of the same colour can be contracted.

Let us emphasise here that the kinetic term in \eqref{action} represents the discrete Laplacian in the Fourier transformed space of the tensor index space.

The generating functional of the model writes:
\begin{equation}
     \mathrm{Z}[J,\bar{J}] = \int \mathcal{D}\varphi\mathcal{D}\bar{\varphi}
     \exp{\left(-N^{\gamma}\mathcal{S}[\varphi,\bar{\varphi}] + N^{\beta}\sum_{\mathbf{x}}
       (\bar{J}_{\mathbf{x}}\varphi^{\mathbf{x}} +
       J_{\mathbf{x}}\bar{\varphi}^{\mathbf{x}}) \right)}.
\end{equation}
Note that we have introduced here the scaling $\beta$ and $\gamma$, for the action and the source terms. Let us also introduce the scaling $\delta$ for the coupling constant $\lambda = N^{\delta}\Tilde{\lambda}$. These scalings will be determined in the sequel, using the SDE.

\begin{figure} 
\centering
\includegraphics[scale=1]{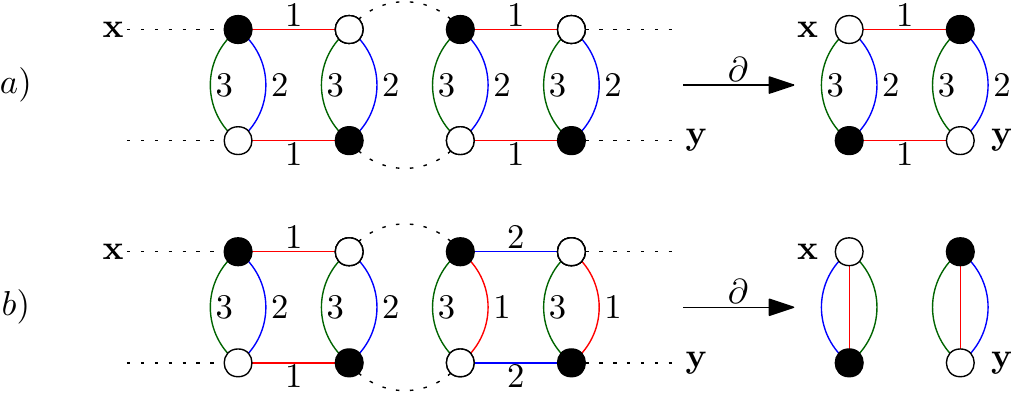}
\caption{Two connected Feynman graphs and the associated boundary graphs in the tensor field theory \eqref{action} for $D=3$. 
In figure a) the boundary graph is connected and in fig. b) the boundary graph is disconnected. \label{fig:boundary}}
\end{figure}

In tensor models, Feynman graphs (see Fig. \ref{fig:boundary}) can be drawn with two types of lines: dotted lines representing the propagator and solid lines which correspond to the contractions of the index of the tensors in the interaction. Hence, each solid line has a colour which correspond to the contracted index of the tensor. A colouring of a graph is then an edge-colouring where the solid lines have colours in $\{1,\hdots,D\}$ and the dotted lines have the colour $0$. The Feynman graphs are then $(D+1)$-coloured graphs. We consider a complex tensor field theory so the graphs are bipartite. 

Moreover each Feynman graph has a boundary graph which is defined as follows: to each external leg of a Feynman graph is associated an external vertex so that the open graph is bipartite. These vertices are exactly the vertices of the boundary graph. An edge of colour $c$ in the boundary graph, corresponds to a path between two external legs in the Feynman graph, which alternates between dotted lines and lines of colour $c$. The boundary graphs are then $D$-coloured graphs, only composed of solid lines. A more detailed exposition of boundary graphs can be found in \cite{Perez-Sanchez:2016zbh} and \cite{Sanchez:2017gxt}.

The connected $2k$-point functions are then split into sectors indexed by a boundary graph $\mathcal{B}$ (see Table \ref{fig:table_CR}), and taken to be
\begin{align}\label{def}
    \mathrm{G}_{\mathcal{B}}^{(2k)}\left(\mathbf{X}\right) =    \displaystyle \left. \frac{N^{-\alpha(\mathcal{B})}}{\mathrm{Z}_0} 
    \frac{\delta}{\delta
      \mathbb{J}(\mathcal{B})(\mathbf{X})}\mathrm{Z}[J,\bar{J}]\right|_{J=\bar{J} =0}=
    \displaystyle \left. \frac{N^{-\alpha(\mathcal{B})}}{\mathrm{Z}_0} \prod_{i=1}^{k}\left(
    \frac{\delta}{\delta
      \bar{J}_{{\mathbf{p}^{i}}}}\frac{\delta}{\delta
      J_{\mathbf{x}^{i}}}\right)\mathrm{Z}[J,\bar{J}]\right|_{J=\bar{J}
      =0}\,,
\end{align}
where ${\mathbf{X}} = (\mathbf{x}^1, \hdots, \mathbf{x}^k) \in \{ \frac{1}{N},\frac{2}{N}, \hdots, 1 \}^{Dk}$ so that for all $c \in \{1,\hdots,D\}$ and $(i,j) \in \{ 1, \hdots, k \}^2$, $x_{c}^{i} \neq x_{c}^{j}$. Moreover  $Z_0=Z[0,0]$ and we note $\mathbb{J}(\mathcal{B})(\mathbf{X}) = J_{\mathbf{x}^1} \hdots J_{\mathbf{x}^k}\bar{J}_{\mathbf{p}^1} \hdots\bar{J}_{\mathbf{p}^k}$. 

The 
$\mathbf{p}^i \in \{ \frac{1}{N},\frac{2}{N}, \hdots, 1 \} ^D$ are momentum $D$-tuples depending on the coordinates $\mathbf{X}$ in a way constrained by the boundary graph $\B$. Hence the $2k$-point functions do not depend on the $\mathbf{p}^i$ but only on $\mathbf{X}$. For instance, for rank-$3$ tensors and for the boundary graph $V_1$ (see Fig. \ref{fig:boundary}), $\mathbb J(V_1)(\mathbf{x}^1,\mathbf{x}^2)=J_{\mathbf{x}^1} J_{\mathbf{x}^2}\bar{J}_{\mathbf{p}^1}
\bar{J}_{\mathbf{p}^2} 
=
J_{\mathbf{x}^1} J_{\mathbf{x}^2}\bar{J}_{x^1_1x^2_2x^2_3}\bar{J}_{x^2_1x^1_2x^1_3}$, where $\mathbf{p}^1=(x^1_1,x^2_2,x^2_3)$ and $\mathbf{p}^2=(x^2_1,x^1_2,x^1_3)$. In the following and for lower point functions, we will prefer the simplified notation $\mathbf{x},\mathbf{y},\mathbf{z}$ instead of $\mathbf{x}^1,\mathbf{x}^2,\mathbf{x}^3$. Let us note that white and black vertices in a boundary graph $\mathcal{B}$, correspond in $\mathbb J(\mathcal{B})$ to the sources $J$ and $\bar{J}$ respectively. 

We also introduce the scalings $\alpha (\mathcal{B})$ for each boundary graph $\mathcal{B}$, note that they do not depend on the choice of colouring of the respective graph $\mathcal{B}$. For example, $\alpha(V_1)=\alpha (V_2)=\alpha (V_3).$

The free energy is written as an expansion over boundary graphs (see again \cite{Perez-Sanchez:2016zbh} for more details):
\begin{equation} \label{eq:W_expansion}
    \mathrm{W}[J,\bar{J}] = \sum \limits_{k=1}^{\infty} \displaystyle\sum_{\substack{\mathcal{B}\in\partial_{\mathcal S_{\mathrm{int}}} \\
   V(\mathcal{B})=2k}} \sum \limits_{\mathbf{X}} \frac{N^{\alpha(\mathcal{B})}}{|\Aut(\mathcal{B})|}
    \mathrm{G}_{\mathcal{B}}^{(2k)}(\mathbf{X})\cdot\mathbb{J}(\mathcal{B})(\mathbf{X}),
\end{equation}
where $\partial_{\mathcal{S}_{\mathrm{int}}}$ is the set of boundary graphs associated to the interaction terms, $V(\mathcal{B})$ is the number of vertices of $\mathcal{B}$. Here $\Aut(\B)$ is the symmetry group of the graph $\mathcal{B}$,
which namely consists of all graph-automorphisms that preserve the bipartiteness 
in a strict sense --- black vertices are mapped to black vertices, white to white --- and respect the colour on edges. 
For instance $\Aut(V_a)=\Z_2$ and $\Aut(Q_a)=\Z_3$
are generated by rotations by angles of $\pi$ or 
$2\pi/3$, respectively
\cite[Def. 7 and examples]{Perez-Sanchez:2016zbh}.

\begin{table}
    \centering
    \includegraphics{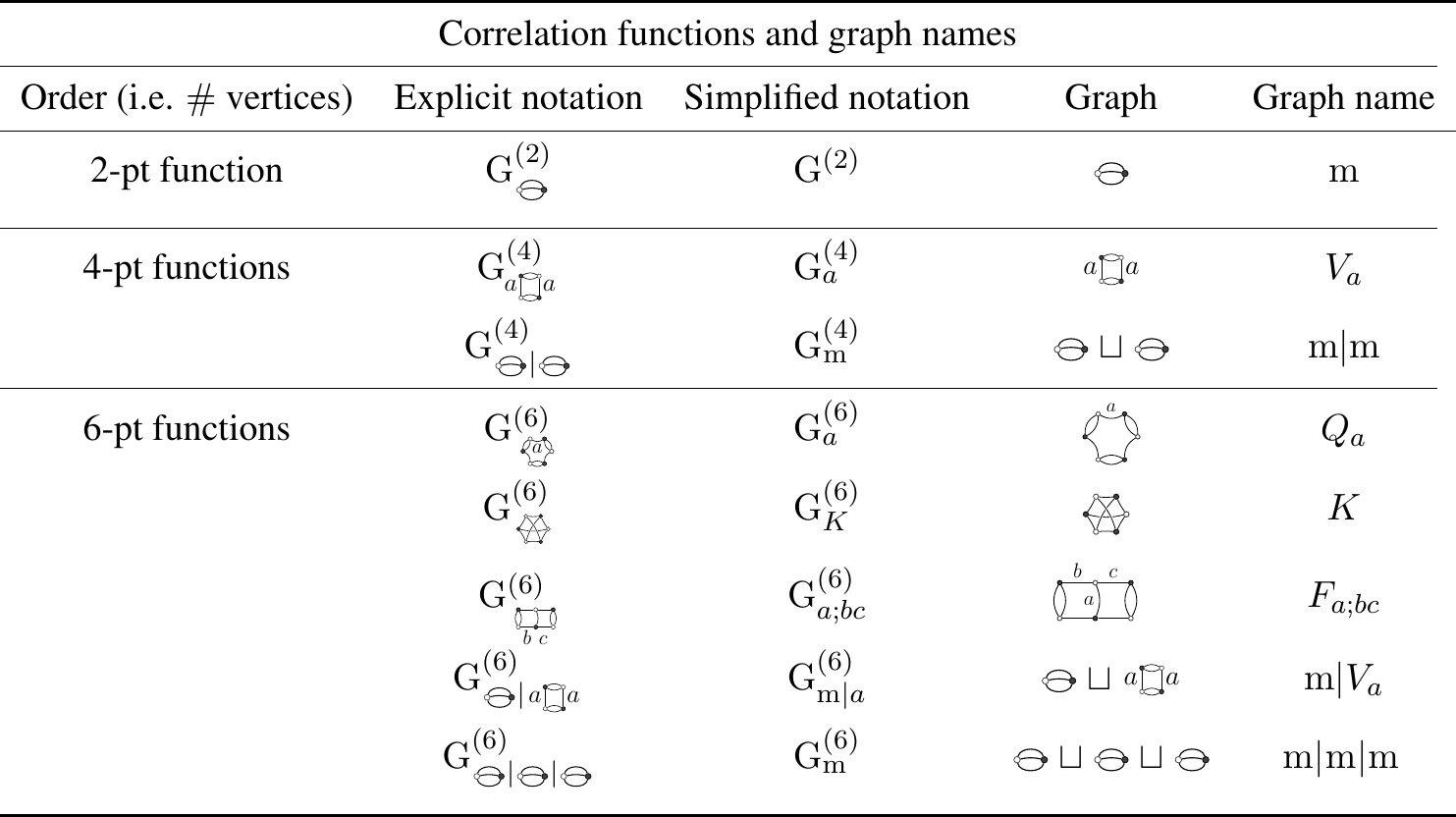}
    \caption{The adopted notation for the correlation
      functions in the tensor field theory \eqref{action} for $D=3$. In this table $a$ is any colour and $b$
      and $c$ are chosen such that  $\{a,b,c\}=\{1,2,3\}$.\label{fig:table_CR} }
    
\end{table}

Expansion \eqref{eq:W_expansion} is derived in \cite[Sec. 5.2]{Perez-Sanchez:2016zbh} and is a consequence 
of  the surjectivity 
of the boundary-map  (see \cite[Thm. 1]{Perez-Sanchez:2016zbh} and  \cite[Lem. 6]{surgery}) which associates to a Feynman graph its boundary graph:
\[\partial: \{\mbox{Feynman graphs of the $\varphi^4_{D,\mathrm{mel.}}$ rank-$D$ theory}\}
\to \{\mbox{possibly disconnected $D$-coloured graphs}\}\,,\]
where the $\varphi^4_{D,\mathrm{mel.}}$-theory refers to the interaction 
$\Sint=\sum_a V_a^D$, $V_a^D$ being the ``pillow'' interaction 
in rank $D$ (e.g. $V_a=  \raisebox{-.294\height}{\includegraphics[height=12pt]{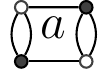}}$ for $D=3$, $
V_4 = \raisebox{-.24\height}{\includegraphics[height=14pt]{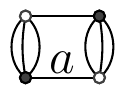}}$ if $D=4$, etc.). Hence every $G_\B\hp{2k}$ we introduced is non-trivial. A second reason to 
label correlation functions with boundary graphs is the fact \cite{GurauRyan}
that taking the boundary of a graph encodes, graph theoretically,
the operation of taking the boundary of a manifold.
Processes with different boundary geometries are therefore considered apart. 

Following \cite{Perez-Sanchez:2016zbh}, we use the WTI, which for rank-$D$ tensors writes:
\begin{equation}
    \sum \limits_{\mathbf{q}_{\hat{a}}} \frac{\delta \,
    \mathrm{Z}[J,\bar{J}]}{\delta J_{\mathbf{q}_{\hat{a}}m_a}\delta \bar{J}_{\mathbf{q}_{\hat{a}}n_a}} - \delta_{m_an_a}\mathrm{Y}_{m_a}^{(a)}[J,\bar{J}]\cdot\mathrm{Z}[J,\bar{J}]
    = \frac{N^{3\beta-2\gamma}}{m_a^2-n_a^2}\sum \limits_{\mathbf{q}_{\hat{a}}}
    \left(\bar{J}_{\mathbf{q}_{\hat{a}}m_a}\frac{\delta}{\delta\bar{J}_{\mathbf{q}_{\hat{a}}n_a}} -
    J_{\mathbf{q}_{\hat{a}}n_a}\frac{\delta}{\delta J_{\mathbf{q}_{\hat{a}}m_a}}\right) \mathrm{Z}[J,\bar{J}], 
\end{equation}
where  $\mathbf{q}_{\hat{a}} =  (q_1, \hdots,q_{a-1},q_{a+1}, \hdots,q_D)$,
and the $\mathrm{Y}$-term is given by
\begin{align}
    \mathrm{Y}_{m_a}^{(a)}[J,\bar{J}] &= \sum \limits_{\mathbf{q}_{\hat{a}}}\frac{\delta^2 \, \mathrm{W}[J,\bar{J}]}{\delta J_{\mathbf{q}_{\hat{a}}m_a} \delta \bar{J}_{\mathbf{q}_{\hat{a}}m_a}} \nonumber \\
    &= \sum_{\mathcal{B}\in\partial_{\mathcal S_{\mathrm{int}}}} \sum \limits_{\mathbf{X}} \mathfrak f_{\mathcal B, m_a}\hp{a}(\mathbf X)\cdot\mathbb{J}(\mathcal{B})(\mathbf{X}),
\end{align}
where $ \mathfrak f_{\mathcal B, m_a}\hp{a}(\mathbf X)$ is the function-coefficient of $\mathbb J (\mathcal B)(\mathbf{X})$ in the graph expansion of the $\mathrm{Y}$-term. 
The $ \mathfrak f\hp a_{B,m_a}$-functions can be computed using a graph algorithm detailed in \cite[Def. 10]{Perez-Sanchez:2016zbh}; since these 
functions were already computed there, this 
algorithm is omitted here and we shall present directly the functions later on.

Moreover, for a graph $\mathcal{B}$ with $2k$ vertices, we define
\begin{align} \label{eq:autom}
    \mathfrak f_a(\mathbf X;m_a; \mathcal B) &= \frac{\delta^{2k} 
    }{\delta \mathbb{J}(\mathcal{B})(\mathbf{X})}\mathrm{Y}_{m_a}^{(a)}[J,\bar{J}]\Bigg|_{J=\bar{J}=0} \nonumber \\
    &=\displaystyle \sum_{\hat \pi \in \mathrm{Aut}(\mathcal B)} 
    (\pi^* \mathfrak f_{\mathcal B, m_a}\hp{a})(\mathbf X) \,,
\end{align}
where $\pi$ is the restriction of the automorphism $\hat \pi$ to the
white vertices of $\mathcal B$, $(\pi^*f)(\mathbf{x}^1,\hdots,\mathbf{x}^k)= f(\mathbf{x}^{\pi^{-1}(1)},\hdots,\mathbf{x}^{\pi^{-1}(k)})$ and the equality between the first and second line was obtained in the Proposition 2.1 of \cite{Sanchez:2017gxt}.
For the pillow graphs $V_a$, equation \eqref{eq:autom} states that 
\begin{align} \label{eq:autom_explicit}
    \mathfrak f_a\big( \mathbf x, \mathbf y; m_a; V_a \big) & = \sum_{\hat \pi \in \mathbb Z_2} (\pi^* \mathfrak f_{V_a, m_a}\hp{a})\big(\mathbf x, \mathbf y\big) \nonumber \\
    &= \mathfrak f_{ V_a, m_a}\hp{a}\big(\mathbf x,\mathbf y \big) + \mathfrak f_{ V_a, m_a}\hp{a}\big(\mathbf y,\mathbf x \big).
\end{align}
Here we are only interested in the explicit coefficients of the graph expansion of the $\mathrm{Y}$-term up to order four in the sources, in the tensor field theory \eqref{action} for $D=3$. 
In the
following equations, $\{a,b,c\}=\{1,2,3\}$, an automatic reordering of the
entries by ascending sub-index is implied, and we omit the powers in $N$ associated to each Green's function.
One thus has:
\allowdisplaybreaks[1]
\begin{subequations}
\begin{align}
&\mathfrak f_{\mathrm{m}, s_a}\hp{a}(\mathbf{x}) = \mathrm{G}\hp{4}_a\big(\mathbf{x},s_a,x_b,x_c\big) + \sum_{c \neq a }\sum_{q_b}\mathrm{G}\hp{4}_c\big(\mathbf{x},s_a,q_b,x_c\big) +\sum_{q_b,q_c} \mathrm{G}_{\mathrm{m}} \hp 4\big(\mathbf{x}, s_a,q_b,q_c\big),  \\
&\mathfrak f_{ V_a, s_a}\hp{a}\big(\mathbf{x},\mathbf{y}\big)
=\frac13 \Big( \mathrm{G}_{a}\hp6\big(s_a,x_b,x_c,\mathbf{x},\mathbf{y}\big)+\mbox{cyclic perm.} \Big) + \frac13 \Big(
\mathrm{G}_{K}\hp6\big(s_a,x_b,y_c,\mathbf{x},\mathbf{y}\big) +
\mbox{cyclic perm.} \Big)\nonumber\\
&+ \sum_{q_b}\mathrm{G}_{b;ac}\hp 6 (\mathbf{x},\mathbf{y},s_a,q_b,y_c) + \displaystyle
\sum_{q_c}\mathrm{G}_{c;ab} \hp 6 (\mathbf{x},\mathbf{y},s_a,q_c,y_b)
+\frac12 \sum_{q_b,q_c}\mathrm{G}_{\mathrm{m}|a}\hp6
\big(s_a,q_b,q_c,\mathbf{x},\mathbf{y}\big), \\
&\mathfrak f_{ V_b, s_a}\hp{a}\big(\mathbf{x},\mathbf{y}\big)= 
\frac13\Big(\sum_{q_b} \mathrm{G}_{b}\hp6\big(s_a,q_b,y_c,\mathbf{x},\mathbf{y}\big) + \mbox{cyclic perm.} \Big)+ \mathrm{G}_{c;ab} \hp 6 \big(s_a,y_b,x_c ,\mathbf{x},\mathbf{y}\big) \nonumber\\
&+ \mathrm{G}_{c;ab} \hp 6 (\mathbf{x} ,s_a,x_b,x_c ,\mathbf{y})
+\sum_{q_b}\mathrm{G}_{a;bc} \hp 6 (\mathbf{x},\mathbf{y},s_a,q_b,y_c)
+\displaystyle\frac12 \sum_{q_b,q_c}\mathrm{G}_{\mathrm{m}|b}\hp6 \big(s_a,q_b,q_c,  \mathbf{x},\mathbf{y}\big), \\
&\mathfrak f_{ V_c, s_a}\hp{a}\big(\mathbf{x},\mathbf{y}\big) = \frac13\Big(\sum_{q_c} \mathrm{G}_{c}\hp6\big(s_a,y_b,q_c,\mathbf{x},\mathbf{y}\big) + \mbox{cyclic perm.} \Big)
+ \mathrm{G}_{c;ab} \hp 6 \big(s_a,y_b,x_c ,\mathbf{x},\mathbf{y}\big) \nonumber \\
&+ \mathrm{G}_{c;ab} \hp 6 (\mathbf{x}, s_a,x_b,x_c ,\mathbf{y}) +\sum_{q_c}\mathrm{G}_{a;bc} \hp 6 (\mathbf{x},s_a,x_b,q_c,\mathbf{y})+\displaystyle\frac12 \sum_{q_b,q_c}\mathrm{G}_{\mathrm{m}|c}\hp6 \big(s_a,q_b,q_c, \mathbf{x},\mathbf{y}\big), \\
&\mathfrak{f}_{\mathrm{m}|\mathrm{m}, s_a}^{(a)}
\Big(\mathbf{x},\mathbf{y}\Big) 
=  \Big( \sum_{q_b,q_c}
\mathrm{G}^{(6)}_{\mathrm{m}}\left(s_a,q_b,q_c,\mathbf{x},\mathbf{y}\right) + \text{cyclic perm.} \Big) + \mathrm{G}^{(6)}_{a;bc}\left(\mathbf{x},s_a,x_b,y_c,\mathbf{y}\right) \nonumber \\
&+ \mathrm{G}^{(6)}_{\mathrm{m}|a}\left(\mathbf{x},s_a,y_b,y_c,\mathbf{y}\right) + \sum \limits_{q_c} \mathrm{G}^{(6)}_{\mathrm{m}|b}\left(\mathbf{x},s_a,y_b,q_c,\mathbf{y}\right) + \sum \limits_{q_b} \mathrm{G}^{(6)}_{\mathrm{m}|c}\left(\mathbf{x},s_a,q_b,y_c,\mathbf{y}\right) \nonumber \\
&+ \sum \limits_{q_c} \mathrm{G}^{(6)}_{\mathrm{m}|b}\left(\mathbf{x},\mathbf{y},s_a,y_b,q_c\right) + \sum \limits_{q_b} \mathrm{G}^{(6)}_{\mathrm{m}|c}\left(\mathbf{x},\mathbf{y},s_a,q_b,y_c\right) + \mathrm{G}^{(6)}_{\mathrm{m}|a}\left(\mathbf{x},\mathbf{y},s_a,y_b,y_c\right). \label{f_m|m}
\end{align}
\end{subequations}
One should keep in mind that $s_a$ is an external data. Also notice
that the super-index $a$ breaks the symmetry between the colours.
These equations follow from the expansion of the $\mathrm{Y}$-term \cite[Eq. (48)]{Perez-Sanchez:2016zbh} and \cite[Lem. 4.1]{Sanchez:2017gxt}. 
Here `cyclic perm.' means the cyclic permutation in the 
3-tuples, e.g. `$F\big(s_a,x_b,x_c,\mathbf{x},\mathbf{y}\big)+ \mbox{cyclic perm.}$' abbreviates $ F\big(s_a,x_b,x_c,\mathbf{x},\mathbf{y}\big)+
F\big(\mathbf{y},s_a,x_b,x_c,\mathbf{x}\big)+
F\big(\mathbf{x},\mathbf{y},s_a,x_b,x_c\big)$.

In the rest of this paper we fix $D=3$.

\section{Constraints on the scalings in \texorpdfstring{$N$}{N}}
\label{constraints}
\subsection{\texorpdfstring{$2$}{2}-point function SDE}

In this subsection, we start with the explicit definition of the $2$-point function, and using the WTI to obtain SDE, we finally get a set of inequalities between the scaling coefficients $\alpha$, $\beta$, $\gamma$ and $\delta$.

The $2$-point function explicitly writes
\begin{align}
    &\mathrm{G}^{(2)}\left(\mathbf{x}\right) = \left. \frac{N^{-\alpha}}{\mathrm{Z}_0}
    \frac{\delta^2 \mathrm{Z}[J,\bar{J}]}{\delta
    \bar{J}_{\mathbf{x}}\delta
    J_{\mathbf{x}}}\right|_{J=\bar{J} =0} \\ 
    &= \left. \frac{N^{-\alpha}}{\mathrm{Z}_0} \frac{\delta^2 }{\delta
    J_{\mathbf{x}}\delta \bar{J}_{\mathbf{x}}} \exp{\left(-N^{\gamma}\Sint\left[\frac{1}{N^{2\beta-\gamma}}\frac{\delta}{\delta J},\frac{1}{N^{2\beta-\gamma}}\frac{\delta}{\delta\bar{J}}\right]\right)}\exp{\left(N^{2\beta-\gamma}\sum\limits_{\mathbf{x}}\frac{J_{\mathbf{x}}\bar{J}_{\mathbf{x}}}{|\mathbf{x}|^2}\right)}\right|_{J=\bar{J} =0} \nonumber \\ 
    &=\left. \frac{N^{2\beta-\gamma-\alpha}}{\mathrm{Z}_0} \frac{\delta }{\delta
    J_{\mathbf{x}}} \exp{\left(- N^{\gamma}\Sint^{\partial}\right)}
    \frac{J_{\mathbf{x}}}{|\mathbf{x}|^2} \exp{\left(N^{2\beta-\gamma}\sum\limits_{\mathbf{x}}\frac{J_{\mathbf{x}}\bar{J}_{\mathbf{x}}}{|\mathbf{x}|^2}\right)}\right|_{J=\bar{J}
    =0} \nonumber \\ 
    &= \frac{N^{2\beta-\gamma-\alpha}}{|\mathbf{x}|^2} + \left. \frac{N^{2\beta-\gamma-\alpha}}{\mathrm{Z}_0}\exp{\left(- N^{\gamma}\Sint^{\partial}\right)}\frac{J_{\mathbf{x}}}{|\mathbf{x}|^2}
    \frac{\delta }{\delta J_{\mathbf{x}}} \exp{\left(N^{2\beta-\gamma}\sum\limits_{\mathbf{x}}\frac{J_{\mathbf{x}}\bar{J}_{\mathbf{x}}}{|\mathbf{x}|^2}\right)}\right|_{J=\bar{J}
    =0} \nonumber \\
    &=\frac{N^{2\beta-\gamma-\alpha}}{|\mathbf{x}|^2}  - \displaystyle \left. \frac{N^{2\beta-\gamma-\alpha}}{\mathrm{Z}_0} \frac{N^{\gamma}}{|\mathbf{x}|^2}\displaystyle
    \left(\bar{\varphi}^{\mathbf{x}}\frac{\partial
    \Sint}{\partial \bar{\varphi}^{\mathbf{x}}}
    \right)\left[\frac{1}{N^{2\beta-\gamma}}\frac{\delta}{\delta J},\frac{1}{N^{2\beta-\gamma}}\frac{\delta}{\delta\bar{J}}\right]
    \mathrm{Z}[J,\bar{J}]\right|_{J=\bar{J} =0}, \nonumber
\end{align}
where we note $\mathrm{F}^{\partial}=\mathrm{F}\left[\frac{1}{N^{2\beta-\gamma}}\frac{\delta}{\delta J},\frac{1}{N^{2\beta-\gamma}}\frac{\delta}{\delta\bar{J}}\right]$. 
In order for the 
free propagator to be dominant in the large $N$ limit, we must fix:
\begin{equation}\label{rel1}
    \alpha = 2\beta - \gamma.
\end{equation}
To simplify the equations, we consider first the contribution of the pillow interaction $V_1$ and we then add the analogous contributions coming from the contributions of the pillow interactions $V_2$ and $V_3$.
One has: 
\begin{equation}
\label{intermediar}
    N^{\gamma}\left(\bar{\varphi}^{\mathbf{x}}\frac{\partial
      \Sint}{\partial \bar{\varphi}^{\mathbf{x}}}
    \right)^{\partial} = 2 \Tilde{\lambda}\frac{N^{5\gamma+\delta}}{N^{8\beta}}\sum
  \limits_{\mathbf{a}} \frac{\delta}{\delta
    J_{x_1x_2x_3}}\frac{\delta}{\delta\bar{J}_{a_1x_2x_3}}\frac{\delta}{\delta
    J_{a_1a_2a_3}}\frac{\delta}{\delta\bar{J}_{x_1a_2a_3}}.
\end{equation}
Using the WTI for the two rightmost derivatives in the expression \eqref{intermediar}
enables us to write:
\begin{align} \label{intWTI}
    & N^{\gamma}\left(\bar{\varphi}^{\mathbf{x}}\frac{\partial
      \Sint}{\partial \bar{\varphi}^{\mathbf{x}}}
    \right)^{\partial} \mathrm{Z}[J,\bar{J}]\bigg|_{J=\bar{J} =0}  =
  \frac{2 \tilde{\lambda}N^{5\gamma+\delta}}{N^{8\beta}}\frac{\delta}{\delta
    J_{\mathbf{x}}}\sum_{a_1}\frac{\delta}{\delta\bar{J}_{a_1x_2x_3}}
  \bigg\{\left( \delta_{x_1 a_1}
  \mathrm{Y}_{a_1}^{(1)}[J,\bar{J}]\right)\cdot\mathrm{Z}[J,\bar{J}]
  \nonumber\\ &+ \sum \limits_{a_2,a_3}
  \frac{N^{3\beta-2\gamma}}{a_1^2-x_1^2}\left(\bar{J}_{a_1a_2a_3}\frac{\delta}{\delta\bar{J}_{x_1a_2a_3}}
  - J_{x_1a_2a_3}\frac{\delta}{\delta
    J_{a_1a_2a_3}}\right) \mathrm{Z}[J,\bar{J}]\bigg\}\bigg|_{J=\bar{J} =0}.
\end{align}
Acting with the two remaining derivatives in \eqref{intermediar}
 on the second term on the RHS of \eqref{intWTI},
and using~\eqref{rel1}, we get:
\begin{equation}\label{difgreen}
  \frac{2\tilde{\lambda}}{N}\sum_{a_1} \frac{N^{2\gamma+\delta+1-3\beta}}{a_1^2-x_1^2}\left(\mathrm{G}^{(2)}(\mathbf{x})-\mathrm{G}^{(2)}(a_1,x_2,x_3)\right).
\end{equation}
For this term to give a well defined large $N$ limit we need the following relation:
\begin{equation}\label{rel2}
    3\beta \geq 2\gamma + \delta + 1.
\end{equation}
Note that if the inequality \eqref{rel2} is taken to be an equality, then the term~\eqref{difgreen} is a leading order term in the large $N$ limit. 

Acting with the remaining derivative on the factor $\mathrm{Z}[J,\bar{J}]$ of the first term of the RHS of \eqref{intWTI} 
gives:
\begin{equation}
   \frac{2\tilde{\lambda}}{N^{6\beta - 4\gamma-\delta}} \delta_{x_1 a_1} \mathrm{Y}_{a_1}^{(1)}[0,0]\mathrm{G}^{(2)}(\mathbf{x}) =  \frac{2\tilde{\lambda}N^{3\gamma + 2 + \delta - 4\beta}}{N^2}\sum \limits_{a_2,a_3} \mathrm{G}^{(2)}(x_1,a_2,a_3)\mathrm{G}^{(2)}(\mathbf{x}).
\end{equation}
This term implies a new inequality on the exponents: 
\begin{equation}\label{rel3}
4\beta \geq 3\gamma + \delta + 2.
\end{equation}
Acting now with these remaining derivatives on the factor $\mathrm{Y}_{a_1}^{(1)}[J,\bar{J}]$
of the first term of the RHS of \eqref{intWTI}
gives:
\begin{equation}
   \delta_{x_1 a_1}
   \left.\frac{\delta\,\mathrm{Y}_{a_1}^{(1)}[J,\bar{J}]}{\delta
     J_{\mathbf{x}}\delta\bar{J}_{a_1x_2x_3}}\right|_{J=\bar{J}
     =0} = N^{\alpha(V_1)}\mathrm{G}^{(4)}_{1}(\mathbf{x},\mathbf{x}) +\frac{N^{\alpha(\mathrm{m}|\mathrm{m})+2}}{N^2}\sum
   \limits_{a_2,a_3}\mathrm{G}^{(4)}_{\mathrm{m}}(\mathbf{x},x_1,a_2,a_3).
\end{equation}
Putting these terms together, we obtain 
the SDE for the 2-point function:
\begin{align}
\label{tower}
   &\mathrm{G}^{(2)}(\mathbf{x}) = \frac{1}{|\mathbf{x}|^2} - \frac{2\tilde{\lambda}}{|\mathbf{x}|^2} \Bigg(\frac{N^{3\gamma + 2 + \delta - 4\beta}}{N^2}\sum \limits_{a_2,a_3} \mathrm{G}^{(2)}(p_1,a_2,a_3) \mathrm{G}^{(2)}(\mathbf{x})+\frac{N^{\alpha(V_1)}}{N^{8\beta-5\gamma-\delta}}\mathrm{G}^{(4)}_1(\mathbf{x},\mathbf{x})
   \nonumber\\
   &  + \frac{N^{\alpha(\mathrm{m}|\mathrm{m})+2}}{N^{8\beta-5\gamma-\delta}}\frac{1}{N^2}\sum \limits_{a_2,a_3} \mathrm{G}^{(4)}_{\mathrm{m}}(x_1,a_2,a_3,\mathbf{x})
   + \frac{1}{N}\sum \limits_{a_1} \frac{N^{2\gamma+\delta+1-3\beta}}{x_1^2 - a_1^2} \left(\mathrm{G}^{(2)}(a_1,x_2,x_3)-\mathrm{G}^{(2)}(\mathbf{x})
  \right)\Bigg). \nonumber\\
\end{align}
For the $4$-point function to be sub-leading in the large $N$ limit taken in \eqref{tower} above,
we need to impose the following two inequalities on the exponents:
\begin{align}
    &\alpha(V_1) < 8\beta-5\gamma-\delta,\label{reldecoupling1} \\
    &\alpha(\mathrm{m}|\mathrm{m}) < 8\beta-5\gamma-\delta-2.\label{reldecoupling2}
\end{align}

As announced above, we now
add the contributions coming from the $2^{nd}$ and $3^{rd}$ pillow
interaction terms, $V_2$ and $V_3$, of the action.
We then get:
\begin{align}
   &\mathrm{G}^{(2)}(\mathbf{x}) = \frac{1}{|\mathbf{x}|^2} - \frac{2\tilde{\lambda}}{|\mathbf{x}|^2} \sum \limits_{a=1}^3 \Bigg(\frac{N^{3\gamma + 2 + \delta - 4\beta}}{N^2} \sum \limits_{\mathbf{q}_{\hat{a}}} \mathrm{G}^{(2)}(\mathbf{q}_{\hat{a}}x_a) \mathrm{G}^{(2)}(\mathbf{x}) +\frac{N^{\alpha(V_1)}}{N^{8\beta-5\gamma-\delta}}\mathrm{G}^{(4)}_a(\mathbf{x},\mathbf{x})
   \nonumber\\
   &  + \frac{N^{\alpha(\mathrm{m}|\mathrm{m})+2}}{N^{8\beta-5\gamma-\delta}}\frac{1}{N^2}\sum \limits_{\mathbf{q}_{\hat{a}}} \mathrm{G}^{(4)}_{\mathrm{m}}(\mathbf{q}_{\hat{a}}x_a,\mathbf{x})
   + \frac{1}{N}\sum \limits_{q_a} \frac{N^{2\gamma+\delta+1-3\beta}}{x_a^2 - q_a^2} \left(\mathrm{G}^{(2)}(\mathbf{x}_{\hat{a}}q_a)-\mathrm{G}^{(2)}(\mathbf{x})\right) \nonumber \\
   &+ \frac{N^{\alpha(V_1)+1}}{N^{8\beta-5\gamma-\delta}}\frac{1}{N}\sum\limits_{c\neq a} \sum \limits_{q_b}\mathrm{G}^{(4)}_{c}(\mathbf{x},\mathbf{x}_{\hat{b}}q_b) \Bigg),
\end{align}
where in the last term $b \neq c$ and $b \neq a$. 
This last term leads to a stronger condition than~\eqref{reldecoupling1}:
\begin{equation}
\alpha(V_1) < 8\beta-5\gamma-\delta-1. \label{reldecoupling3} 
\end{equation}
Moreover for $a_1 \neq x_1$ and using~\eqref{rel1}, the WTI implies
\begin{align}
&N^{5\beta-3\gamma}\frac{\mathrm{G}^{(2)}(a_1,x_2,x_3)-\mathrm{G}^{(2)}(\mathbf{x})}{x_1^2 - a_1^2} = N^{4\beta-2\gamma} \mathrm{G}^{(2)}(a_1,x_2,x_3)\mathrm{G}^{(2)}(\mathbf{x}) + \frac{N^{\alpha(V_1)+2}}{N^2}\sum \limits_{a_2,a_3} \mathrm{G}^{(4)}_{1}(\mathbf{a},\mathbf{x}) \nonumber\\
&+\frac{N^{\alpha(V_1)+1}}{N}\left(\sum
  \limits_{a_3}\mathrm{G}^{(4)}_{2}(\mathbf{x},a_1,x_2,a_3) + \sum \limits_{a_2}
  \mathrm{G}^{(4)}_{3}(\mathbf{x},a_1,a_2,x_3)\right).
\end{align}
This identity rewrites as
\begin{align}
    &\frac{\mathrm{G}^{(2)}(a_1,x_2,x_3)-\mathrm{G}^{(2)}(\mathbf{x})}{x_1^2 - a_1^2} = \frac{1}{N^{\beta-\gamma}} \mathrm{G}^{(2)}(a_1,x_2,x_3)\mathrm{G}^{(2)}(\mathbf{x}) + \frac{N^{\alpha(V_1)+2}}{N^{5\beta-3\gamma}}\frac{1}{N^2}\sum \limits_{a_2,a_3} \mathrm{G}^{(4)}_{1}(\mathbf{a},\mathbf{x}) \nonumber \\
    &+\frac{N^{\alpha(V_1)+1}}{N^{5\beta-3\gamma}}\left(\frac{1}{N}\sum
    \limits_{a_3}\mathrm{G}^{(4)}_{2}(\mathbf{x},a_1,x_2,a_3) +\frac{1}{N} \sum \limits_{a_2}\mathrm{G}^{(4)}_{3}(\mathbf{x},a_1,a_2,x_3)\right),
\end{align}
which implies the following two inequalities:
\begin{align}
    &\beta \geq \gamma, \label{b>=g}\\
    &\alpha(V_1) \leq 5\beta-3\gamma-2. \label{a1<}
\end{align}

\subsection{\texorpdfstring{$2k$}{2k}-point function SDE for connected boundary graphs}

In this subsection we start with the definition of the $2k$-point function, and as above, we use
WTI to obtain the SDE. This finally leads to a new inequality between the scaling coefficients.

\medskip

\begin{figure} 
\centering
\includegraphics[scale=1.3]{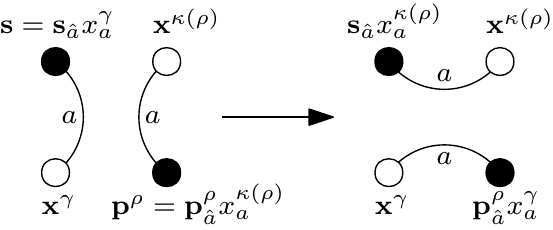}
\caption{Swapping of two $a$-coloured edges.
\label{fig:swap}}
\end{figure}

\begin{figure} 
\centering
\includegraphics[scale=1]{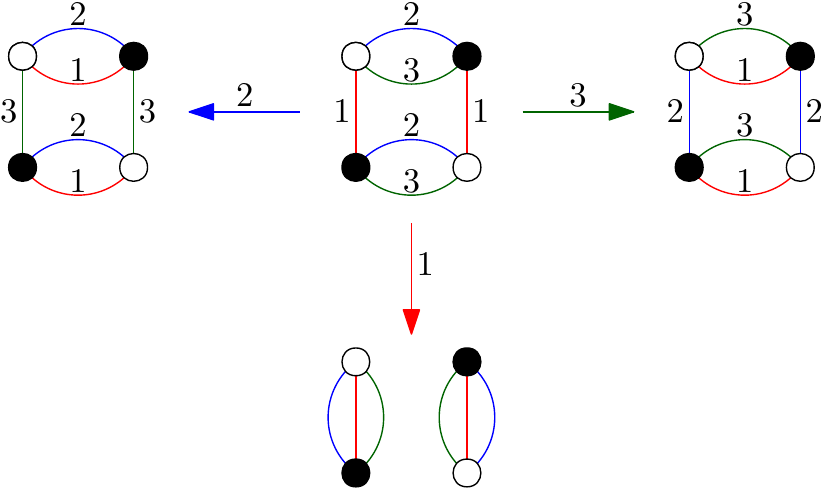}
\caption{Swapping of the three different colours edges, starting from the pillow graph $V_1$.
\label{fig:swapping}}
\end{figure}

From now on we consider altogether the contributions coming from the three pillow interactions $V_1$, $V_2$ and $V_3$. Let us start from the definition of a $2k$-point function with a connected boundary graph given in \eqref{def}.
Following \cite{Sanchez:2017gxt}, in order to obtain then SDE, we first consider the term:
\begin{align}\label{SDE}
    \frac{\delta \mathrm{W}[J,\bar{J}]}{\delta
    \bar{J}_{\mathbf{s}}} &=
    \frac{N^{2\beta-\gamma}}{\mathrm{Z}[J,\bar{J}]}\exp{\left(-N^{\gamma}\Sint^{\partial}\right)}   \frac{J_{\mathbf{s}}}{|\mathbf{s}|^2}\exp{\left(N^{2\beta-\gamma} \sum
    \limits_{\mathbf{a}}\frac{\bar{J}_{\mathbf{a}}J_{\mathbf{a}}}{|\mathbf{a}|^2} \right)} \nonumber \\ 
     &=  N^{2\beta-\gamma}\frac{J_{\mathbf{s}}}{|\mathbf{s}|^2} -\frac{1}{|\mathbf{s}|^2}\frac{N^{2\beta}}{\mathrm{Z}[J,\bar{J}]} \left(\frac{\delta \Sint}{\delta
      \bar{\varphi}^{\mathbf{s}}}\right)^{\partial}
    \mathrm{Z}[J,\bar{J}].
\end{align}
Note that here $\bf s$ is an unspecified vector of indices. The WTI enables us to write:
\begin{align}\label{SDEWTI}
    &N^{2\beta}\left(\frac{\delta \Sint}{\delta
      \bar{\varphi}^{\mathbf{s}}}\right)^{\partial} \mathrm{Z}[J,\bar{J}]
    = \frac{2\tilde{\lambda}N^{4\gamma+\delta}}{N^{6\beta}} \sum_{a=1}^3\sum_{b_a}
    \frac{\delta}{\delta\bar{J}_{\mathbf{s}_{\hat{a}}b_a}}\sum \limits_{\mathbf{b}_{\hat{a}}}\frac{\delta}{\delta
     J_{\mathbf{b}}}\frac{\delta}{\delta\bar{J}_{\mathbf{b}_{\hat{a}}s_a}}\mathrm{Z}[J,\bar{J}] \nonumber\\ 
     &= \frac{2\tilde{\lambda}N^{4\gamma+\delta}}{N^{6\beta}}\sum_{a=1}^3 \left(\frac{\delta \left(
     \mathrm{Y}_{\mathrm{s}_a}^{(a)}[J,\bar{J}]\cdot\mathrm{Z}[J,\bar{J}]
     \right)}{\delta \bar{J}_{\mathbf{s}}} + \sum_{b_a}
    \frac{N^{3\beta-2\gamma}}{b_a^2-s_a^2}
    \frac{\delta\mathrm{Z}[J,\bar{J}]}{\delta
     \bar{J}_{\mathbf{s}}} \right.\nonumber\\ 
     &\left. - N^{3\beta-2\gamma}\sum_{\mathbf{b}}
    \frac{J_{\mathbf{b}_{\hat{a}}s_a}}{b_a^2-s_a^2}
    \frac{\delta^2\mathrm{Z}[J,\bar{J}]}{\delta\bar{J}_{\mathbf{s}_{\hat{a}}b_a}\delta
    J_{\mathbf{b}}} + N^{3\beta-2\gamma}
    \sum_{\mathbf{b}}\frac{\bar{J}_{\mathbf{b}}}{b_a^2-s_a^2}\frac{\delta^2\mathrm{Z}[J,\bar{J}]}{\delta\bar{J}_{\mathbf{s}_{\hat{a}}b_a}\delta
    \bar{J}_{\mathbf{b}_{\hat{a}}s_a}} \right). 
\end{align}
For $\mathbf{s}=\mathbf{p}^1$, recalling that
\begin{equation}
    \mathrm{Y}_{p^1_a}^{(a)}[0,0] =
    N^{\alpha}\sum_{\mathbf{q}_{\hat{a}}}\mathrm{G}^{(2)}(\mathbf{q}_{\hat{a}}p^1_a),
\end{equation}
we apply the remaining $2k-1$ derivatives of \eqref{def} to~\eqref{SDEWTI}. This leads to the SDE for a $2k$-point function with a connected boundary graph:
\begin{align} \label{SDE2k}
    &\mathrm{G}^{(2k)}_{\mathcal{B}}(\mathbf{X}) = -\frac{2\Tilde{\lambda}}{|\mathbf{p}^1|^2}
    \sum_a \Bigg\{\frac{N^{3\gamma+2+\delta-4\beta}}{N^2}
    \sum_{\mathbf{q}_{\hat{a}}}\mathrm{G}^{(2)}(\mathbf{q}_{\hat{a}}p^1_a) \mathrm{G}^{(2k)}_{\mathcal{B}}(\mathbf{X}) +\frac{N^{4\gamma+\delta-6\beta}}{N^{\alpha(\mathcal{B})}}\mathfrak{f}_a\left(\mathbf{X};p^1_a;\mathcal{B}\right)
   \nonumber\\ 
    & + \frac{1}{N}\sum_{b_a}
    \frac{N^{2\gamma+\delta+1-3\beta}}{b_a^2-(x^{\gamma}_a)^2}\left(\mathrm{G}^{(2k)}_{\mathcal{B}}(\mathbf{X})-\mathrm{G}^{(2k)}_{\mathcal{B}}(\left.\mathbf{X}\right|_{x_a^{\gamma}\rightarrow
    b_a})\right)\nonumber \\
    &+ \frac{N^{2\gamma+\delta-3\beta}}{N^{\alpha(\mathcal{B})}}\sum_{\rho =2}^{k}\frac{1}{(p_a^{\rho})^2-(p_a^1)^2}\frac{1}{\mathrm{Z}_0}\Bigg[\frac{\partial\mathrm{Z}[J,\bar{J}]}{\partial\zeta_a(\mathcal{B};1,\rho)}(\mathbf{X})-\frac{\partial\mathrm{Z}[J,\bar{J}]}{\partial\zeta_a(\mathcal{B};1,\rho)}(\left.\mathbf{X}\right|_{x_a^{\gamma}\rightarrow
    p^{\rho}_a})\Bigg] \Bigg\},
\end{align}
where $\mathbf{x}^{\gamma}$ corresponds to the only white vertex such that $x^{\gamma}_{a}=s_a$ and $\zeta_a(\mathcal{B};1,\rho)$ is the graph obtained by swapping the a-coloured lines between $\mathbf{p}^1$ and $\mathbf{p}^{\rho}$ in a graph $\mathcal{B}$ (see figure \ref{fig:swap}). Similarly to $\mathbf{x}^{\gamma}$, we note $\mathbf{x}^{\kappa(\rho)}$ the only white vertex such that $x^{\kappa(\rho)}_{a}=p^{\rho}_a$. An explicit example of this operation is given in figure \ref{fig:swapping}. Starting from the pillow graph $V_1$, swapping edges of colour $2$ and resp. $3$ gives the graphs $V_3$ and resp. $V_2$; however swapping edges of colour $1$ gives the disconnected graph $\mathrm{m}|\mathrm{m}$.

The first term of the RHS of \eqref{SDE2k} gives a well defined large $N$ limit if~\eqref{rel3} is satisfied.
The terms of the second line  of \eqref{SDE2k} 
require~\eqref{rel2}.
The terms contributing to $\mathfrak{f}_a\left(\mathbf{X};p^1_a;\mathcal{B}\right)$ for $V(\mathcal{B})=2k$ are $2(k+1)$-point functions with at most two sums on dummy variables. 
Hence to get a well defined large $N$ limit we need:
\begin{equation}\label{genrel}
\alpha(\mathcal{B}) \geq \alpha(\mathcal{B}')+2+4\gamma+\delta-6\beta,
\end{equation}
with $V(\mathcal{B}')=2k+2$. If the inequality is strict, the $2(k+1)$-point function terms in the SDE for the $2k$-point function are sub-leading and the tower of SDE decouples at leading order.

\medskip


For the $4$-point function and for $\mathbf{s} = (x_1,y_2,y_3)$ , the general equation \eqref{SDE2k} gives
\begin{align}\label{SDE4}
    &\mathrm{G}^{(4)}_1(\mathbf{x},\mathbf{y}) =-\frac{2\tilde{\lambda}}{|\mathbf{s}|^2}\left\{\frac{N^{4\gamma+\delta-6\beta}}{N^{\alpha(V_1)}}\sum_{a=1}^3
  \mathfrak{f}_a(\mathbf{x},\mathbf{y};s_a; V_a)
  +\frac{N^{3\gamma+2+\delta-4\beta}}{N^2}\sum_{a=1}^3\sum_{\mathbf{q}_{\hat{a}}}\mathrm{G}^{(2)}(\mathbf{q}_{\hat{a}}s_a)\mathrm{G}^{(4)}_1(\mathbf{x},\mathbf{y}) \right. \nonumber 
  \\ &+\frac{1}{N} \sum_{b_1}
  \frac{N^{2\gamma+\delta+1-3\beta}}{b_1^2-x_1^2}\left(\mathrm{G}^{(4)}_1(\mathbf{x},\mathbf{y}) -\mathrm{G}^{(4)}_1(b_1,x_2,x_3,\mathbf{y})\right) \nonumber 
  \\ &+\frac{N^{2\gamma+\delta-3\beta}}{y_2^2-x_2^2}\left(\mathrm{G}^{(4)}_3(\mathbf{x},y_1,x_2,y_3)-\mathrm{G}^{(4)}_3(\mathbf{x},\mathbf{y})\right)\nonumber 
  \\ &+ \frac{1}{N} \sum_{b_2}
  \frac{N^{2\gamma+\delta+1-3\beta}}{b_2^2-y_2^2}\left(\mathrm{G}^{(4)}_1(\mathbf{x},\mathbf{y}) -
  \mathrm{G}^{(4)}_1(\mathbf{x},y_1,b_2,y_3)\right) \nonumber 
  \\ &+ \frac{N^{2\gamma+\delta-3\beta}}{y_3^2-x_3^2}\left(\mathrm{G}^{(4)}_2(\mathbf{x},y_1,y_2,x_3)-
  \mathrm{G}^{(4)}_2(\mathbf{x},\mathbf{y})\right) \nonumber
  \\ &+\frac{1}{N} \sum_{b_3}
  \frac{N^{2\gamma+\delta+1-3\beta}}{b_3^2-y_3^2}\left(\mathrm{G}^{(4)}_1(\mathbf{x},\mathbf{y}) -
  \mathrm{G}^{(4)}_1(\mathbf{x},y_1,y_2,b_3)\right)  \nonumber  
  \\ & +\frac{N^{2\gamma+\delta-3\beta}}{y_1^2-x_1^2}\frac{N^{2\alpha}}{N^{\alpha(V_1)}}
  \mathrm{G}^{(2)}(\mathbf{y})\left(\mathrm{G}^{(2)}(\mathbf{x})-\mathrm{G}^{(2)}(y_1,x_2,x_3)\right)\nonumber  
  \\ &+\frac{N^{2\gamma+\delta-3\beta}}{y_1^2-x_1^2}\frac{N^{\alpha(\mathrm{m}|\mathrm{m})}}{N^{\alpha(V_1)}}\left(\mathrm{G}^{(4)}_{\mathrm{m}}(\mathbf{x},\mathbf{y})-\mathrm{G}^{(4)}_{\mathrm{m}}(x_1,x_2,x_3,\mathbf{y})\right) \Bigg\}.
\end{align}

Using~\eqref{rel1}, 
the eighth term in \eqref{SDE4} leads to 
\begin{equation}
\alpha(V_1) \geq \beta+\delta, \label{a1>=}
\end{equation}
The last term of \eqref{SDE4} leads to
\begin{equation}
\alpha(V_1) \geq \alpha(\mathrm{m}|\mathrm{m}) + 2\gamma+\delta-3\beta.
\end{equation}
Moreover the fourth and sixth terms imply
\begin{equation}
    3\beta \geq 2\gamma + \delta,
\end{equation}
which must be a strict inequality to be consistent with~\eqref{rel2}. Hence these two terms are sub-leading in the large $N$ limit.
\\
\\
\noindent
\textit{Remark.} 
The colour-$a$ edge swapping $\zeta_a$ appeared naturally in 
\cite{Sanchez:2017gxt} while describing the SDE. Leaving the restriction to boundary graphs, in general, 
when $\zeta_a$ is applied 
to a connected coloured graph (i.e. when it is a unary operation), $\zeta_a$ is known as  \textit{flip}\footnote{C.I.P.S thanks Paola Cristofori for 
her explanation of this terminology in the GEM and crystallisation contexts.} and this terminology traces back to the theory of graph encoded manifolds (GEM)  \cite{linsmulazzani} or, equivalently, the crystallisation of manifolds (see \cite[Def. 25]{PaolaRita} as well). 
Flips and \textit{blobs} (in the tensor model context known as \textit{melonic insertion})
are two fundamental operations in the sense that coloured 
graphs representing the same piece-wise manifold might differ only by a finite sequence of flips and blobs. 
The binary version of $\zeta_a (\B_1\sqcup \B_2;w,v)$ (when the
argument is a two-component graph and the two vertices $v$ and $w$ are in different components) has been explored in \cite{surgery}
and in the tensor model context represents the graph
theoretical connected sum\footnote{The virtue of this 
connected sum is being a binary 
operation on the set of Feynman diagrams of a tensor model $\Sint$ , $\zeta_{a=0}: \mathrm{Feyn}(\Sint)\times\mathrm{Feyn}(\Sint)\to \mathrm{Feyn}(\Sint)$ (by 
preserving the interaction vertices that would be 
destroyed by the old connected sum of the ``crystallisation theory'' that consist of the removal of two graph-vertices).
 This is seen from the fact that propagators are 
 represented by the $0$ colour; therefore $\zeta_{a=0}$
 only swaps two ends of two propagators inside
 a Feynman graph, leaving untouched the interaction vertices 
 of the model $\Sint$ in question.}.

\section{The \texorpdfstring{$4$}{4}-point function SDE with disconnected boundary graph}
\label{disco4pt}

In this section, we apply the same approach for the $4$-point function SDE with disconnected boundary graph. As already mentioned in the introduction, this case was not considered in \cite{Sanchez:2017gxt}.

The $4$-point function with a disconnected boundary graph writes
\begin{equation}
    \mathrm{G}^{(4)}_{\mathrm{m}}(\mathbf{x},\mathbf{y}) =\frac{1}{N^{\alpha(\mathrm{m}|\mathrm{m})}}
    \left. \frac{\delta^4
      \mathrm{W}[J,\bar{J}]}{\delta\bar{J}_{\mathbf{y}}\delta
      J_{\mathbf{y}}\delta\bar{J}_{\mathbf{x}} \delta
      J_{\mathbf{x}}}\right|_{J=\bar{J}=0},
\end{equation}
Let us start from~\eqref{SDE} with $\mathbf{s}=\mathbf{x}$ and where we applied the three remaining derivatives
\begin{equation}
    \frac{\delta^4 \mathrm{W}[J,\bar{J}]}{\delta\bar{J}_{\mathbf{y}}\delta
    J_{\mathbf{y}}\delta\bar{J}_{\mathbf{x}} \delta
    J_{\mathbf{x}}}=  -\frac{N^{2\beta}}{|\mathbf{x}|^2}\frac{\delta^2}{\delta\bar{J}_{\mathbf{y}}\delta
    J_{\mathbf{y}}}\left(\frac{1}{\mathrm{Z}[J,\bar{J}]} \frac{\delta}{\delta J_{\mathbf{x}} }\left(\frac{\delta \Sint}{\delta\bar{\varphi}^{\mathbf{x}}}\right)^{\partial}
    \mathrm{Z}[J,\bar{J}]\right).
\end{equation}
For a connected boundary graph, all the derivatives give a vanishing contribution when applied to $\frac{1}{\mathrm{Z}[J,\bar{J}]}$. 
For the disconnected boundary graph case we treat here, one has: 
\begin{align}\label{SDEdisc}
    \frac{\delta^4 \mathrm{W}[J,\bar{J}]}{\delta\bar{J}_{\mathbf{y}}\delta
    J_{\mathbf{y}}\delta\bar{J}_{\mathbf{x}} \delta
    J_{\mathbf{x}}} &= -\frac{N^{2\beta}}{|\mathbf{x}|^2}\frac{1}{\mathrm{Z}[J,\bar{J}]} \frac{\delta^3}{\delta J_{\mathbf{x}}\delta\bar{J}_{\mathbf{y}}\delta
    J_{\mathbf{y}}}\left(\frac{\delta \Sint}{\delta\bar{\varphi}^{\mathbf{x}}}\right)^{\partial}
    \mathrm{Z}[J,\bar{J}] \nonumber \\
    &+ \frac{N^{2\beta}}{|\mathbf{x}|^2}\frac{1}{\mathrm{Z}^2[J,\bar{J}]}\frac{\delta^2\mathrm{Z}[J,\bar{J}]}{\delta\bar{J}_{\mathbf{y}}\delta
    J_{\mathbf{y}}} \frac{\delta}{\delta J_{\mathbf{x}} }\left(\frac{\delta \Sint}{\delta\bar{\varphi}^{\mathbf{x}}}\right)^{\partial}
    \mathrm{Z}[J,\bar{J}].
\end{align}
The first line is the same as in the case of a connected boundary graph, the second line is a new type of term.
As above, the WTI leads to the first term below:
\begin{align}
    &\frac{1}{\mathrm{Z}_0}\left.\frac{\delta^4 \left(
     \mathrm{Y}_{x^1_a}^{(a)}[J,\bar{J}]\cdot\mathrm{Z}[J,\bar{J}]
     \right)
    }{\delta\bar{J}_{\mathbf{y}}\delta
    J_{\mathbf{y}}\delta\bar{J}_{\mathbf{x}} \delta
    J_{\mathbf{x}}}\right|_{J=\bar{J}=0}
    = N^{\alpha}
    \sum_{\mathbf{q}_{\hat{a}}}\mathrm{G}^{(2)}(\mathbf{q}_{\hat{a}}x_a)\left(N^{\alpha(\mathrm{m}|\mathrm{m})}\mathrm{G}^{(4)}_{\mathrm{m}}(\mathbf{x},\mathbf{y}) + N^{2\alpha}
    \mathrm{G}^{(2)}(\mathbf{x})\mathrm{G}^{(2)}(\mathbf{y}) \right) \nonumber\\ 
    &+N^{\alpha}\mathrm{G}^{(2)}(\mathbf{x})
    \displaystyle \left.\frac{\delta^2
    \mathrm{Y}_{x_a}^{(a)}[J,\bar{J}]}{\delta\bar{J}_{\mathbf{y}}\delta
    J_{\mathbf{y}}}\right|_{J=\bar{J}=0}+ N^{\alpha}\mathrm{G}^{(2)}(\mathbf{y})
    \displaystyle \left.\frac{\delta^2
    \mathrm{Y}_{x_a}^{(a)}[J,\bar{J}]}{\delta\bar{J}_{\mathbf{x}}\delta
    J_{\mathbf{x}}}\right|_{J=\bar{J}=0} + \left.\frac{\delta^4
    \mathrm{Y}_{x_a}^{(a)}[J,\bar{J}]}{\delta
    (\mathrm{m}|\mathrm{m})}\right|_{J=\bar{J}=0},
\end{align}
where
\begin{align}
    \left.\frac{\delta^4 \mathrm{Y}_{x_a}^{(a)}[J,\bar{J}]}{\delta
      (\mathrm{m}|\mathrm{m})}\right|_{J=\bar{J}=0} &= \textstyle
    \mathfrak{f}_{\mathrm{m}|\mathrm{m},
      x_a}^{(a)}\left(\mathbf{x},\mathbf{y}\right) + \mathfrak{f}_{\mathrm{m}|\mathrm{m},
      x_a}^{(a)}\left(\mathbf{y},\mathbf{x}\right),
    \\ \left.\frac{\delta^2
      \mathrm{Y}_{x_a}^{(a)}[J,\bar{J}]}{\delta\bar{J}_{\mathbf{x}}\delta
      J_{\mathbf{x}}}\right|_{J=\bar{J}=0} &= \textstyle
    \mathfrak{f}_{\mathrm{m},
      x_a}^{(a)}\left(\mathbf{x}\right),
    \\ \left.\frac{\delta^2
      \mathrm{Y}_{x_a}^{(a)}[J,\bar{J}]}{\delta\bar{J}_{\mathbf{y}}\delta
      J_{\mathbf{y}}}\right|_{J=\bar{J}=0} &= \textstyle
    \mathfrak{f}_{\mathrm{m},
      x_a}^{(a)}\left(\mathbf{y}\right).
\end{align}
This term corresponds to the first term in \eqref{SDEWTI}.
We also need to compute the contribution from the swapping (the term corresponding to the last term of \eqref{SDEWTI}). This writes
\begin{equation}
    \frac{1}{\mathrm{Z}_0}\left.\frac{\partial \mathrm{Z}[J,\bar{J}]}{\partial \zeta_a(\mathrm{m}|\mathrm{m},1,2)(\mathbf{x},\mathbf{y})}\right|_{J=\bar{J} =0} =N^{\alpha(V_1)}\mathrm{G}^{(4)}_a(\mathbf{x},\mathbf{y}),
\end{equation}
Finally, the contribution from the two remaining terms of~\eqref{SDEWTI} writes
\begin{align}
\label{intermediar2}
    &-N^{3\beta-2\gamma}\sum_{\mathbf{b}}\frac{\delta^3}{\delta\bar{J}_{\mathbf{y}}\delta
    J_{\mathbf{y}}\delta J_{\mathbf{x}}}\left( 
    \frac{J_{\mathbf{b}_{\hat{a}}x_a}}{b_a^2-x_a^2}
    \frac{\delta^2\mathrm{Z}[J,\bar{J}]}{\delta\bar{J}_{\mathbf{x}_{\hat{a}}b_a}\delta
    J_{\mathbf{b}}}\right)\bigg|_{J=\bar{J}=0}= \nonumber \\
    &-\frac{1}{N}\sum_{b_a}\frac{N^{3\beta-2\gamma+1}}{b_a^2-x_a^2}\left(N^{\alpha(\mathrm{m}|\mathrm{m})}\mathrm{G}^{(4)}_{\mathrm{m}}(\mathbf{x}_{\hat{a}}b_a,\mathbf{y}) \right. +N^{2\alpha}\left.\mathrm{G}^{(2)}(\mathbf{x}_{\hat{a}}b_a)\mathrm{G}^{(2)}(\mathbf{y}) + N^{\alpha(a)}\delta^{b_a}_{y_a}\mathrm{G}^{(4)}_a(\mathbf{x}_{\hat{a}}y_a,\mathbf{y})\right), \\  
    &\sum_{b_a}
    \frac{N^{3\beta-2\gamma}}{b_a^2-x_a^2}\frac{1}{\mathrm{Z}_0}\left.\frac{\delta^4
   \mathrm{Z}[J,\bar{J}]}{\delta\bar{J}_{\mathbf{y}}\delta
      J_{\mathbf{y}}\delta
     \bar{J}_{\mathbf{x}_{\hat{a}}b_a} \delta
      J_{\mathbf{x}}}\right|_{J=\bar{J}=0} 
    \nonumber \\
    &=\frac{1}{N}\sum_{b_a}\frac{N^{3\beta-2\gamma+1}}{b_a^2-x_a^2}\left(N^{\alpha(\mathrm{m}|\mathrm{m})}\mathrm{G}^{(4)}_{\mathrm{m}}(\mathbf{x},\mathbf{y}) + N^{2\alpha} \mathrm{G}^{(2)}(\mathbf{x})\mathrm{G}^{(2)}(\mathbf{y}) \right).
\end{align}
Let us note here that in \eqref{intermediar2}, we obtain not only a contribution coming from the
disconnected $4$-point function, but also a supplementary contribution
as a
product of $2$-point functions. These products of $2$-point functions and the term
\begin{equation}
    \mathrm{G}^{(2)}(\mathbf{y})
    \displaystyle\frac{\delta^2
    \mathrm{Y}_{x_a}^{(a)}[J,\bar{J}]}{\delta\bar{J}_{\mathbf{x}}\delta
    J_{\mathbf{x}}},
\end{equation}
give rise to disconnected Feynman graphs because the dependence in momenta factorises. They should not appear in a connected Green's function, hence they need to be compensated. 

They will be cancelled by the term coming from the second line of~\eqref{SDEdisc}. 
This will give us new relations on the exponents of $N$. Noting that
\begin{equation}
    \frac{\delta}{\delta J_{\mathbf{x}} }\left(\frac{\delta \Sint}{\delta\bar{\varphi}^{\mathbf{x}}}\right)^{\partial}
    \mathrm{Z}[J,\bar{J}] = \left(\bar{\varphi}^{\mathbf{x}}\frac{\partial
      \Sint}{\partial \bar{\varphi}^{\mathbf{x}}}
    \right)^{\partial} \mathrm{Z}[J,\bar{J}],
\end{equation}
we already have computed these terms in the SDE for the $2$-point function. Indeed, all the terms proportional to $\tilde{\lambda}$ in the SDE for the $2$-point function are multiplied by
\begin{equation}
    -\frac{N^{2\beta-\gamma}}{N^{\alpha(\mathrm{m}|\mathrm{m})}}\frac{\delta^2\mathrm{Z}[J,\bar{J}]}{\delta\bar{J}_{\mathbf{y}}\delta
    J_{\mathbf{y}}} 
\end{equation}
to obtain the contribution from the second line of~\eqref{SDEdisc} in the SDE for the $4$-point function with a disconnected boundary graph. This writes:
\begin{align}
    &\frac{2\tilde{\lambda}}{|\mathbf{x}|^2} \frac{\mathrm{G}^{(2)}(\mathbf{y})}{N^{\alpha(\mathrm{m}|\mathrm{m})}}\sum \limits_{a=1}^3\Bigg(\frac{N^{3\gamma + 2 + \delta - 4\beta}}{N^2} \sum \limits_{\mathbf{q}_{\hat{a}}} \mathrm{G}^{(2)}(\mathbf{q}_{\hat{a}}x_a) \mathrm{G}^{(2)}(\mathbf{x}) + \frac{\mathfrak{f}_{\mathrm{m},
      x_a}^{(a)}\left(\mathbf{x}\right)}{N^{8\beta-5\gamma-\delta}} \\
   &+ \frac{1}{N}\sum \limits_{q_a} \frac{N^{2\gamma+\delta+1-3\beta}}{(x_a)^2 - q_a^2} \left(\mathrm{G}^{(2)}(\mathbf{x}_{\hat{a}}q_a)-\mathrm{G}^{(2)}(\mathbf{x})\right) \nonumber \Bigg).
\end{align}
Collecting all the terms above and again making use of~\eqref{rel1}, we get
\begin{align}
    &\mathrm{G}^{(4)}_{\mathrm{m}}(\mathbf{x},\mathbf{y}) = -\frac{2\tilde{\lambda}}{|\mathbf{x}|^2}  \sum_{a=1}^3 \Bigg\{\frac{1}{N^2}\sum_{\mathbf{q}_{\hat{a}}}\mathrm{G}^{(2)}(\mathbf{q}_{\hat{a}}x_a)\left(\frac{\mathrm{G}^{(4)}_{\mathrm{m}}(\mathbf{x},\mathbf{y})}{N^{4\beta-3\gamma-\delta-2}} + \frac{N^{\gamma + \delta + 2}}{N^{\alpha(\mathrm{m}|\mathrm{m})}}
    \mathrm{G}^{(2)}(\mathbf{x})\mathrm{G}^{(2)}(\mathbf{y}) \right) \nonumber\\ 
    &-\frac{1}{N}\sum_{q_a}\frac{1}{q_a^2-x_a^2}\left(N^{2\gamma+\delta-3\beta+1}\mathrm{G}^{(4)}_{\mathrm{m}}(\mathbf{x}_{\hat{a}}q_a,\mathbf{y}) + \displaystyle \frac{N^{\beta+\delta+1}}{N^{\alpha(\mathrm{m}|\mathrm{m})}}
    \mathrm{G}^{(2)}(\mathbf{x}_{\hat{a}}q_a)\mathrm{G}^{(2)}(\mathbf{x}) \right) \nonumber \\
    &+\frac{1}{N} \sum_{q_a}\frac{1}{q_a^2-x_a^2}\left(N^{2\gamma+\delta-3\beta+1}\mathrm{G}^{(4)}_{\mathrm{m}}(\mathbf{x},\mathbf{y}) + \displaystyle \frac{N^{\beta+\delta+1}}{N^{\alpha(\mathrm{m}|\mathrm{m})}}
    \mathrm{G}^{(2)}(\mathbf{x})\mathrm{G}^{(2)}(\mathbf{y}) \right) \nonumber \\ 
    &+\frac{N^{\alpha(V_1)+2\gamma+\delta-3\beta}}{N^{\alpha(\mathrm{m}|\mathrm{m})}}
    \frac{1}{y_a^2-x_a^2}\left(\mathrm{G}^{(4)}_a(\mathbf{x},\mathbf{y})-\mathrm{G}^{(4)}_a(\mathbf{x}_{\hat{a}}y_a,\mathbf{x}^2)\right)+ \frac{N^{3\gamma+\delta-4\beta}}{N^{\alpha(\mathrm{m}|\mathrm{m})}}\mathrm{G}^{(2)}(\mathbf{x})\mathfrak{f}_{\mathrm{m},x_a}^{(a)}\left(\mathbf{y}\right)\nonumber \\ 
    &+ \frac{N^{4\gamma+\delta-6\beta}}{N^{\alpha(\mathrm{m}|\mathrm{m})}} \left(\mathfrak{f}_{\mathrm{m}|\mathrm{m},
    x^1_a}^{(a)}\left(\mathbf{x},\mathbf{y}\right)+\mathfrak{f}_{\mathrm{m}|\mathrm{m},
    x^1_a}^{(a)}\left(\mathbf{y},\mathbf{x}\right)\right)
    + \frac{N^{3\gamma+\delta-4\beta}}{N^{\alpha(\mathrm{m}|\mathrm{m})}}\mathrm{G}^{(2)}(\mathbf{y}) \mathfrak{f}_{\mathrm{m},
    x_a}^{(a)}\left(\mathbf{x}\right) \nonumber \\
    &- \frac{N^{5\gamma+\delta-8\beta}}{N^{\alpha(\mathrm{m}|\mathrm{m})}}\mathrm{G}^{(2)}(\mathbf{y})\mathfrak{f}^{(a)}_{\mathrm{m},x_a}(\mathbf{x})  - \frac{N^{3\gamma + 2 + \delta - 4\beta}}{N^{\alpha(\mathrm{m}|\mathrm{m})}}\frac{\mathrm{G}^{(2)}(\mathbf{y})}{N^2} \sum \limits_{\mathbf{q}_{\hat{a}}} \mathrm{G}^{(2)}(\mathbf{q}_{\hat{a}}x_a) \mathrm{G}^{(2)}(\mathbf{x}) \nonumber \\
    &- \frac{N^{2\gamma+\delta+1-3\beta}}{N^{\alpha(\mathrm{m}|\mathrm{m})}}\frac{\mathrm{G}^{(2)}(\mathbf{y})}{N}\sum \limits_{q_a} \frac{1}{x_a^2-q_a^2} \left(\mathrm{G}^{(2)}(\mathbf{x}_{\hat{a}}q_a)-\mathrm{G}^{(2)}(\mathbf{x})\right) \Bigg\}.
\end{align}
Let us determine the conditions on the exponents for the disconnected term to be cancelled. We have the following three identities:
\begin{align}
    \frac{N^{3\gamma+\delta-4\beta}}{N^{\alpha(\mathrm{m}|\mathrm{m})}}G^{(2)}(\mathbf{y}) \mathfrak{f}_{\mathrm{m},
    x_a}^{(a)}\left(\mathbf{x}\right) &= \frac{N^{5\gamma+\delta-8\beta}}{N^{\alpha(\mathrm{m}|\mathrm{m})}}G^{(2)}(\mathbf{y})\mathfrak{f}^{(a)}_{\mathrm{m},x_a}(\mathbf{x}),  \\
    \frac{N^{\gamma + \delta + 2}}{N^{\alpha(\mathrm{m}|\mathrm{m})}}
    G^{(2)}(\mathbf{x})G^{(2)}(\mathbf{y}) &= \frac{N^{3\gamma + 2 + \delta - 4\beta}}{N^{\alpha(\mathrm{m}|\mathrm{m})}} G^{(2)}(\mathbf{x}) G^{(2)}(\mathbf{y}), \\
    \frac{N^{\beta+\delta+1}}{N^{\alpha(\mathrm{m}|\mathrm{m})}}G^{(2)}(\mathbf{y}) \frac{G^{(2)}(\mathbf{x}_{\hat{a}}q_a)-G^{(2)}(\mathbf{x})}{x_a^2-q_a^2} &= \frac{N^{2\gamma+\delta+1-3\beta}}{N^{\alpha(\mathrm{m}|\mathrm{m})}}G^{(2)}(\mathbf{y}) \frac{G^{(2)}(\mathbf{x}_{\hat{a}}q_a)-G^{(2)}(\mathbf{x})}{x_a^2-q_a^2}.
\end{align}
Each of these identities leads to the condition:
\begin{equation}
\label{ineg}
    2\beta = \gamma.
\end{equation}
The SDE for the $4$-point function with a disconnected boundary graph then writes:
\begin{align}\label{SDEmm}
    &\mathrm{G}^{(4)}_{\mathrm{m}}(\mathbf{x},\mathbf{y}) = -\frac{2\tilde{\lambda}}{|\mathbf{x}|^2}  \sum_{a=1}^3 \Bigg\{\frac{1}{N^2}\sum_{\mathbf{q}_{\hat{a}}}\mathrm{G}^{(2)}(\mathbf{q}_{\hat{a}}x_a)\left(\frac{\mathrm{G}^{(4)}_{\mathrm{m}}(\mathbf{x},\mathbf{y})}{N^{4\beta-3\gamma-\delta-2}} \right)  \nonumber \\
    &+ \frac{N^{4\gamma+\delta-6\beta}}{N^{\alpha(\mathrm{m}|\mathrm{m})}} \left(\mathfrak{f}_{\mathrm{m}|\mathrm{m},
    x_a}^{(a)}\left(\mathbf{x},\mathbf{y}\right) + \mathfrak{f}_{\mathrm{m}|\mathrm{m},
    x_a}^{(a)}\left(\mathbf{y},\mathbf{x}\right)\right)\nonumber\\
    &+\frac{1}{N} \sum_{q_a}\frac{N^{2\gamma+\delta-3\beta+1}}{x_a^2-q_a^2}\left(\mathrm{G}^{(4)}_{\mathrm{m}}(\mathbf{x}_{\hat{a}}q_a,\mathbf{y})-\mathrm{G}^{(4)}_{\mathrm{m}}(\mathbf{x},\mathbf{y}) \right) \nonumber \\
    &+\frac{N^{\alpha(V_1)+2\gamma+\delta-3\beta}}{N^{\alpha(\mathrm{m}|\mathrm{m})}}
    \frac{1}{y_a^2-x_a^2}\left(\mathrm{G}^{(4)}_a(\mathbf{x},\mathbf{y})-\mathrm{G}^{(4)}_a(\mathbf{x}_{\hat{a}}y_a,\mathbf{y})\right) + \frac{N^{3\gamma+\delta-4\beta}}{N^{\alpha(\mathrm{m}|\mathrm{m})}}\mathrm{G}^{(2)}(\mathbf{x})\mathfrak{f}_{\mathrm{m},x_a}^{(a)}\left(\mathbf{y}\right)\Bigg\}.
\end{align}
The first term of the RHS requires again~\eqref{rel3}; the third term gives again~\eqref{rel2}. Then, the fourth term gives the relation: 
\begin{equation}
     \alpha(\mathrm{m}|\mathrm{m}) \geq \alpha(V_1) + 2\gamma + \delta -3\beta.
\end{equation}
To obtain relations on the exponents from the last term we need the following expression
\begin{align}
    \mathfrak f_{\mathrm{m}, x_a}^{(a)}(\mathbf{y})
    &= N^{\alpha(V_1)}\mathrm{G}^{(4)}_a(\mathbf{y};x_a,y_b,y_c) +\frac{ N^{\alpha(V_1)+1}}{N} \sum\limits_{c\neq a}\sum\limits_{q_b}\mathrm{G}^{(4)}_c(\mathbf{y};x_a,q_b,y_c) \nonumber \\
    &+ \frac{ N^{\alpha(\mathrm{m}|\mathrm{m})+2}}{N^2}\sum\limits_{q_b,q_c}\mathrm{G}^{(4)}_{\mathrm{m}}(\mathbf{y};x_a,q_b,q_c).
\end{align}
From the first term of this equation we recover the same relation between $\alpha(\mathrm{m}|\mathrm{m})$ and $\alpha(V_1)$ as above, but we also have a stronger condition from the second term. This condition writes:
\begin{equation}
\alpha(\mathrm{m}|\mathrm{m}) \geq \alpha(V_1) + 2\gamma + \delta -3\beta+1,
\end{equation}
which becomes an equality if one wants the second order graphs in the perturbation expansion (which are the lowest order graphs) to be leading order. The last term requires again~\eqref{rel3}. Finally, the terms in $ \mathfrak{f}_{\mathrm{m}|\mathrm{m},x_a}^{(a)}$ give the same type of relations as~\eqref{genrel}.

\section{The SDE in the large \texorpdfstring{$N$}{N} limit}
\label{largeN}
In this section we find appropriate scalings which allow us to obtain a well defined SDE in the large $N$ limit.

\subsection{\texorpdfstring{$2$}{2}- and \texorpdfstring{$4$}{4}-point functions}

Using \eqref{rel1} and
\eqref{ineg} one has:
\begin{equation}
\alpha = 0. \\
\end{equation}
In the large $N$ limit, we need the $2$-point function SDE to have the following form:
\begin{equation}
    \mathrm{G}^{(2)}(\mathbf{x}) = \frac{1}{|\mathbf{x}|^2} - \frac{2\tilde{\lambda}}{|\mathbf{x}|^2} \sum \limits_{a=1}^3\sum \limits_{\mathbf{q}_{\hat{a}}} \mathrm{G}^{(2)}(\mathbf{q}_{\hat{a}}x_a) \mathrm{G}^{(2)}(\mathbf{x}).
\end{equation}
We need $4\beta = 3\gamma + \delta + 2$. Using \eqref{rel2}, we get:
\begin{align}
    &\delta = -2-2\beta,\label{delta}\\
    &\beta > -1.
\end{align}
The relations \eqref{b>=g} and \eqref{ineg} between $\beta$ and $\gamma$ lead to:
\begin{equation}
    0 > \beta > \gamma, \,\, \text{or} \,\, \beta = \gamma = 0.
\end{equation}
From the inequalities \eqref{a1<} and \eqref{a1>=} on $\alpha(V_1)$, we get:
\begin{equation}
\label{interm2}
    \alpha(V_1) = -2-\beta.
\end{equation}

From now on we chose $\beta=\gamma=0$.
Note that we could chose $0 > \beta > -1$. However, this would change the value of the exponents $\alpha(\mathcal{B})$ but would give the same SDE. 
Equations \eqref{delta} and \eqref{interm2} thus become:
\begin{equation}
\alpha(V_1) = -2 = \delta.
\end{equation}

Assuming that $\alpha(V_1) >  \alpha(\mathrm{m}|\mathrm{m})$ leads to:
\begin{equation}
    -2 > \alpha(\mathrm{m}|\mathrm{m}) \geq -3,
\end{equation}
When choosing
\begin{equation}
     \alpha(\mathrm{m}|\mathrm{m}) = -3.
\end{equation}
we have a well defined large $N$ limit.
Moreover, we can see that in general we need that $\alpha(\mathcal{B})$ decreases strictly with the number of points of the Green function and the number of connected components of $\mathcal{B}$. Hence at this point we can conjecture that
\begin{equation}
\alpha(\mathcal{B}) = 3-B-2k,
\end{equation}
where $2k$ is the number of vertices of $\mathcal{B}$, $B$ is its number of connected components. 

With the scalings above, 
the SDE in the large $N$ limit writes
\begin{align}
    &\mathrm{G}^{(2)}(\mathbf{x}) = \left(|\mathbf{x}|^2 + 2\tilde{\lambda}\sum \limits_{a=1}^3\int\mathrm{d}\mathbf{q}_{\hat{a}} \mathrm{G}^{(2)}(\mathbf{q}_{\hat{a}}x_a) \right)^{-1}, \label{SDE2} \\
    &\mathrm{G}^{(4)}_1(\mathbf{x},\mathbf{y}) = -2\tilde{\lambda} \mathrm{G}^{(2)}(x_1,y_2,y_3)\mathrm{G}^{(2)}(\mathbf{y})\frac{\mathrm{G}^{(2)}(\mathbf{x})-\mathrm{G}^{(2)}(y_1,x_2,x_3)}{y_1^2-x_1^2}, \\
    &\mathrm{G}^{(4)}_{\mathrm{m}}(\mathbf{x},\mathbf{y}) =- 2\tilde{\lambda}(\mathrm{G}^{(2)}(\mathbf{x}))^2\sum_{a=1}^3\Bigg\{\sum\limits_{c\neq a}\int\mathrm{d}q_b\mathrm{G}^{(4)}_c(x_a,q_b,y_c,\mathbf{y}) + \int\mathrm{d}\mathbf{q}_{\hat{a}}\mathrm{G}^{(4)}_{\mathrm{m}}(\mathbf{q}_{\hat{a}}x_a,\mathbf{y})\Bigg\},
\end{align}
where we used the SDE for the 2-point function to rewrite the SDE for the 4-point functions and where $\mathrm{d}\mathbf{q}_{\hat{a}} = \mathrm{d}q_b\mathrm{d}q_c $ for $a \neq b,c$. 

\subsection{Higher-point functions}

Let us now look at the SDE for the higher-point functions with a connected boundary graph in the large $N$ limit, and in particular to the 6-point functions. 

From~\eqref{SDE2k}, we get 
\begin{align}
     &\mathrm{G}^{(2k)}_{\mathcal{B}}(\mathbf{X}) = -\frac{2\Tilde{\lambda}}{|\mathbf{s}|^2}
    \sum_{a=1}^3 \Bigg\{
    \int \mathrm{d}\mathbf{q}_{\hat{a}}\mathrm{G}^{(2)}(\mathbf{q}_{\hat{a}}s_a) \mathrm{G}^{(2k)}_{\mathcal{B}}(\mathbf{X}) 
   \nonumber\\ 
    &+ N^{-\alpha(\mathcal{B})-2}\sum_{\rho =2}^{k}\frac{1}{(p_a^{\rho})^2-s_a^2}\frac{1}{\mathrm{Z}_0}\Bigg[\frac{\partial\mathrm{Z}[J,\bar{J}]}{\partial\zeta_a(\mathcal{B};1,\rho)}(\mathbf{X})-\frac{\partial\mathrm{Z}[J,\bar{J}]}{\partial\zeta_a(\mathcal{B};1,\rho)}(\left.\mathbf{X}\right|_{x_a^{\gamma}\rightarrow
    p^{\rho}_a})\Bigg] \Bigg\}.
\end{align}
Let us analyse the large $N$ limit of this equation. 
The first term in the RHS is always present in the large $N$ limit, but the terms coming from the swappings can be of leading order or be neglected. Indeed, a swapping can add at most one more connected component (see figure \ref{fig:swapping}), then the second term of the RHS can give differences of three type of terms: $N^{\alpha(m|\mathcal{B}')}\mathrm{G}^{(2k)}_{m|\mathcal{B}'}$, $ N^{\alpha(\mathcal{B}')}\mathrm{G}^{(2)}\mathrm{G}^{(2(k-1))}_{\mathcal{B}'}$ and $N^{\alpha(\mathcal{B}'')}\mathrm{G}^{(2k)}_{\mathcal{B}''}$. The first type of term is neglected, since in the large $N$ limit, we took $\alpha(m|\mathcal{B}')=\alpha(\mathcal{B}')-1$ and $\alpha(\mathcal{B}') = \alpha(\mathcal{B}) + 2 $, hence $\alpha(m|\mathcal{B}')-\alpha(\mathcal{B})-2 = - 1 $. However, the second 
type of term 
is of leading order since $\alpha(\mathcal{B}') - \alpha(\mathcal{B}) - 2 = 0 $. 
Let us now analyse the last term, which is more involved. 
From the study of the $4$-point functions 
one could think that $\alpha(\mathcal{B}) = \alpha(\mathcal{B}'')$ for all connected boundary graphs $\mathcal{B}$ and $\mathcal{B}''$ with $2k$ vertices.
Nevertheless, this does not hold.
This follows from the analysis of the $6$-point functions
and in particular of $\mathrm{G}^{(6)}_{K}$. 
In fact, applying the swapping procedure to $K$ can only give 
$F_{a;bc}$ for $\{a,b,c\} = \{1,2,3\}$,
which has six vertices.
Hence if we take $\alpha(K)=\alpha(F_{a;bc})$ and from the previous discussion, we get, for $\mathbf{s} = (x_1,y_2,z_3)$, the SDE
\begin{equation}
    \mathrm{G}^{(6)}_{K}(\mathbf{x},\mathbf{y},\mathbf{z})=-\frac{2\Tilde{\lambda}}{x_1^2+y_2^2+z_3^2}
    \sum_{a=1}^3
    \int \mathrm{d}\mathbf{q}_{\hat{a}}\mathrm{G}^{(2)}(\mathbf{q}_{\hat{a}}s_a) \mathrm{G}^{(6)}_{K}(\mathbf{x},\mathbf{y},\mathbf{z}).
\end{equation}
However, this equation does not give any information on the $6$-point function $\mathrm{G}^{(6)}_{K}$. This implies that we need to define $\alpha(K)$ such that the terms in $\mathrm{G}^{(6)}_{a;bc}$ are also of leading order. We thus need to have the following scaling:
\begin{equation}
\alpha(K) = \alpha(F_{a;bc})-2.
\end{equation}
This gives the following SDE, for $\mathbf{s}=(x_1,y_2,z_3)$ and where we used equation \eqref{SDE2}:
\begin{align}
    &\mathrm{G}^{(6)}_{K}(\mathbf{x},\mathbf{y},\mathbf{z})=-2\Tilde{\lambda}\mathrm{G}^{(2)}(x_1,y_2,z_3)\Bigg\{\frac{\mathrm{G}^{(6)}_{{1;23}}(\mathbf{x},\mathbf{z},\mathbf{y})-\mathrm{G}^{(6)}_{{1;23}}(y_1,x_2,x_3,\mathbf{z},\mathbf{y})}{y_1^2-x_1^2} \nonumber \\
    &+ \frac{\mathrm{G}^{(6)}_{{1;23}}(\mathbf{z},\mathbf{x},\mathbf{y})-\mathrm{G}^{(6)}_{{1;23}}(\mathbf{z},\mathbf{y},z_1,x_2,x_3)}{z_1^2-x_1^2} + \frac{\mathrm{G}^{(6)}_{{2;13}}(\mathbf{z},\mathbf{x},\mathbf{y})-\mathrm{G}^{(6)}_{{2;13}}(\mathbf{z},\mathbf{x},y_1,z_2,y_3)}{z_2^2-y_2^2} \nonumber \\
    &+ \frac{\mathrm{G}^{(6)}_{{1;23}}(\mathbf{y},\mathbf{z},\mathbf{x})-\mathrm{G}^{(6)}_{{1;23}}(y_1,x_2,y_3,\mathbf{z},\mathbf{x})}{x_2^2-y_2^2} + \frac{\mathrm{G}^{(6)}_{{3;12}}(\mathbf{z},\mathbf{y},\mathbf{x})-\mathrm{G}^{(6)}_{{2;13}}(z_1,z_2,x_3\mathbf{z},\mathbf{x})}{x_3^2-z_3^2} \nonumber \\ 
    &+ \frac{\mathrm{G}^{(6)}_{{3;12}}(\mathbf{y},\mathbf{x},\mathbf{z})-\mathrm{G}^{(6)}_{{1;23}}(\mathbf{y},\mathbf{x},z_1,z_2,y_3)}{y_3^2-z_3^2}  \Bigg\}. 
\end{align}
Note that this could be expected because $K$ is the first non-planar graph which appears in our analysis.
Moreover, in the large $N$ limit and using \eqref{SDE2}, the SDE for the other $6$-point functions with connected boundary graphs (see table \ref{fig:table_CR}) are
\begin{align}
    &\mathrm{G}^{(6)}_{1}(\mathbf{x},\mathbf{y},\mathbf{z})=-2\Tilde{\lambda}\mathrm{G}^{(2)}(x_1,y_2,y_3)\Bigg\{
    \mathrm{G}^{(2)}(\mathbf{y})\frac{\mathrm{G}^{(4)}_{1}(\mathbf{x},\mathbf{z})-\mathrm{G}^{(4)}_{1}(y_1,x_2,x_3,\mathbf{z})}{y_1^2-x_1^2} \nonumber \\
    &+ \mathrm{G}^{(4)}_{1}(\mathbf{y},\mathbf{z})\frac{\mathrm{G}^{(2)}(\mathbf{x})-\mathrm{G}^{(2)}(z_1,x_2,x_3)}{z_1^2-x_1^2} \Bigg\}, 
\end{align}
for $\mathbf{s}=(x_1,y_2,y_3)$, and 
\begin{equation}
    \mathrm{G}^{(6)}_{{1;23}}(\mathbf{x},\mathbf{y},\mathbf{z})=-2\Tilde{\lambda}\mathrm{G}^{(2)}(x_1,y_2,x_3)\mathrm{G}^{(2)}(\mathbf{x})\frac{\mathrm{G}^{(4)}_{3}(\mathbf{y},\mathbf{z})-\mathrm{G}^{(4)}_{3}(y_1,x_2,y_3,\mathbf{z})}{x_2^2-y_2^2}, 
\end{equation}
for $\mathbf{s}=(x_1,y_2,x_3)$. 

We can see that 
all these equations are algebraic. For a connected boundary graph of degree zero, the SDE depends only on lower-point functions with a connected boundary graph. However, the $K$ SDE  depends only on the other $6$-point functions and the $2$-point function. 

Finally, from the previous discussions, we can conjecture a general formula for the scaling
\begin{equation}
\label{conjectura}
\alpha(\mathcal{B}) = 3-B-2g-2k,
\end{equation}
where $2k$ is the number of vertices of $\mathcal{B}$, $B$ its number of connected components and $g$ its genus. Note that, since we deal in this paper with rank three tensors, for boundary graphs (where one colour is lost) the degree is the genus \cite{gs} and \cite{book}. 

\section{Concluding remarks}

In this paper we have used the WTI to study the large $N$ limit of SDE of tensor field theory. This allowed us to obtain explicit values for the scalings of the various terms appearing in the action of the model studied here.

\medskip

The first perspectives of this work are the proof of the conjecture \eqref{conjectura} and the generalisation of our results for the case when any boundary graph can be disconnected \cite{Perez-Sanchez:2018qkd}.

A second perspective is to solve the SDE in the large $N$ limit. One could initially tackle this hard task using numerical methods, as it was done in \cite{Samary:2014oya} for a $\phi^4_5$ just renormalizable tensor model.
Another way to tackle this, is to use the analytic method implemented in 
\cite{Panzer:2018tvy}
for non-commutative quantum field theory - see \cite{pascalie_solvable_2020}.



A third perspective appears to us to be the implementation of the analytic methods used in this paper for the study of SYK-like tensor models such as the ones of \cite{Witten} and \cite{Klebanov:2016xxf}. The main difficulty here would come from the fact that one would then need to take into consideration an additional time coordinate (SYK models being $(0+1)$-dimensional models, and not $0$-dimensional models such as the model studied in this paper).

\section*{Acknowledgement}
The authors thank Thomas Krajewski for helpful discussions throughout this project.
R. Pascalie and A. Tanasa are partially supported by the CNRS Infiniti ModTens grant. The research of 
R. Wulkenhaar and C.I. Perez-Sanchez are supported by the Deutsche 
Forschungsgemeinschaft,SFB 878 (Mathematical Institute, University of M\"unster, Germany). C.I.P-S. thanks moreover the Faculty of Physics, Astronomy and Applied Computer Science,  Jagiellonian University (Cracow, Poland) for hospitality and acknowledges the Short-Term Scientific Mission program of the COST Action MP 1405 for this mobility opportunity.
Towards the end of this work, one of the authors (C.I.P.S.) has been supported by the TEAM programme of the Foundation for Polish Science co-financed by the European Union under the
European Regional Development Fund (POIR.04.04.00-00-5C55/17-00). 
The authors acknowledge the remarks of the referee that helped 
to improve this article.

\appendix 
\renewcommand{\thesection}{\Alph{section}} 


\section{Perturbative expansion}

In this appendix, we perform a perturbative check of the SDE up to second order of the coupling constant, before and after taking the large $N$ limit. For simplicity, we do not write the powers in $N$ in the equations.

\subsection*{\texorpdfstring{$2$}{2}-point function}

The SDE for the $2$-point function is
\begin{align}
   &\mathrm{G}^{(2)}(\mathbf{x}) = \frac{1}{|\mathbf{x}|^2} - \frac{2\lambda}{|\mathbf{x}|^2} \sum \limits_{c=1}^3\Bigg(\sum \limits_{\mathbf{q}_{\hat{c}}} \mathrm{G}^{(2)}(\mathbf{q}_{\hat{c}}x_c) \mathrm{G}^{(2)}(\mathbf{x}) +\mathrm{G}^{(4)}_c(\mathbf{x},\mathbf{x})
  \\
   &  + \sum \limits_{\mathbf{q}_{\hat{c}}} \mathrm{G}^{(4)}_{\mathrm{m}}(\mathbf{q}_{\hat{c}}x_c,\mathbf{x})
   + \sum \limits_{q_c} \frac{1}{x_c^2 - q_c^2} \left(\mathrm{G}^{(2)}(\mathbf{x}_{\hat{c}}q_c)-\mathrm{G}^{(2)}(\mathbf{x})\right) + \sum\limits_{d \neq c} \sum \limits_{q_c}\mathrm{G}^{(4)}_{d}(\mathbf{x},\mathbf{x}_{\hat{c}}q_c) \Bigg),  \nonumber
\end{align}
Let us look at the perturbative equation up to $2^{nd}$ order in the coupling constant. We can first remark that the term with $\lambda\mathrm{G}^{(4)}_{\mathrm{m}}$ will only start contributing at order $\lambda^3$. The other terms give
\begin{align}
    &-\frac{2\lambda}{|\mathbf{x}|^2} \sum \limits_{c=1}^3\sum \limits_{\mathbf{q}_{\hat{c}}} \mathrm{G}^{(2)}(\mathbf{q}_{\hat{c}}x_c) \mathrm{G}^{(2)}(\mathbf{x}) = 2\sum \limits_{c=1}^3
    \includegraphics[scale=0.25]{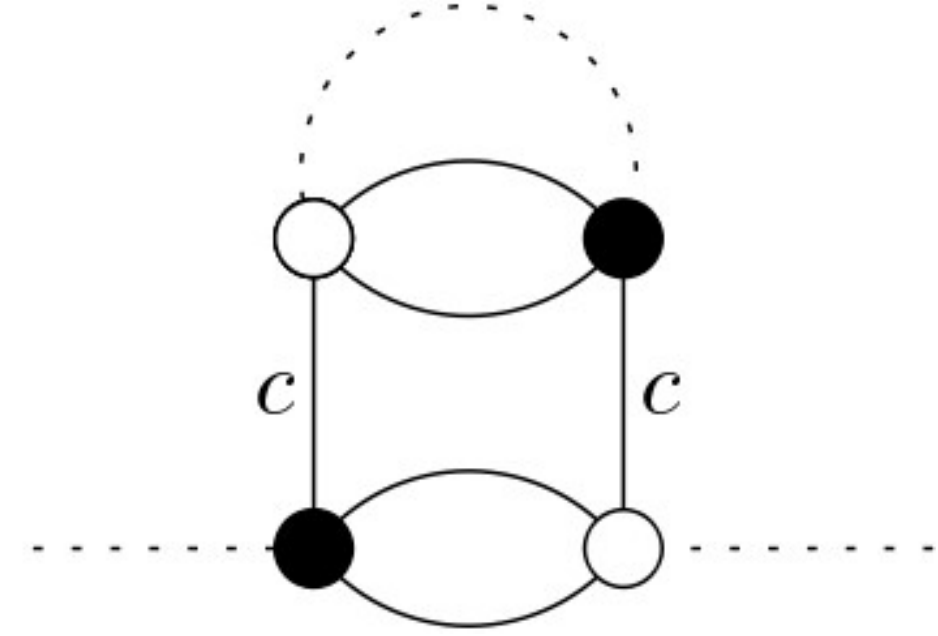} \\
    &+ 4\sum \limits_{c=1}^3\sum \limits_{d=1}^3 \Bigg(\includegraphics[scale=0.25]{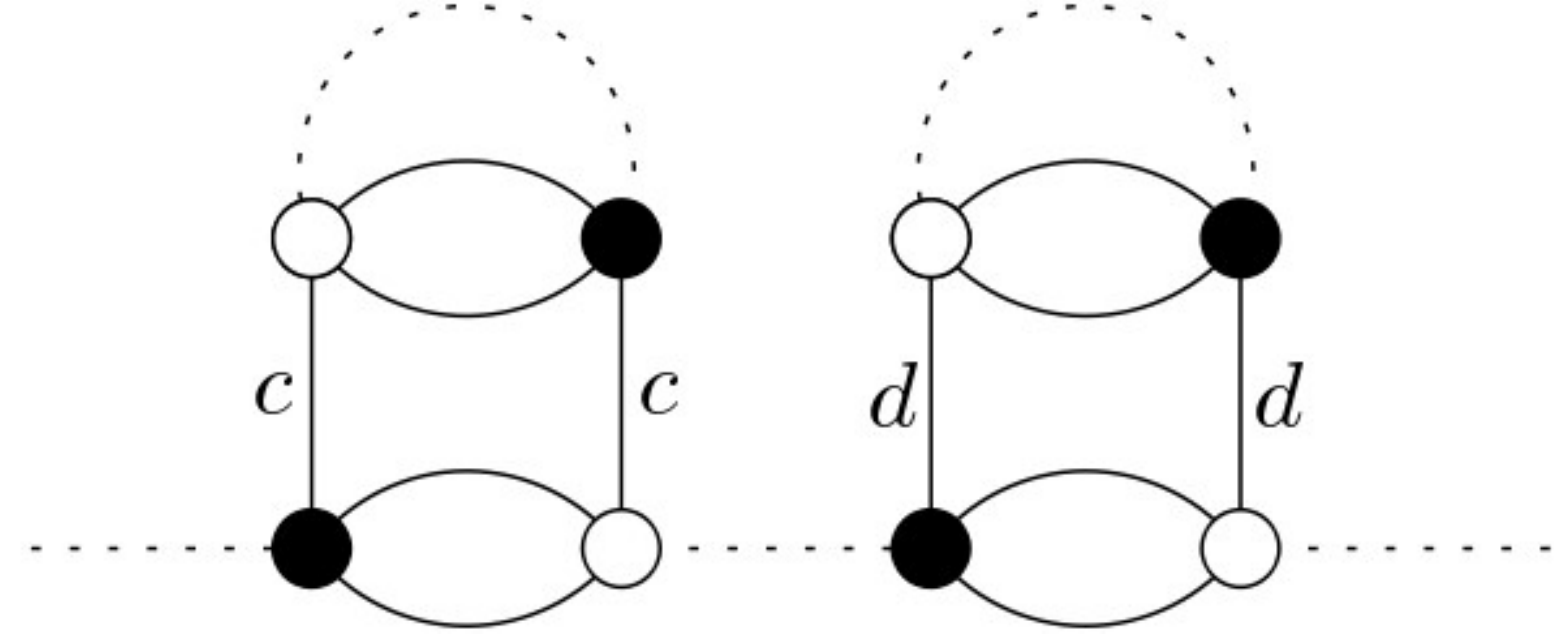} + \includegraphics[scale=0.25]{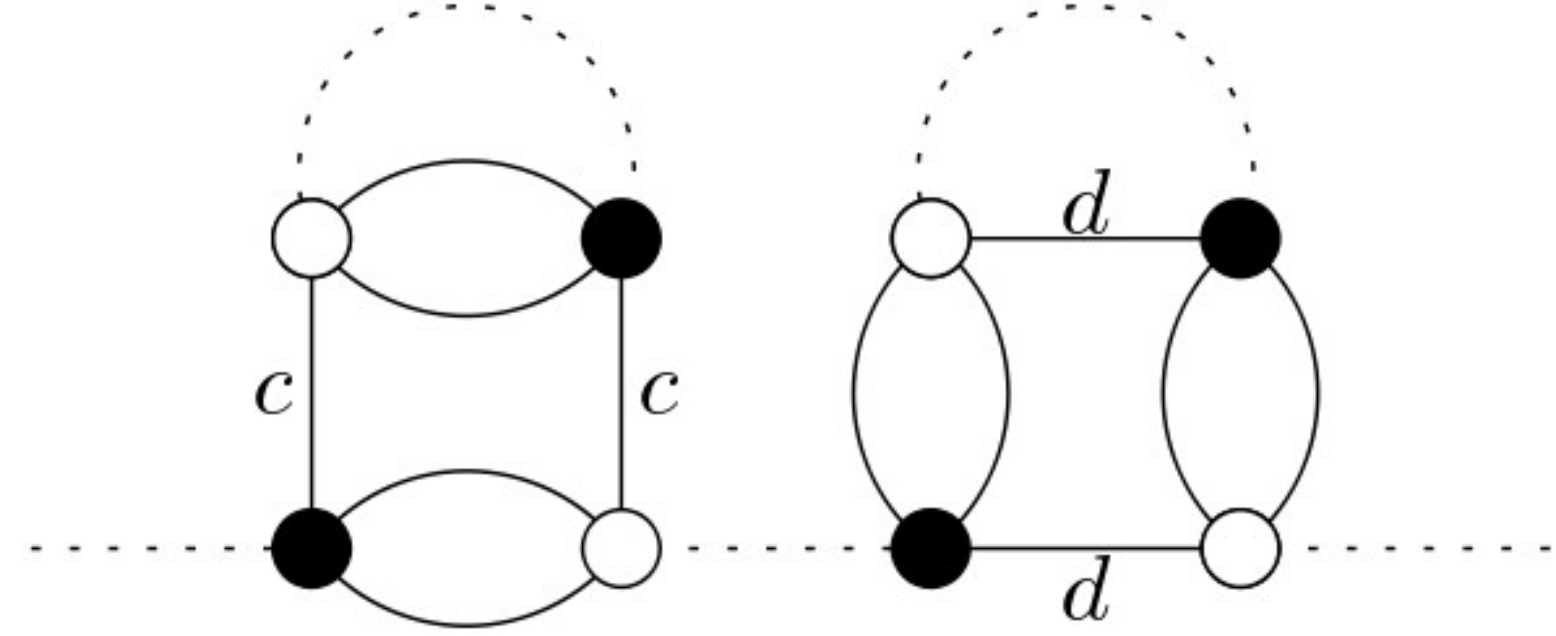}\Bigg) \nonumber \\
    &+4\sum \limits_{c=1}^3\sum \limits_{d=1}^3\Bigg(\includegraphics[scale=0.25]{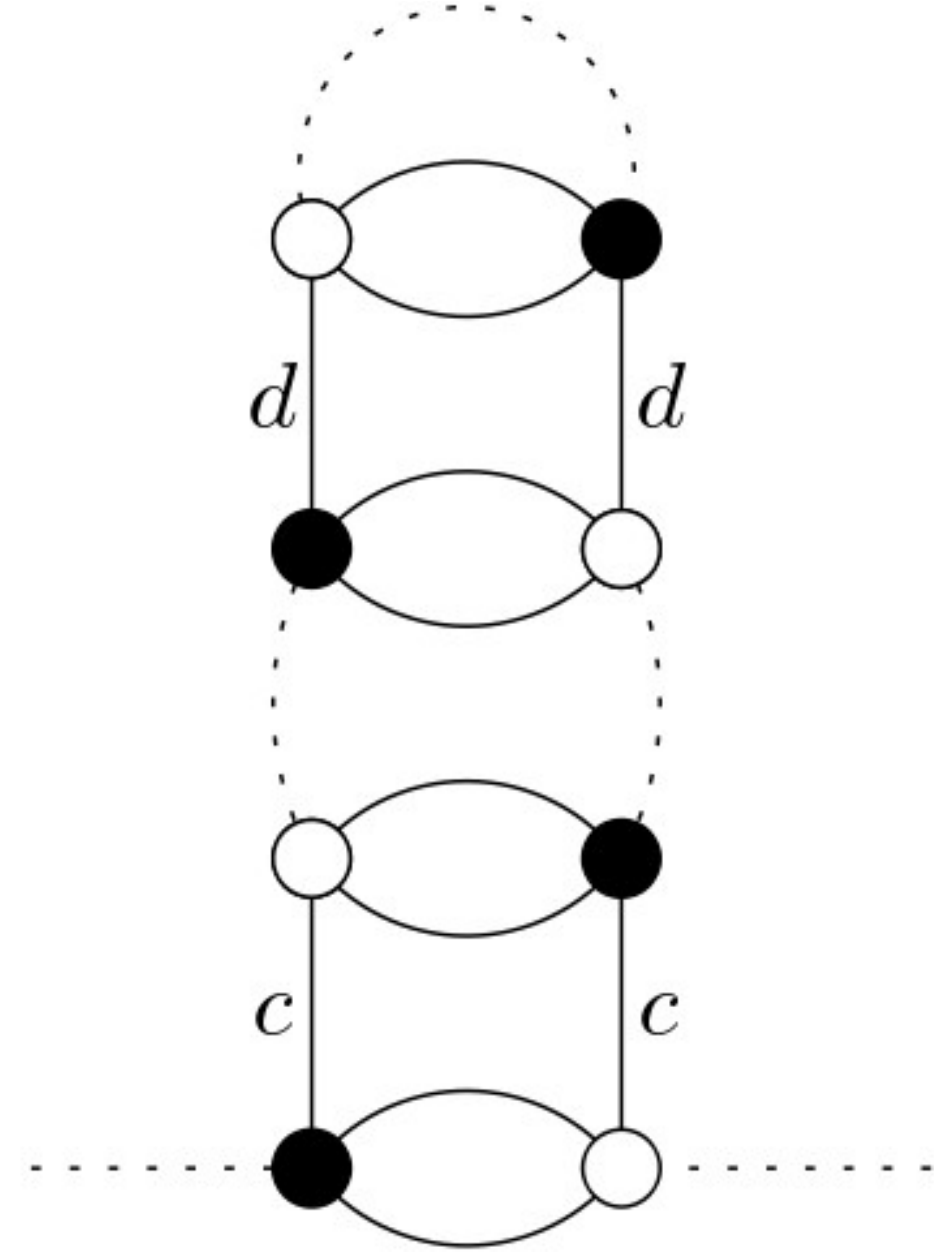} + \includegraphics[scale=0.25]{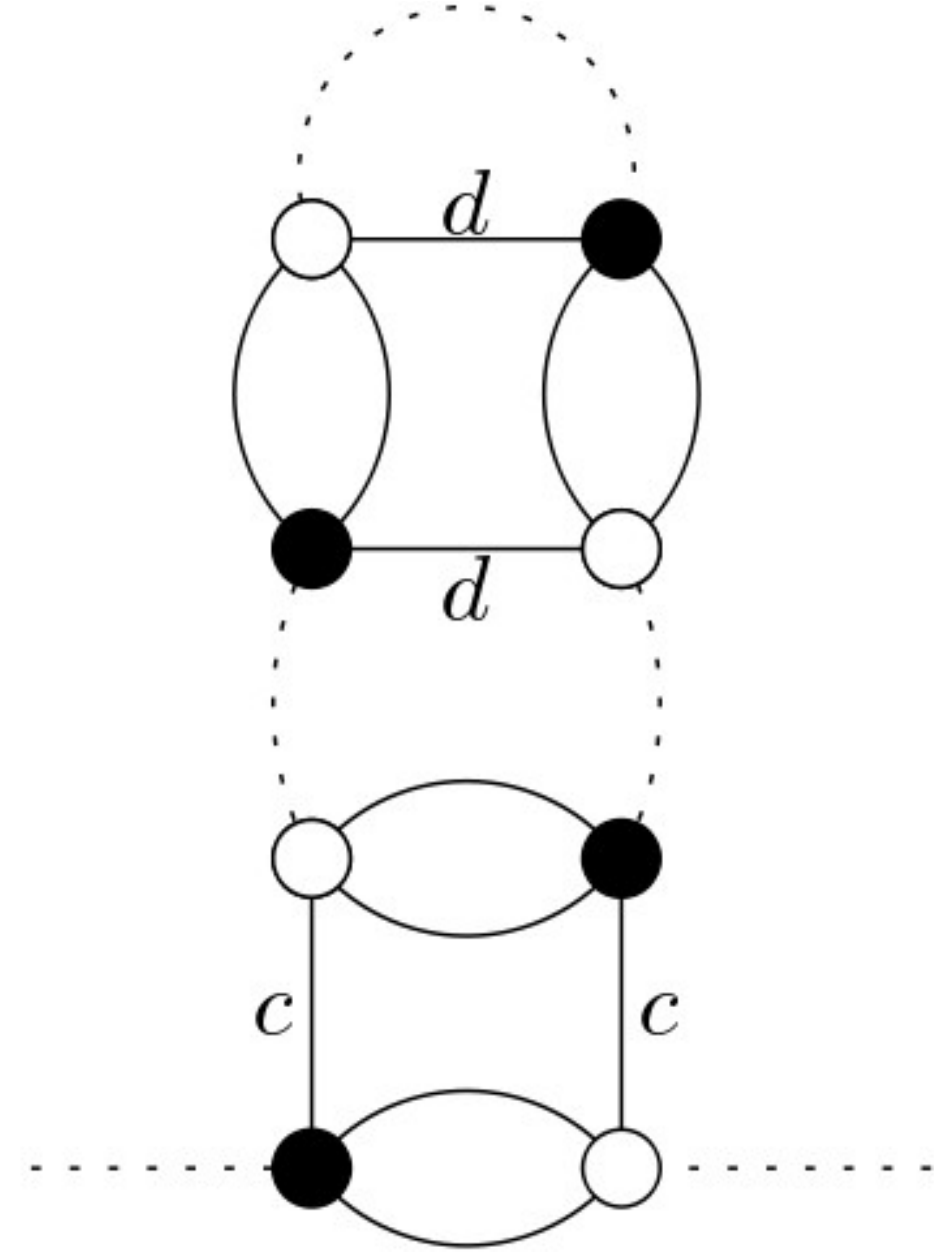}\Bigg) + O(\lambda^3), \nonumber 
\end{align}
\begin{equation}
   -\frac{2\lambda}{|\mathbf{x}|^2}\sum \limits_{c=1}^3 \mathrm{G}^{(4)}_c(\mathbf{x},\mathbf{x}) = 4\sum \limits_{c=1}^3\includegraphics[scale=0.25]{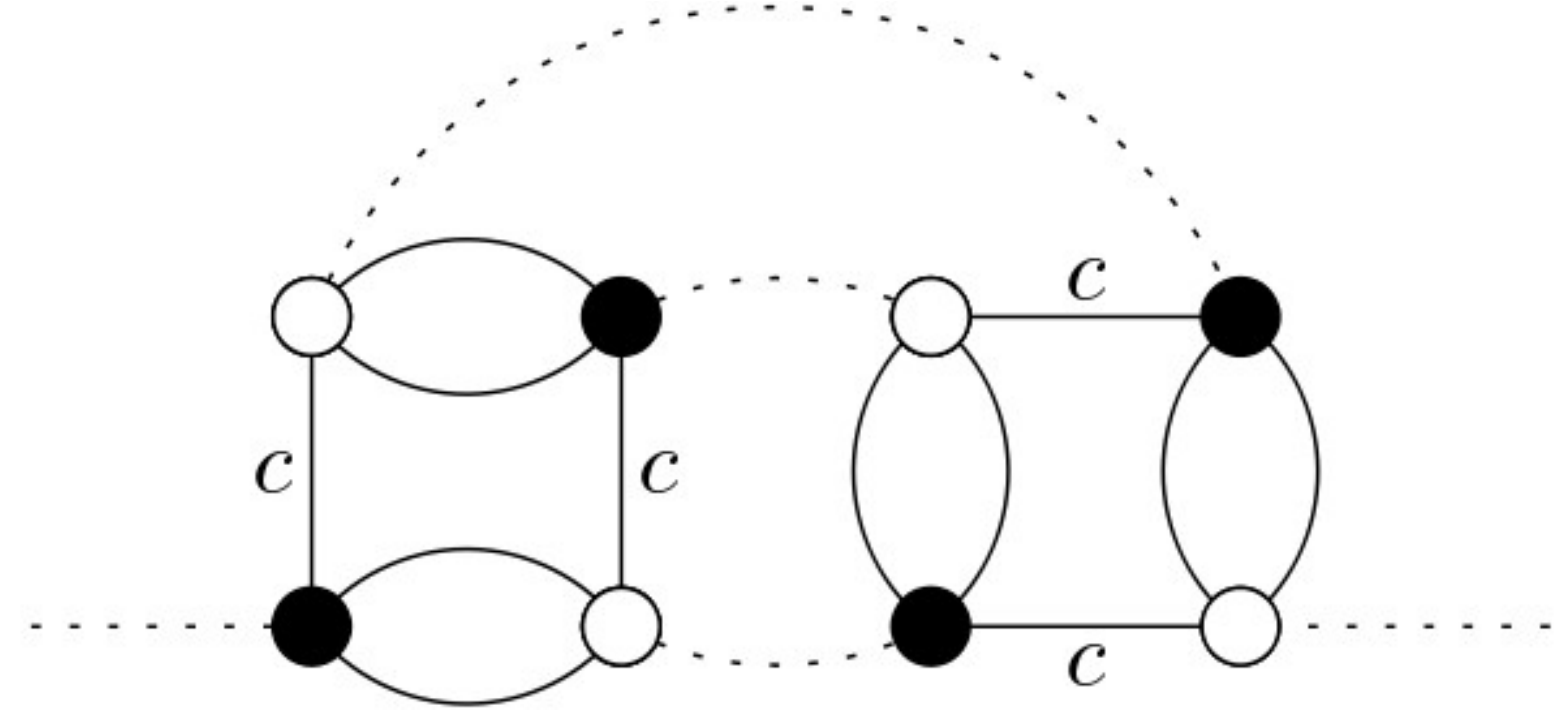} + O(\lambda^3),
\end{equation}
\begin{equation}
   -\frac{2\lambda}{|\mathbf{x}|^2}\sum \limits_{c=1}^3 \sum\limits_{c\neq d} \sum \limits_{q_c}\mathrm{G}^{(4)}_{d}(\mathbf{x},\mathbf{x}_{\hat{c}}q_c)  = 4\sum \limits_{c=1}^3\sum\limits_{c\neq d} \includegraphics[scale=0.25]{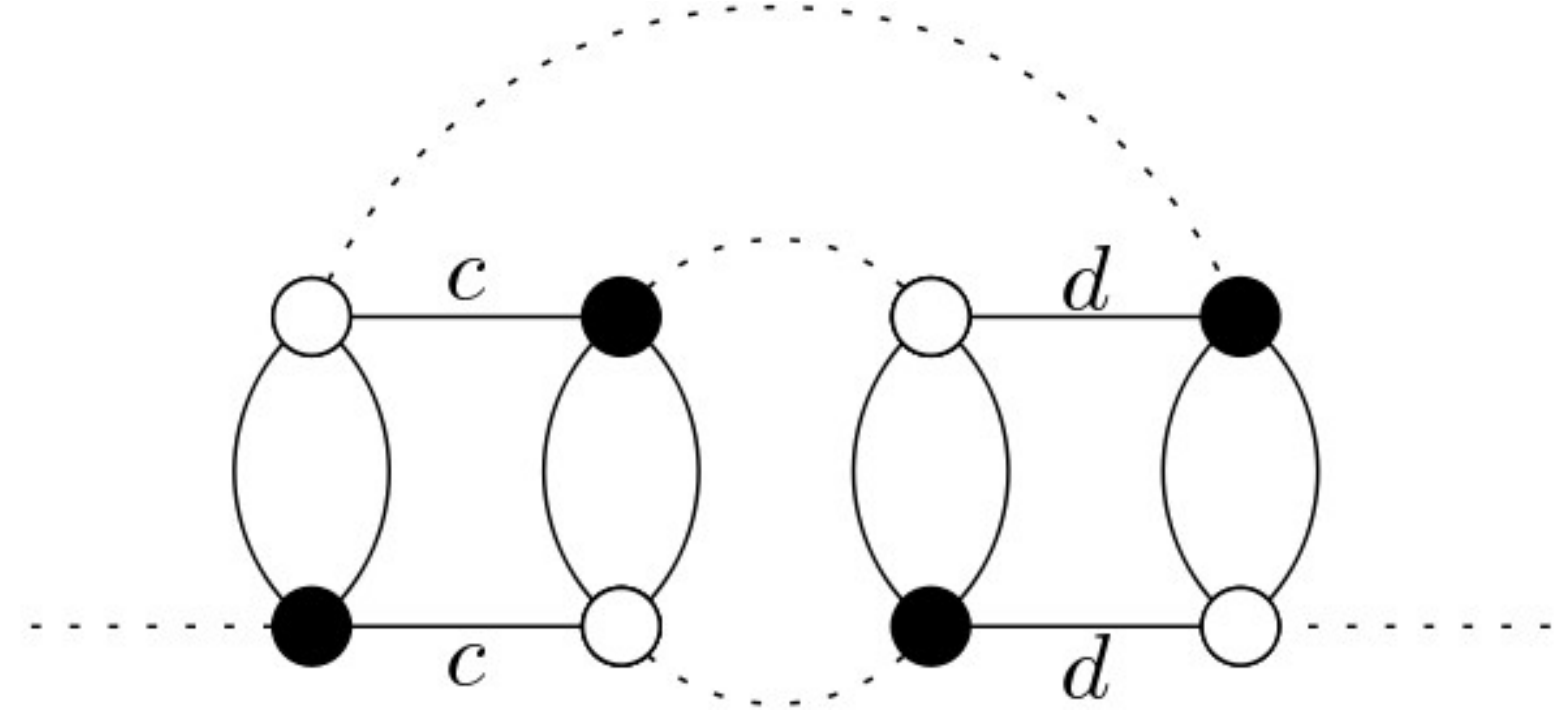} + O(\lambda^3),
\end{equation}
It is more involved to obtain the perturbative expansion from the difference of 2-point functions. At first order, we have
\begin{equation}
    - \frac{2\lambda}{|\mathbf{x}|^2} \sum \limits_{c=1}^3 \sum \limits_{q_c} \frac{1}{x_c^2 - q_c^2} \left(\mathrm{G}^{(2)}(\mathbf{x}_{\hat{c}}q_c)-\mathrm{G}^{(2)}(\mathbf{x})\right) =2\sum \limits_{c} \includegraphics[scale=0.25]{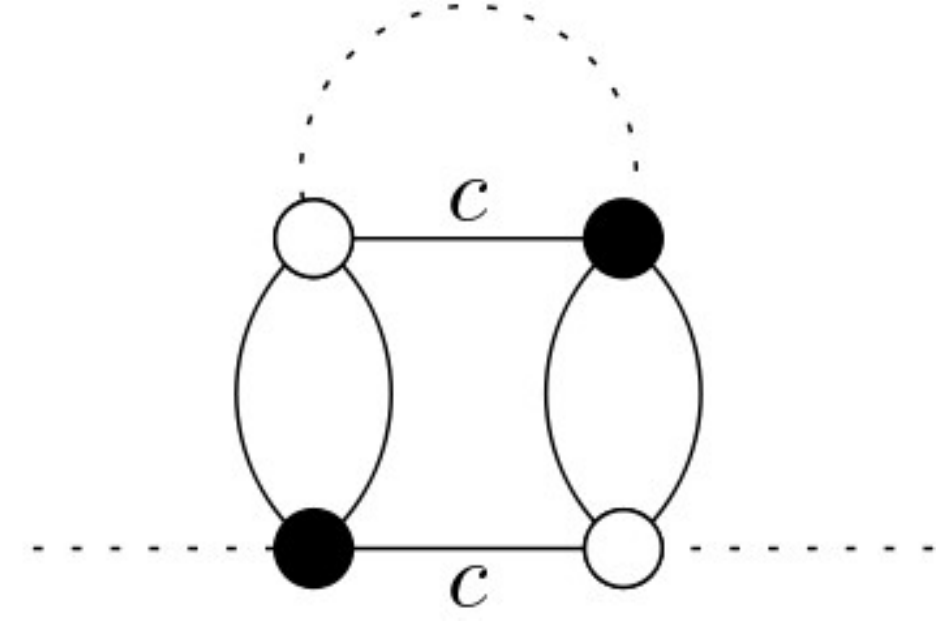} + O(\lambda^2).
\end{equation}
We are going to take the example of $c=1$ and compute explicitly the diagrams at $2^{\text{nd}}$ order in the coupling constant.
\begin{align}
    &-\frac{2\lambda}{|\mathbf{x}|^2}\sum \limits_{a_1} \frac{1}{x_1^2 - a_1^2} \left(\mathrm{G}^{(2)}(a_1,x_2,x_3)-\mathrm{G}^{(2)}(\mathbf{x})
    \right)\Bigg|_{\lambda^2} = \\
    &\frac{4\lambda^2}{|\mathbf{x}|^2}\sum \limits_{a_1}\frac{1}{x_1^2 - a_1^2}\Bigg\{\frac{1}{(a_1^2+x_2^2+x_3^2)^2}\Bigg[\sum \limits_{b_1,b_2}\frac{1}{b_1^2+b_2^2+x_3^2}+\sum \limits_{b_1,b_3}\frac{1}{b_1^2+x_2^2+b_3^2}+\sum \limits_{b_2,b_3}\frac{1}{a_1^2+b_2^2+b_3^2} \nonumber \\
    &+\sum \limits_{b_1}\frac{1}{b_1^2+x_2^2+x_3^2}+\sum \limits_{b_2}\frac{1}{a_1^2+b_2^2+x_3^2}+\sum \limits_{b_3}\frac{1}{a_1^2+x_2^2+b_3^2}\Bigg]-\frac{1}{|\mathbf{x}|^4}\Bigg[\sum \limits_{b_1,b_2}\frac{1}{b_1^2+b_2^2+x_3^2}\nonumber \\
    &+\sum \limits_{b_1,b_3}\frac{1}{b_1^2+x_2^2+b_3^2}+\sum \limits_{b_2,b_3}\frac{1}{x_1^2+b_2^2+b_3^2} +\sum \limits_{b_1}\frac{1}{b_1^2+x_2^2+x_3^2}+\sum \limits_{b_2}\frac{1}{x_1^2+b_2^2+x_3^2}+\sum \limits_{b_3}\frac{1}{x_1^2+x_2^2+b_3^2}\Bigg]\Bigg\}. \nonumber
\end{align}
89Half of the terms are straightforward to combine, let us look first at
\begin{align}
    &\frac{4\lambda^2}{|\mathbf{x}|^2}\sum \limits_{b_1}\frac{1}{b_1^2+x_2^2+x_3^2}\sum \limits_{a_1}\frac{1}{a_1^2 - x_1^2}\Bigg(\frac{1}{|\mathbf{x}|^4}-\frac{1}{(a_1^2+x_2^2+x_3^2)^2}\Bigg) \\
    &= \frac{4\lambda^2}{|\mathbf{x}|^4}\sum \limits_{a_1,b_1}\frac{1}{(b_1^2+x_2^2+x_3^2)(a_1^2+x_2^2+x_3^2)}\Bigg(\frac{1}{|\mathbf{x}|^2}+\frac{1}{a_1^2+x_2^2+x_3^2}\Bigg) \nonumber \\
    &= 2\includegraphics[scale=0.3]{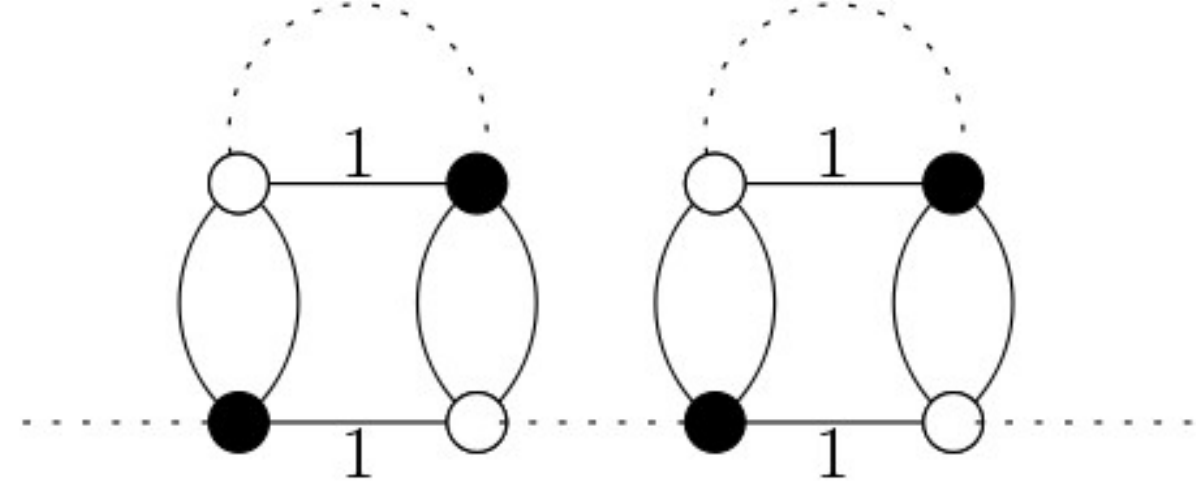} + 2\includegraphics[scale=0.25]{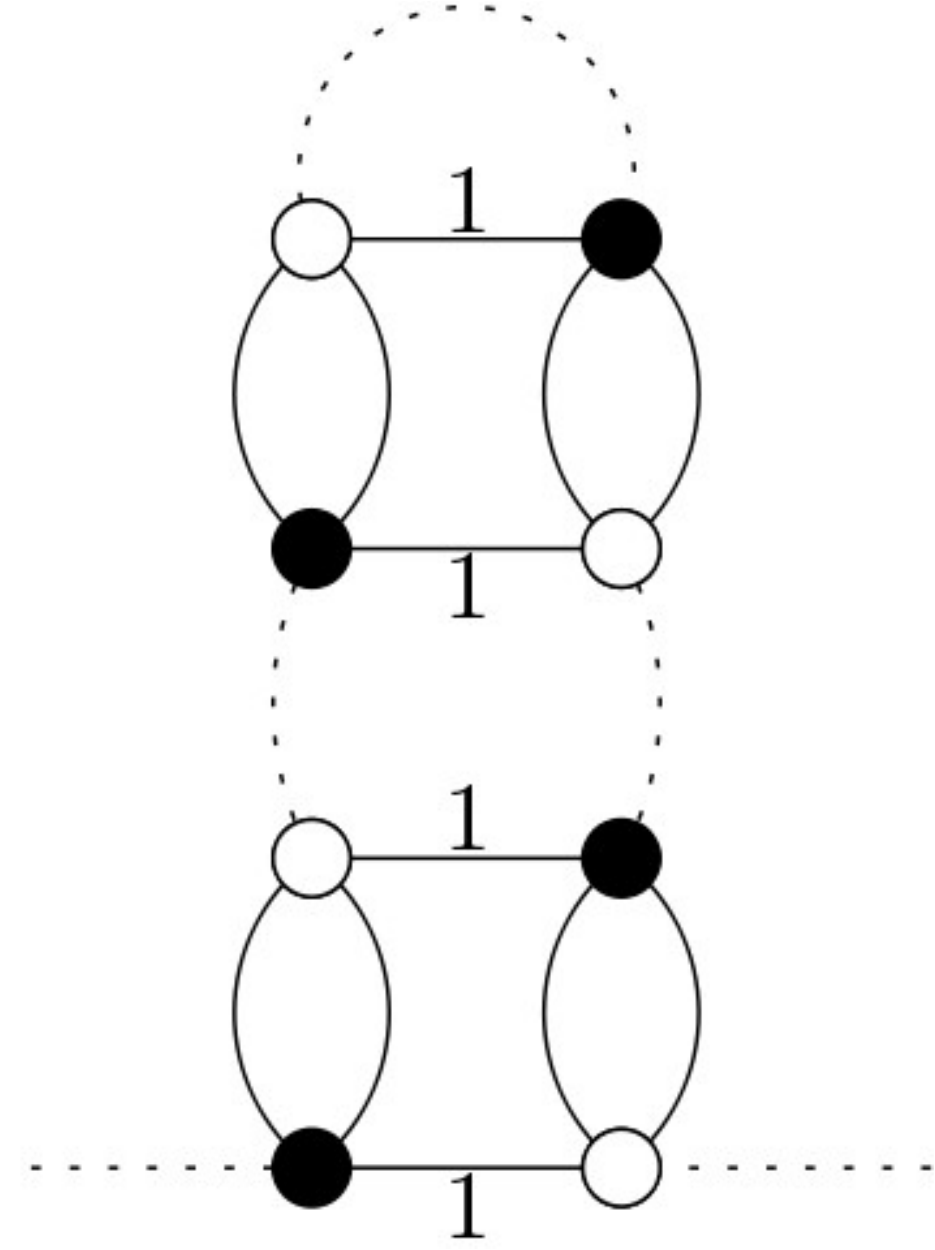}.
\end{align}
And combining the terms with sums on $b_1,b_2$ and $b_1,b_3$ gives
\begin{equation}
  4\includegraphics[scale=0.25]{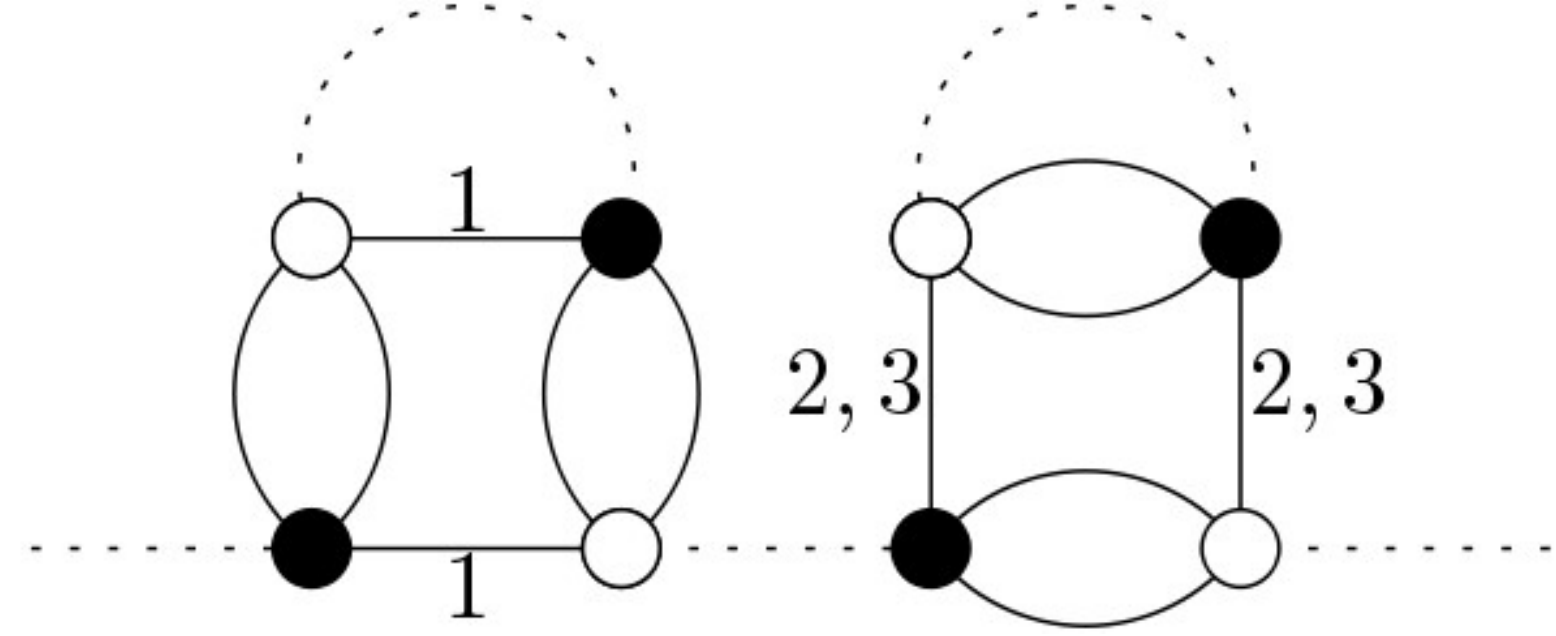} + 4\includegraphics[scale=0.25]{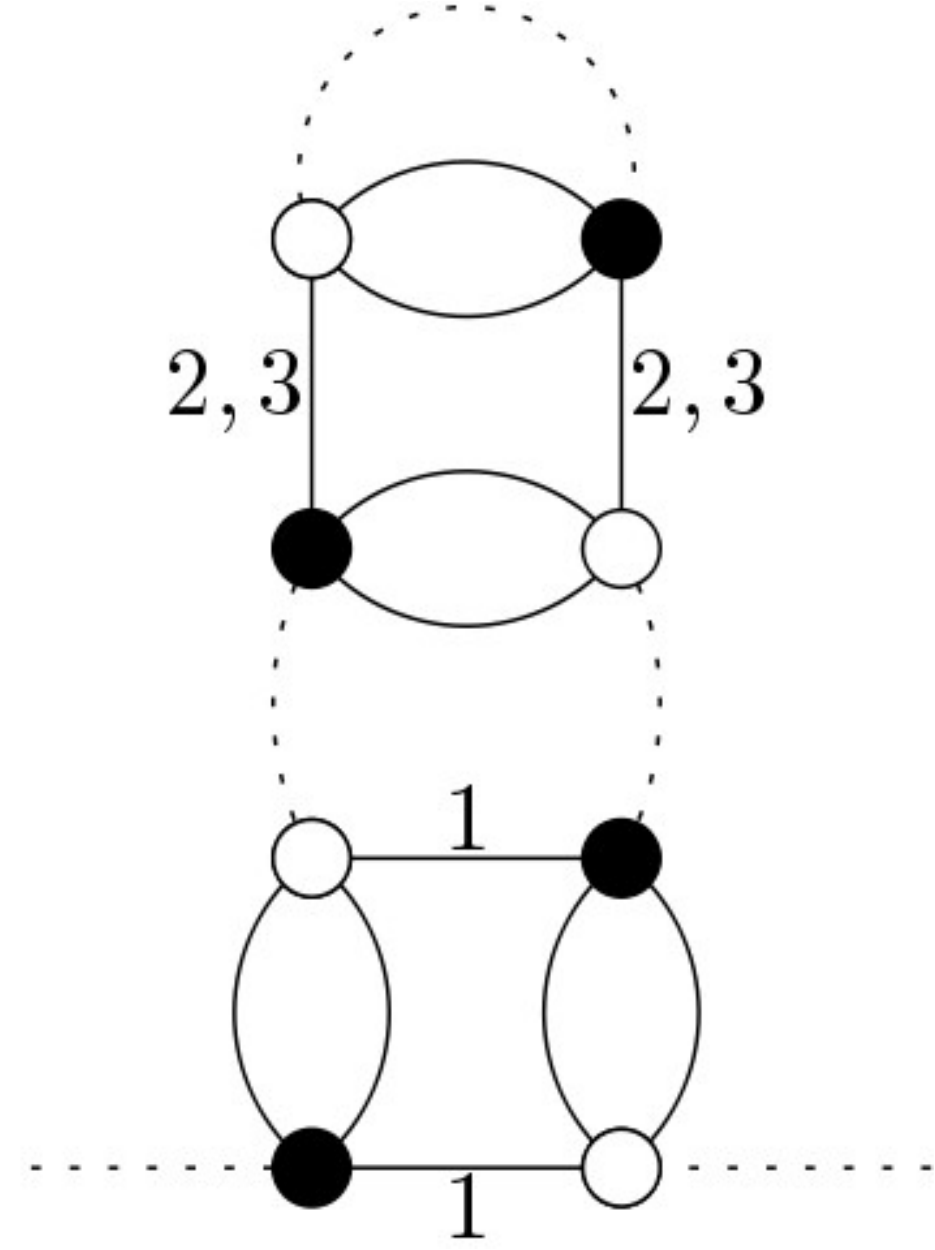}.
\end{equation}
Now let us look at the two terms
\begin{equation}
    \frac{4\lambda^2}{|\mathbf{x}|^2}\sum \limits_{a_1}\frac{1}{a_1^2 - x_1^2}\Bigg(\frac{1}{|\mathbf{x}|^4}\sum\limits_{b_2,b_3}\frac{1}{x_1^2+b_2^2+b_3^2} -\frac{1}{(a_1^2+x_2^2+x_3^2)^2}\sum\limits_{b_2,b_3}\frac{1}{a_1^2+b_2^2+b_3^2}\Bigg),
\end{equation}
and compute
\begin{align}
    &(a_1^2+x_2^2+x_3^2)^2(a_1^2+b_2^2+b_3^2) - |\mathbf{x}|^4(x_1^2+b_2^2+b_3^2) = \nonumber \\
    &a_1^6-x_1^6 + 2(a_1^4-x_1^4)(x_2^2+x_3^2) + (a_1^2-x_1^2)(x_2^4+x_3^4+2(b_2^2+b_3^2)(x_2^2+x_3^2)).
\end{align}
By writing 
\begin{align}
    &a_1^6-x_1^6 = (a_1^2-x_1^2)(a_1^4+x_1^4) + x_1^2a_1^4 -x_1^4a_1^2 = (a_1^2-x_1^2)(a_1^4+x_1^4+a_1^2x_1^2), \\
    &2(a_1^4-x_1^4)(x_2^2+x_3^2) = 2(a_1^2-x_1^2)(a_1^2+x_1^2)(x_2^2+x_3^2),
\end{align}
we get
\begin{equation}
    \frac{4\lambda^2}{|\mathbf{x}|^2}\sum \limits_{a_1,b_2,b_3}\frac{a_1^4+x_1^4+a_1^2x_1^2+2(x_2^2+x_3^2)(a_1^2+x_1^2)+x_2^4+x_3^4+2(b_2^2+b_3^2)(x_2^2+x_3^2)}{|\mathbf{x}|^4(x_1^2+b_2^2+b_3^2)(a_1^2+x_2^2+x_3^2)^2(a_1^2+b_2^2+b_3^2)}.
\end{equation}
Now we can factorise
\begin{align}
    &a_1^4+x_1^4+a_1^2x_1^2+2(x_2^2+x_3^2)(a_1^2+x_1^2)+x_2^4+x_3^4+2(b_2^2+b_3^2)(x_2^2+x_3^2) \nonumber \\
    &= |\mathbf{x}|^2(a_1^2 +x_2^2 + x_3^2)+(a_1^2 +x_2^2 + x_3^2)(a_1^2+b_2^2+b_3^2) + |\mathbf{x}|^2(x_1^2+b_2^2+b_3^2),
\end{align}
which gives
\begin{align}
    &\frac{4\lambda^2}{|\mathbf{x}|^2}\sum \limits_{a_1,b_2,b_3}\Bigg(\frac{1}{|\mathbf{x}|^2(x_1^2+b_2^2+b_3^2)(a_1^2+x_2^2+x_3^2)^(a_1^2+b_2^2+b_3^2)} \nonumber \\
    &+ \frac{1}{|\mathbf{x}|^4(x_1^2+b_2^2+b_3^2)(a_1^2+x_2^2+x_3^2)} + \frac{1}{|\mathbf{x}|^2(a_1^2+x_2^2+x_3^2)^2(a_1^2+b_2^2+b_3^2)}\Bigg) \nonumber \\
    &= 4\includegraphics[scale=0.25]{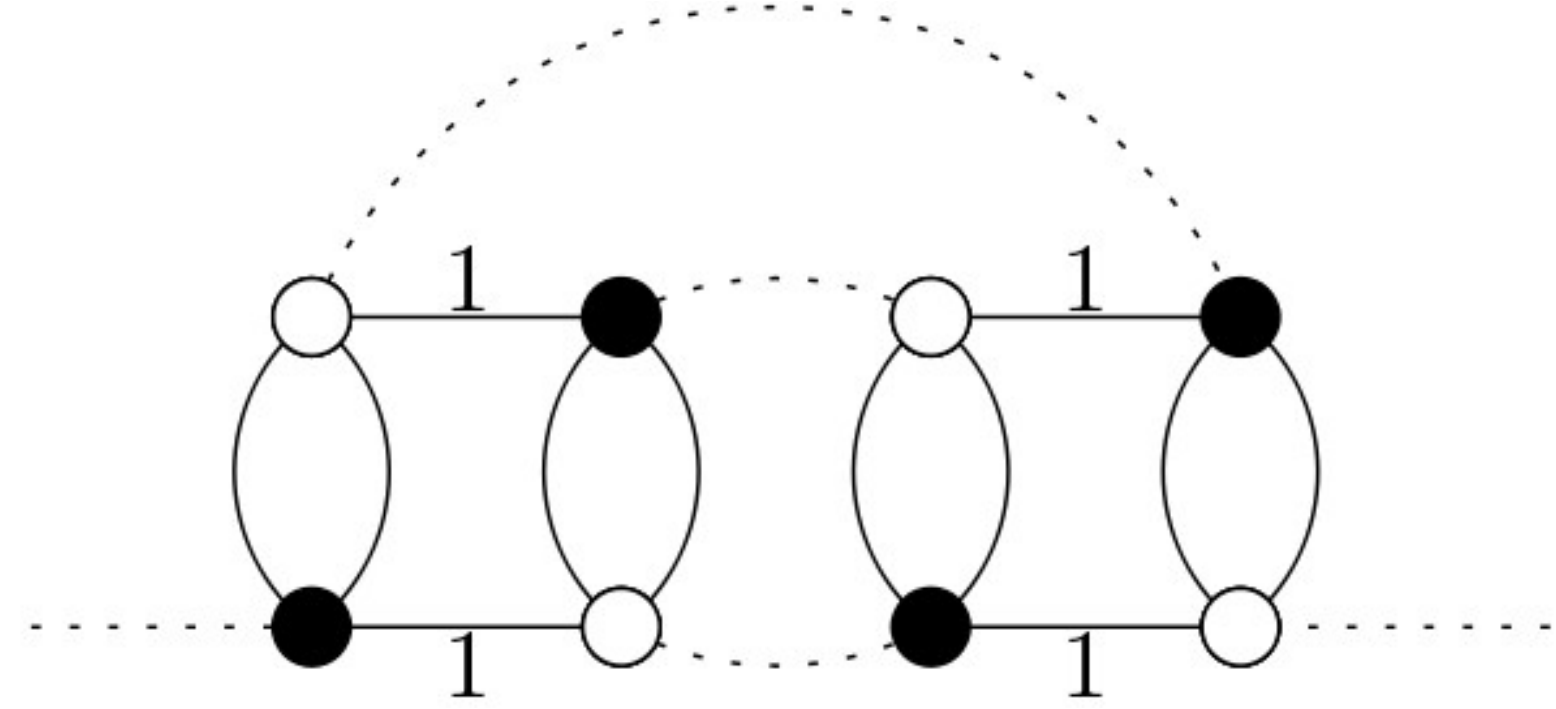} + 4\includegraphics[scale=0.25]{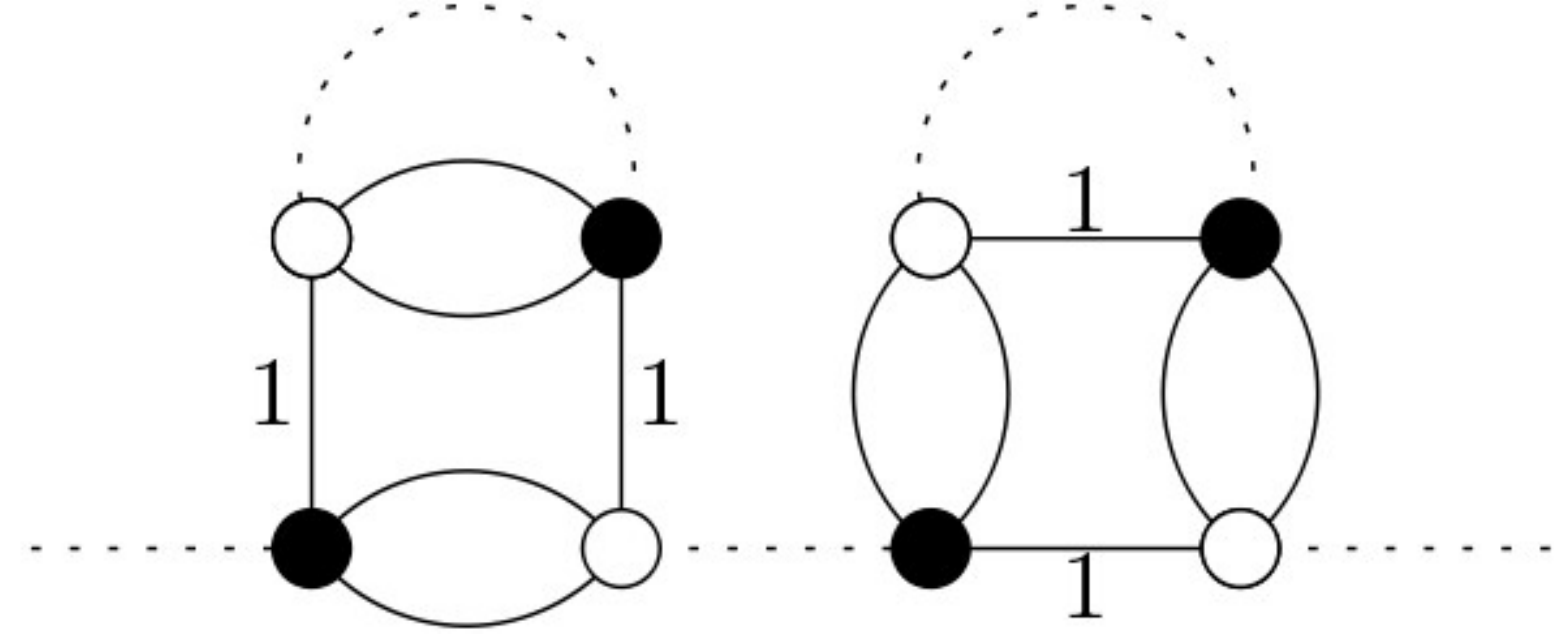} + 4\includegraphics[scale=0.25]{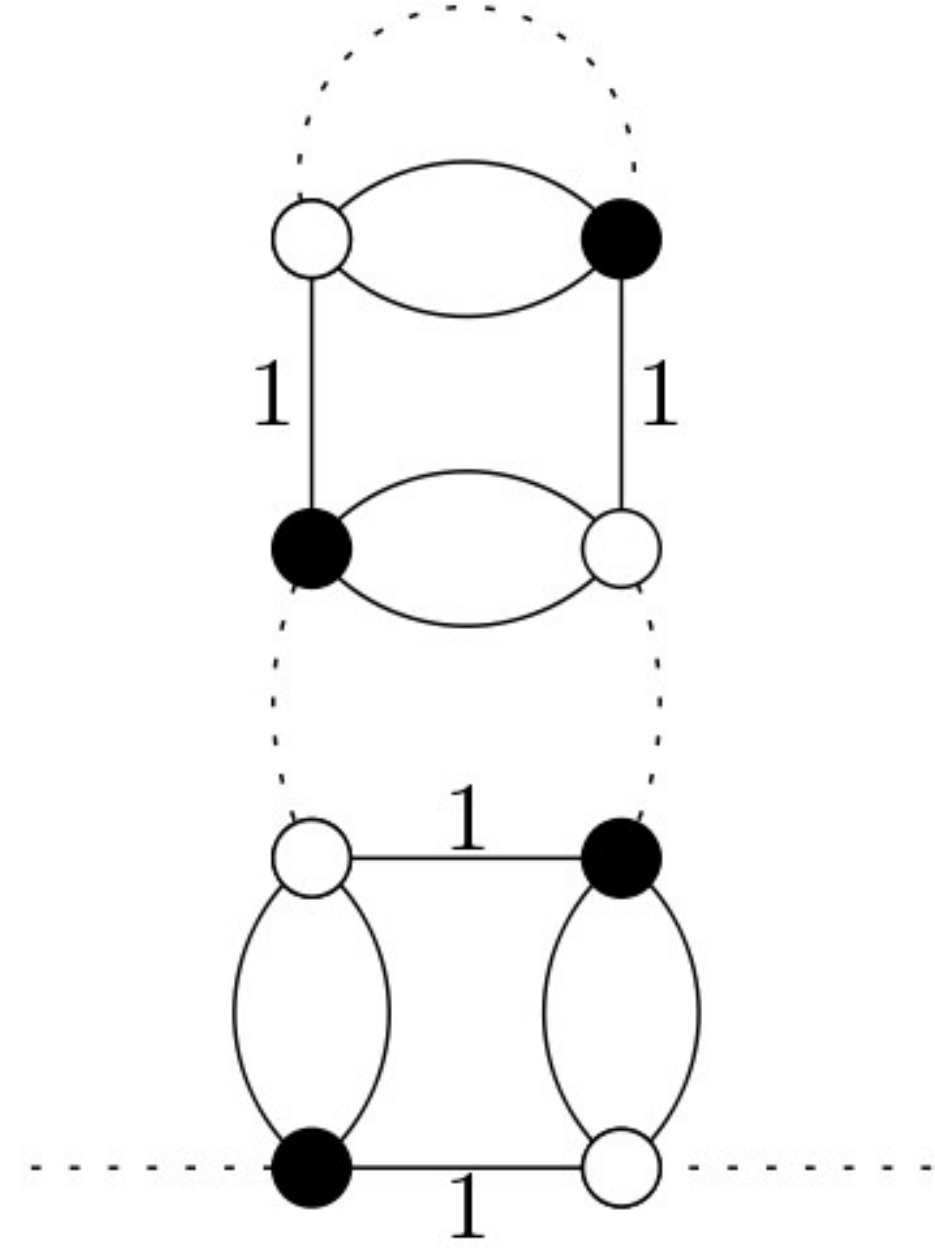}.
\end{align}
Then by combining the terms with a sum on $b_2$ or on $b_3$, we get an analogous result which correspond to replace $b_3$ by $x_3$ or $b_2$ by $x_2$ in the previous equation. And we obtain the following diagrams
\begin{equation}
    4\includegraphics[scale=0.25]{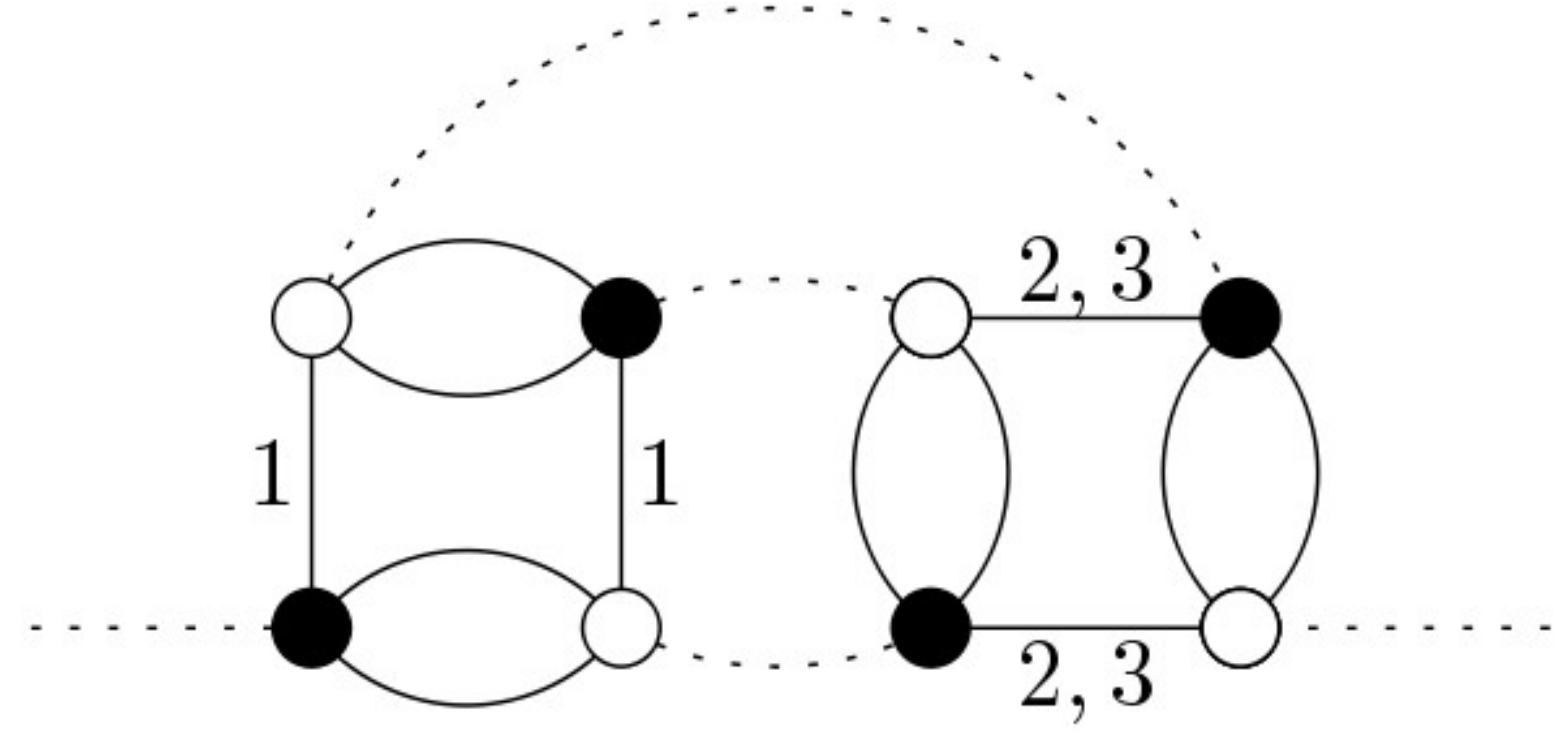} + 4\includegraphics[scale=0.25]{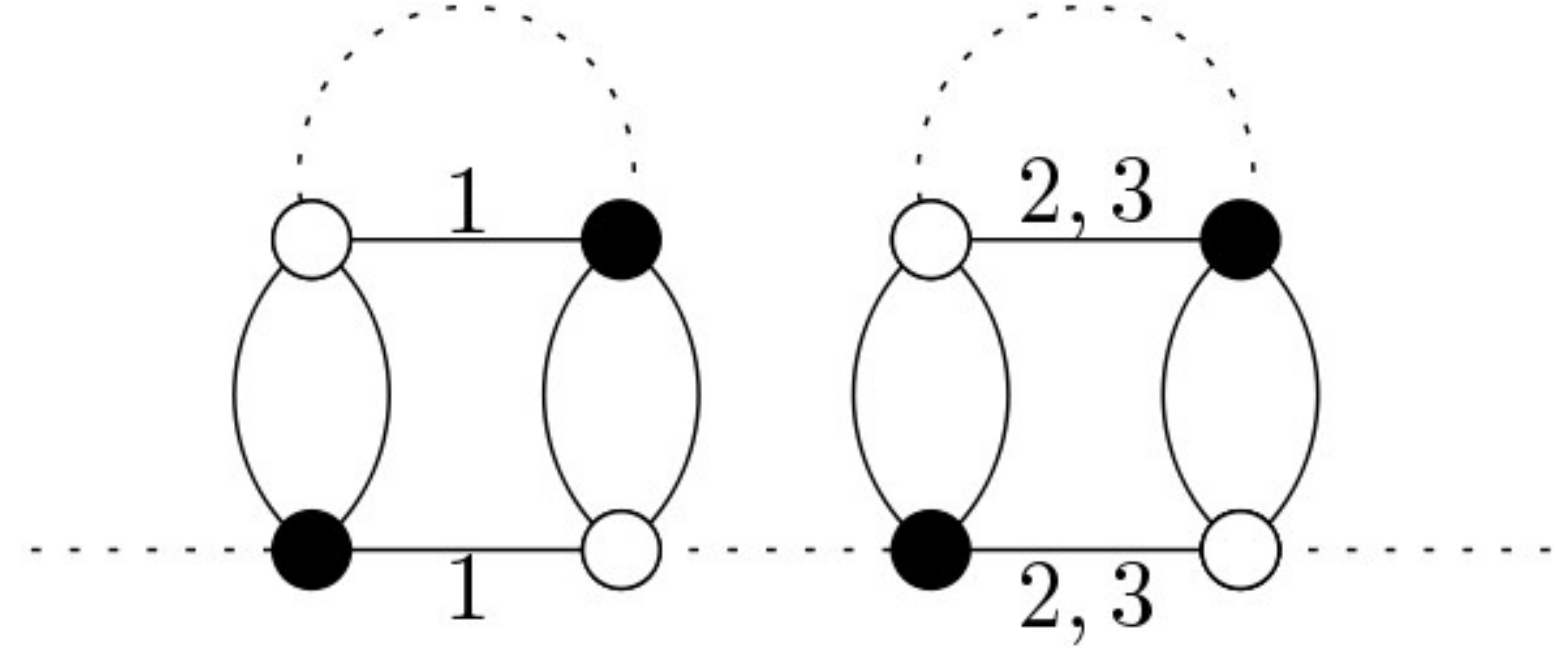} + 4\includegraphics[scale=0.25]{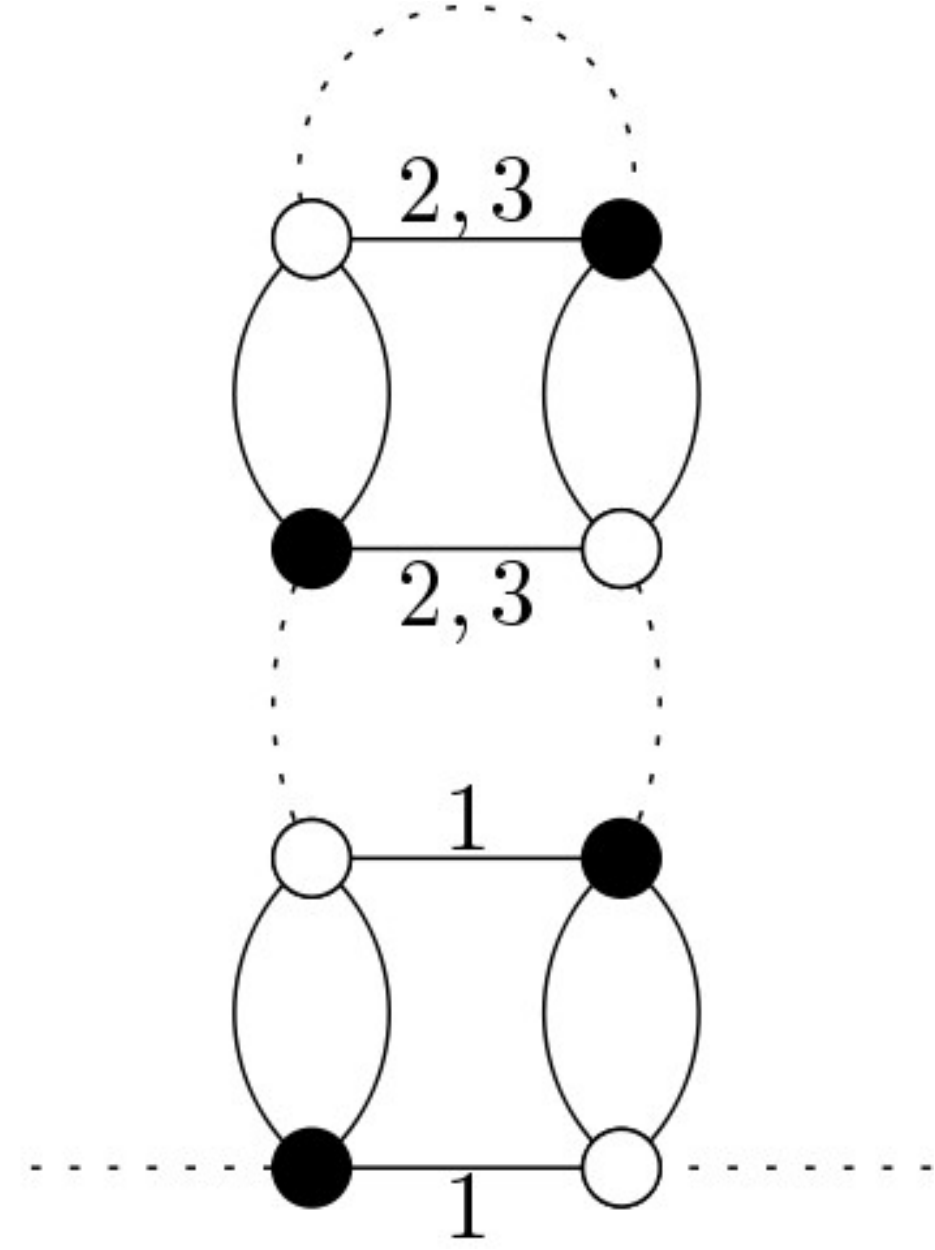}.
\end{equation}
This computation is completely analogous for $c=2,3$. Collecting all the diagrams we get
\begin{align}
    &\mathrm{G}^{(2)}(\mathbf{x}) = \includegraphics[scale=0.25]{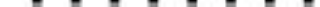} + \sum \limits_{c=1}^3 \Bigg\{2\includegraphics[scale=0.25]{1pillow2.pdf} + 2\includegraphics[scale=0.25]{1pillow1c.pdf} \\
    &+ \sum \limits_{d=1}^3 \Bigg[ 4 \includegraphics[scale=0.25]{2pillow2.pdf} + 4 \includegraphics[scale=0.16]{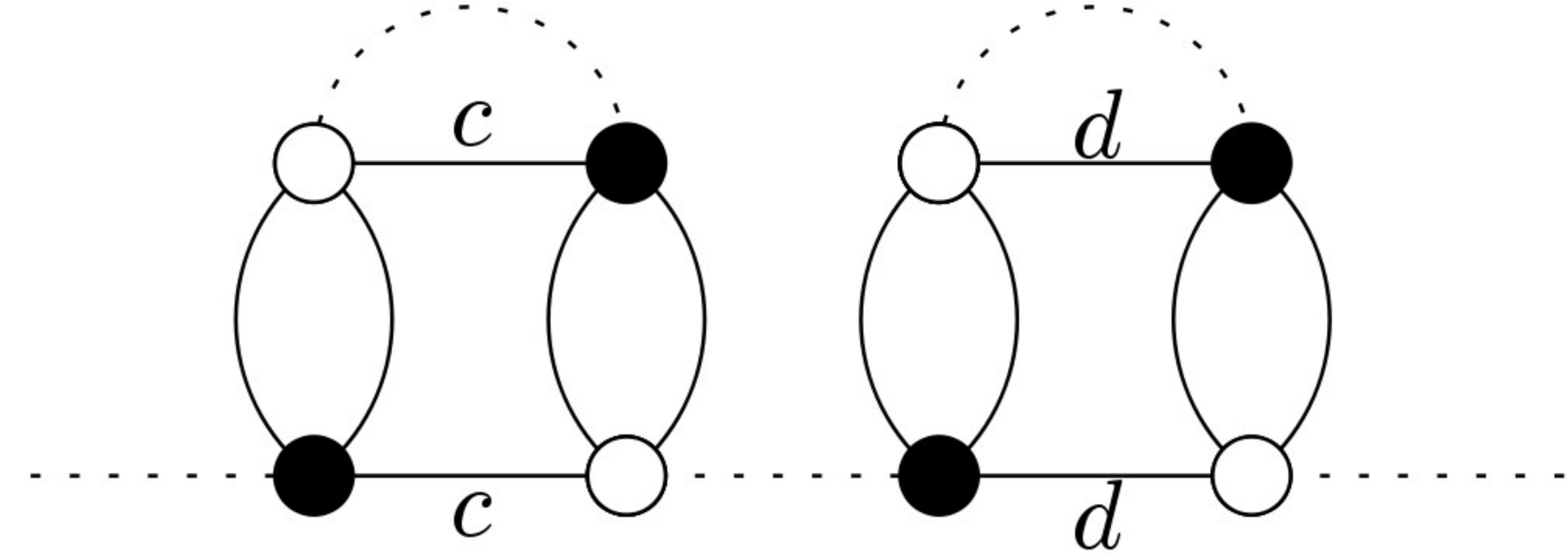} + 8 \includegraphics[scale=0.25]{2pillow1.pdf} \nonumber \\
    &+ 4 \includegraphics[scale=0.25]{1PI2pillow2.pdf} + 4 \includegraphics[scale=0.25]{1PI2pillow1.pdf} + 4 \includegraphics[scale=0.20]{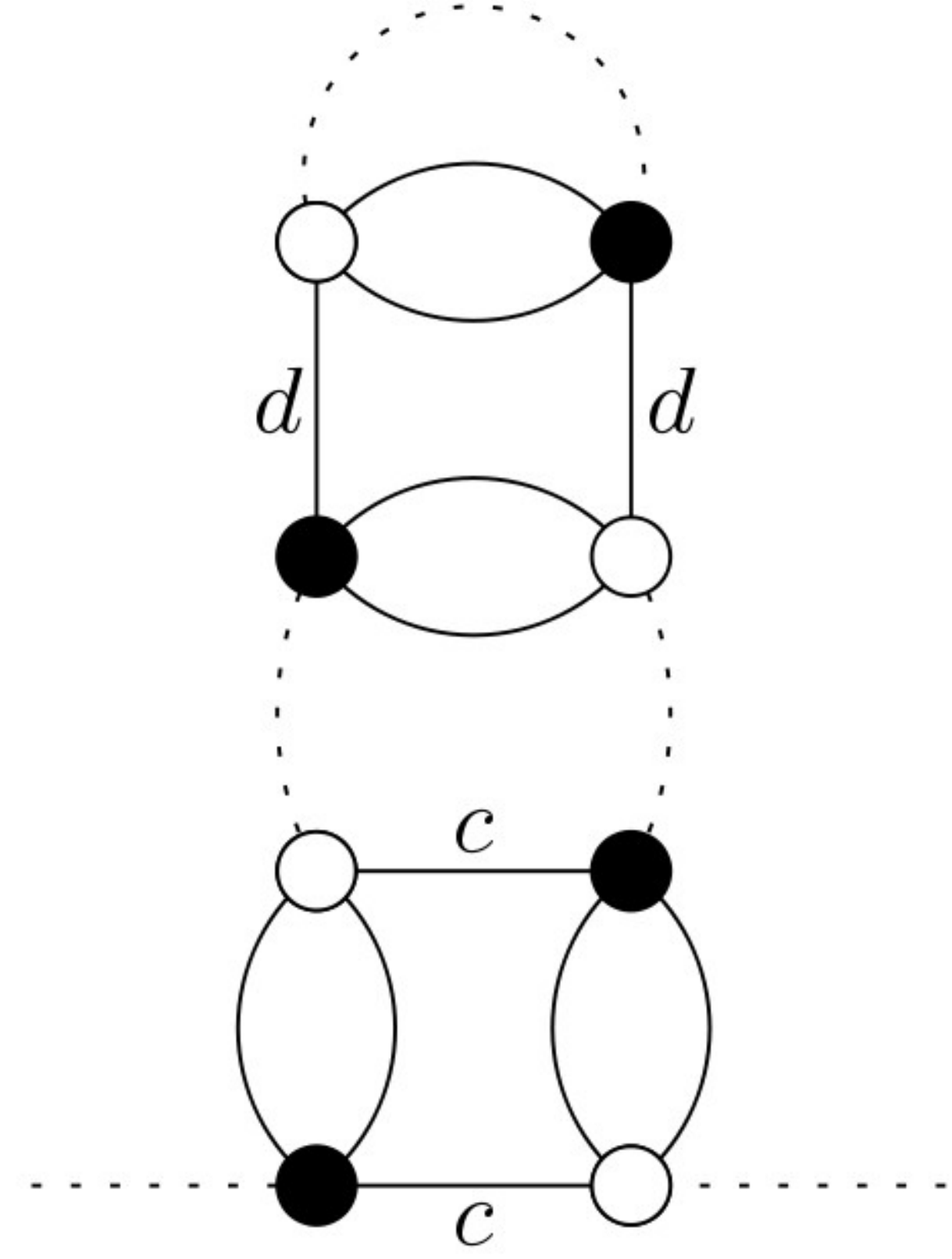} + 4 \includegraphics[scale=0.20]{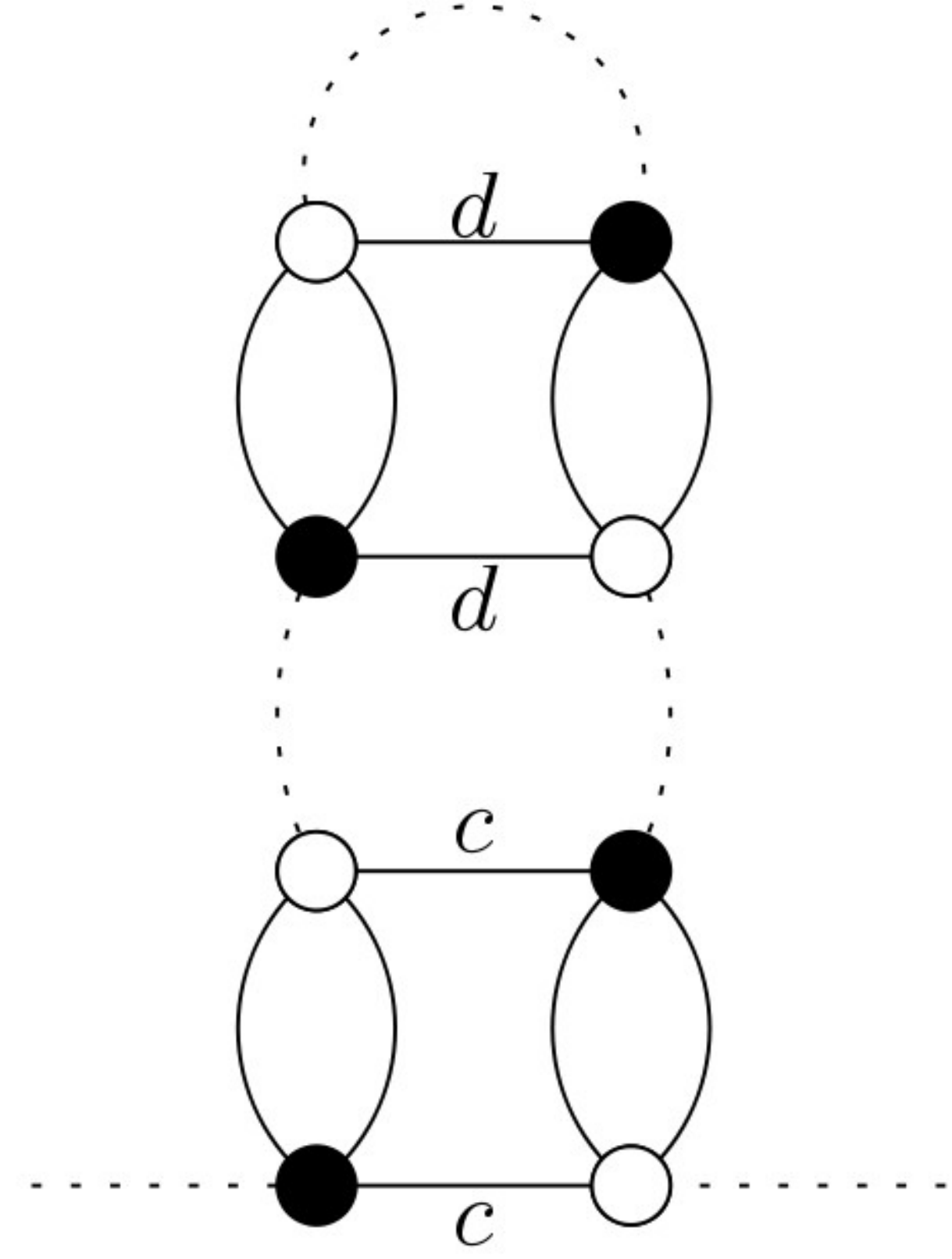} \nonumber \\
    &+ 4 \includegraphics[scale=0.25]{2ptGc4.pdf} + 4 \includegraphics[scale=0.20]{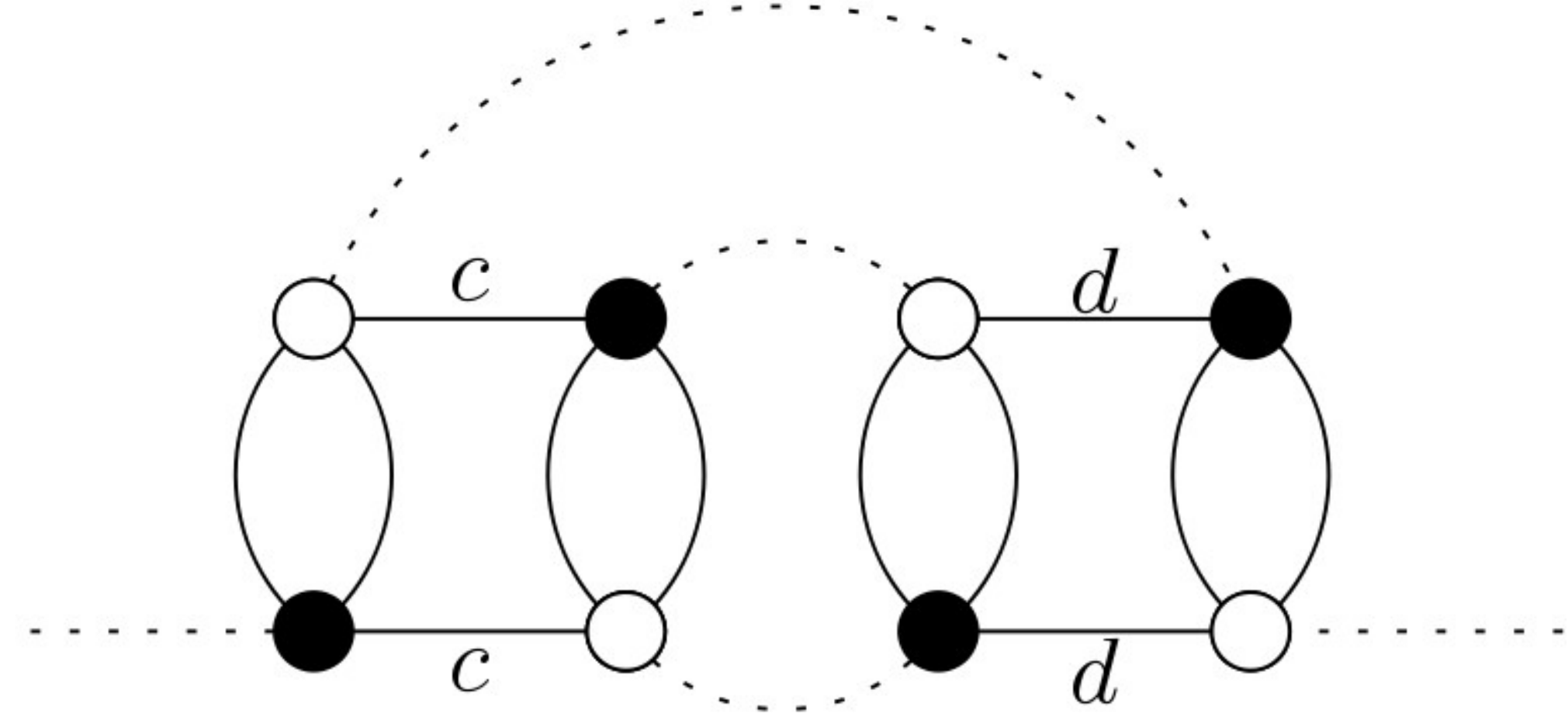} \Bigg] \Bigg\}  + O(\lambda^3).\nonumber
\end{align}
Once we take the large $N$ limit, we get the following expansion and SDE
\begin{align}\label{SDE2N}
    \mathrm{G}^{(2)}(\mathbf{x}) 
    &= \includegraphics[scale=0.23]{free.pdf} + \sum \limits_{c=1}^3 \Bigg\{2\includegraphics[scale=0.23]{1pillow2.pdf} + 4\sum \limits_{d=1}^3 \Bigg[  \includegraphics[scale=0.23]{2pillow2.pdf} +   \includegraphics[scale=0.23]{1PI2pillow2.pdf} \Bigg]\Bigg\} + O(\lambda^3) \nonumber \\
     &= \left(|\mathbf{x}|^2 + 2\tilde{\lambda}\sum \limits_{c=1}^3\int\mathrm{d}\mathbf{q}_{\hat{c}} \mathrm{G}^{(2)}(\mathbf{q}_{\hat{c}}x_c) \right)^{-1}.
\end{align}

\subsection*{\texorpdfstring{$4$}{4}-point function with connected boundary}
The full SDE for the $4$-point function is
\begin{align}
    &\mathrm{G}^{(4)}_1(\mathbf{x},\mathbf{y}) =-\frac{2\lambda}{x_1^2+y_2^2+y_3^2}\left\{\sum_{c=1}^3\mathfrak{f}_c(\mathbf{x},\mathbf{y};s_c; V_c)
  +\sum_{c=1}^3\sum_{\mathbf{q}_{\hat{c}}}\mathrm{G}^{(2)}(\mathbf{q}_{\hat{c}}s_c)\mathrm{G}^{(4)}_1(\mathbf{x},\mathbf{y}) \right. \nonumber \\ 
  &+\sum_{b_1}
  \frac{1}{b_1^2-x_1^2}\left(\mathrm{G}^{(4)}_1(\mathbf{x},\mathbf{y}) -\mathrm{G}^{(4)}_1(b_1,x_2,x_3,\mathbf{y})\right)+ \sum_{b_2}
  \frac{1}{b_2^2-y_2^2}\left(\mathrm{G}^{(4)}_1(\mathbf{x},\mathbf{y}) -
  \mathrm{G}^{(4)}_1(\mathbf{x},y_1,b_2,y_3)\right) \nonumber \\ 
  &+ \sum_{b_3}
  \frac{1}{b_3^2-y_3^2}\left(\mathrm{G}^{(4)}_1(\mathbf{x},\mathbf{y}) -\mathrm{G}^{(4)}_1(\mathbf{x},y_1,y_2,b_3)\right)  +\frac{1}{y_2^2-x_2^2}\left(\mathrm{G}^{(4)}_3(\mathbf{x},y_1,x_2,y_3)-
  \mathrm{G}^{(4)}_3(\mathbf{x},\mathbf{y})\right) \nonumber \\ 
  &+ 
  \frac{1}{y_3^2-x_3^2}\left(\mathrm{G}^{(4)}_2(\mathbf{x},y_1,y_2,x_3)-
  \mathrm{G}^{(4)}_2(\mathbf{x},\mathbf{y})\right) +\frac{\mathrm{G}^{(2)}(\mathbf{y})}{y_1^2-x_1^2}
  \left(\mathrm{G}^{(2)}(\mathbf{x})-\mathrm{G}^{(2)}(y_1,x_2,x_3)\right) \nonumber \\
  &+ \frac{1}{y_1^2-x_1^2}\left(\mathrm{G}^{(4)}_{\mathrm{m}}(\mathbf{x},\mathbf{y})-\mathrm{G}^{(4)}_{\mathrm{m}}(y_1,x_2,x_3,\mathbf{y})\right)\Bigg\},
\end{align}
where $\mathbf{s} = (x_1,y_2,y_3)$. We can remark that the terms in $\lambda\mathfrak{f}_c$ involve only 6-point functions and start to contribute to the perturbative expansion only at order $\lambda^3$, and so does the terms in $\lambda\mathrm{G}^{(4)}_{\mathrm{m}}$. Hence up to the $2^{\text{nd}}$ order in the coupling constant the other terms give
\begin{align}
 &\sum_{d=1}^3\sum_{\mathbf{q}_{\hat{d}}}\mathrm{G}^{(2)}(\mathbf{q}_{\hat{d}}s_d)\mathrm{G}^{(4)}_1(\mathbf{x},\mathbf{y}) = 4 \sum_{d=1}^3 \includegraphics[scale=0.25]{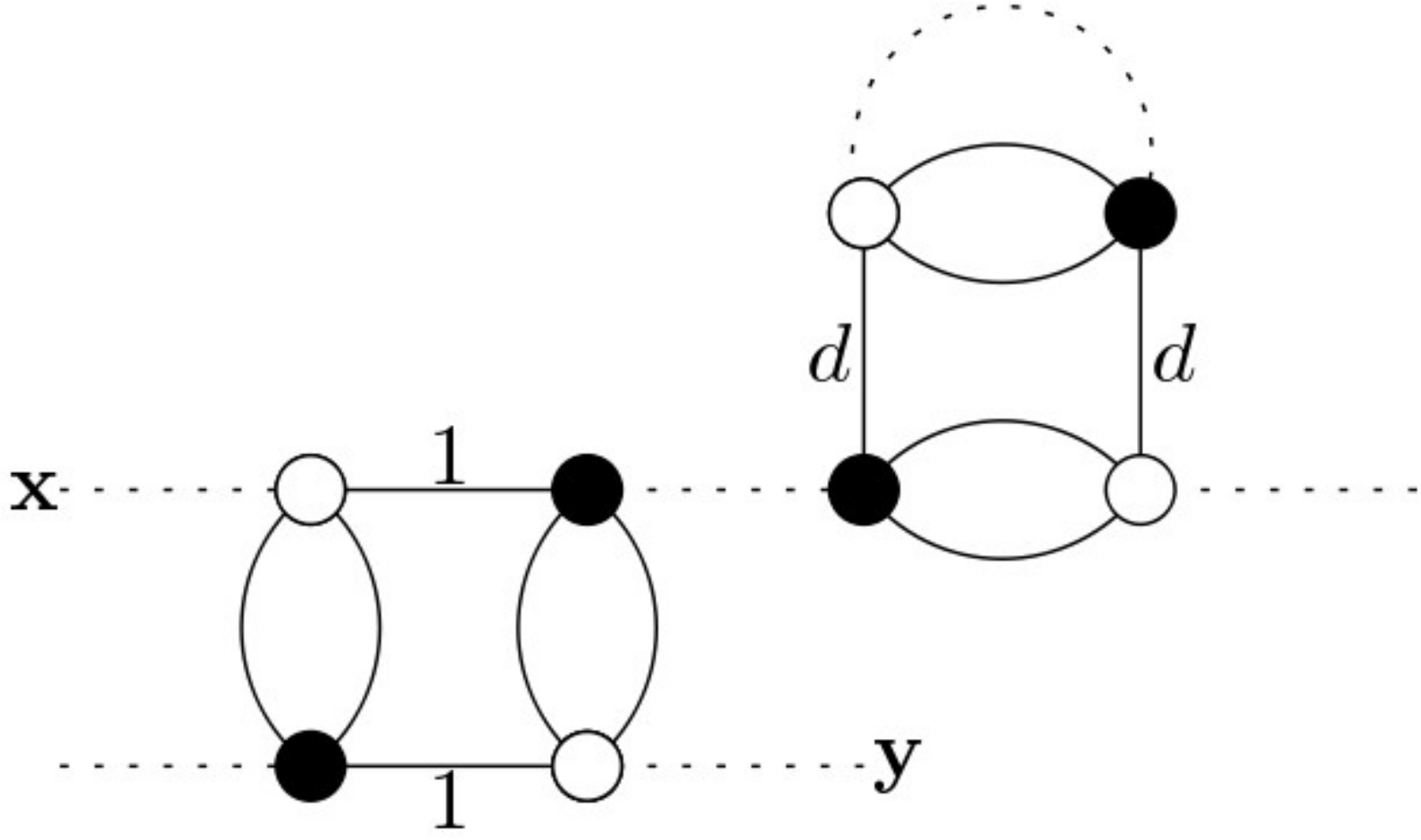}  + O(\lambda^3), \\
 &-\frac{2\lambda}{x_1^2+y_2^2+y_3^2} \mathrm{G}^{(2)}(\mathbf{y})\frac{\mathrm{G}^{(2)}(\mathbf{x})-\mathrm{G}^{(2)}(y_1,x_2,x_3)}{y_1^2-x_1^2} = 2 \includegraphics[scale=0.25]{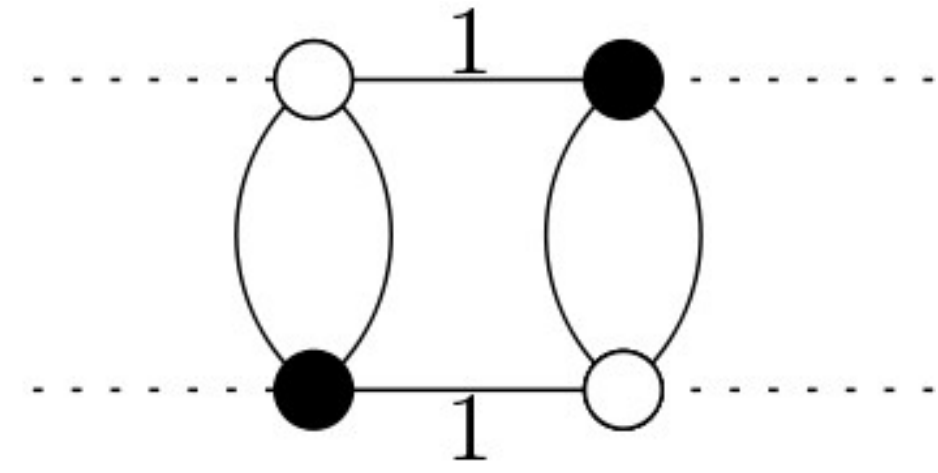} + 4 \includegraphics[scale=0.25]{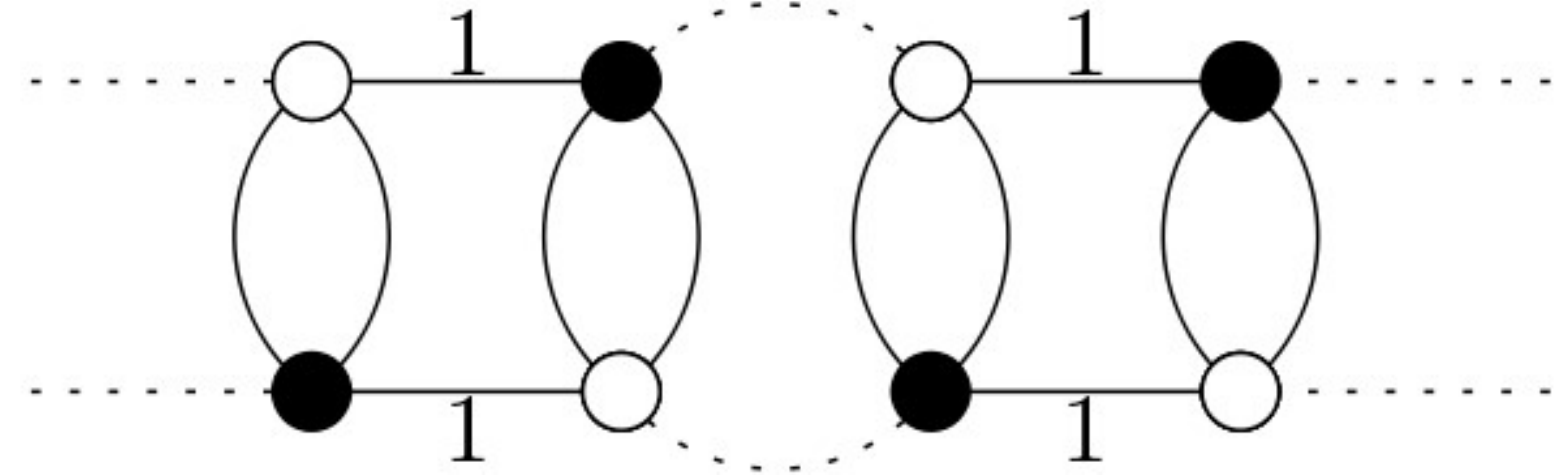} \nonumber \\
 &+ 4 \includegraphics[scale=0.2]{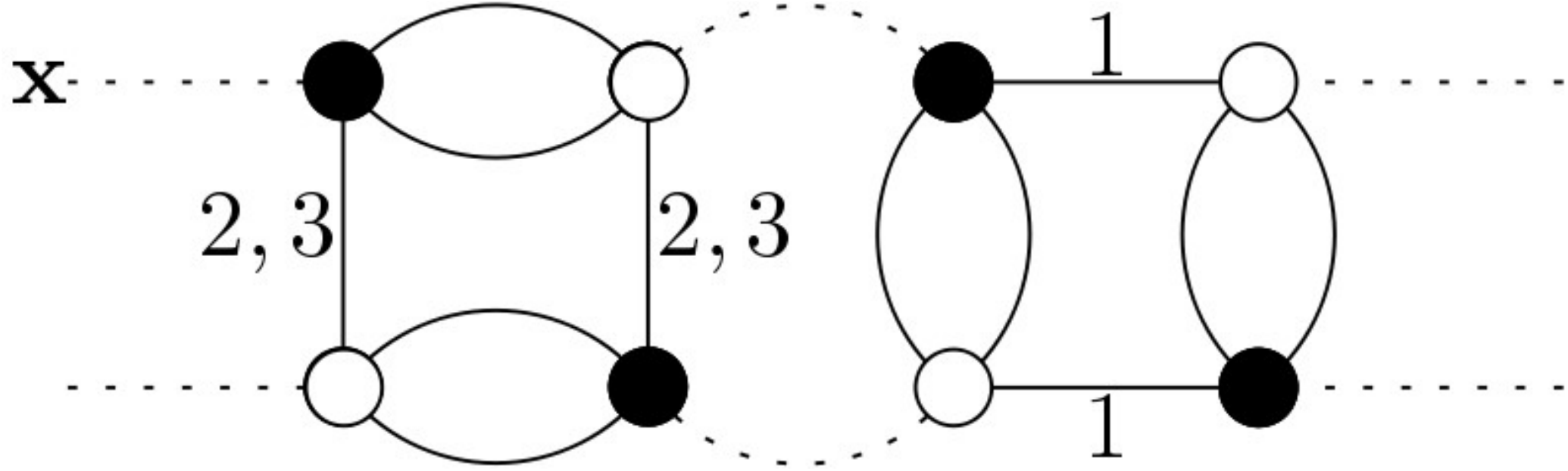} + 4 \sum \limits_{d=1}^3 \Bigg(\includegraphics[scale=0.25]{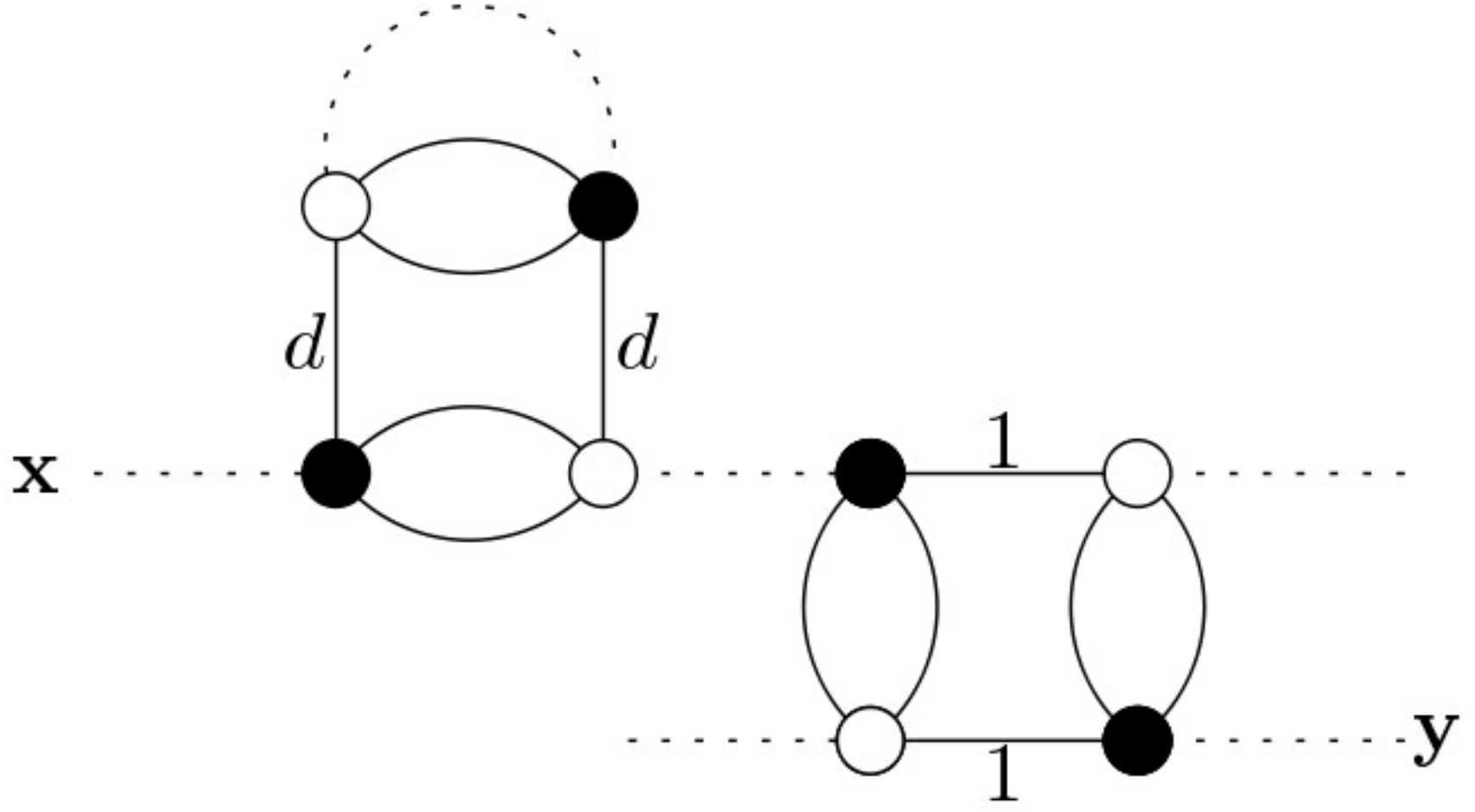} + \includegraphics[scale=0.25]{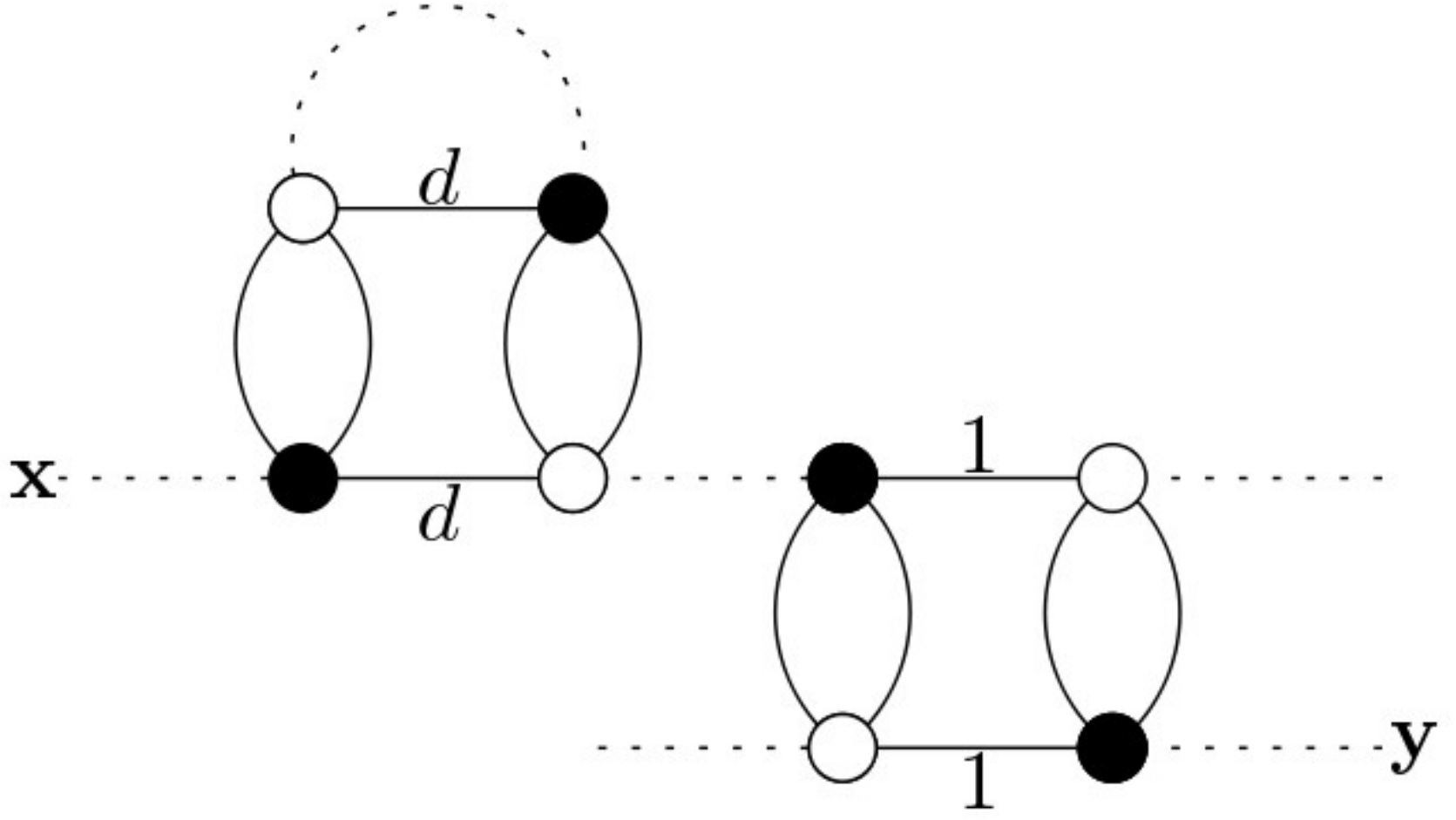} \nonumber\\
 & + \includegraphics[scale=0.25]{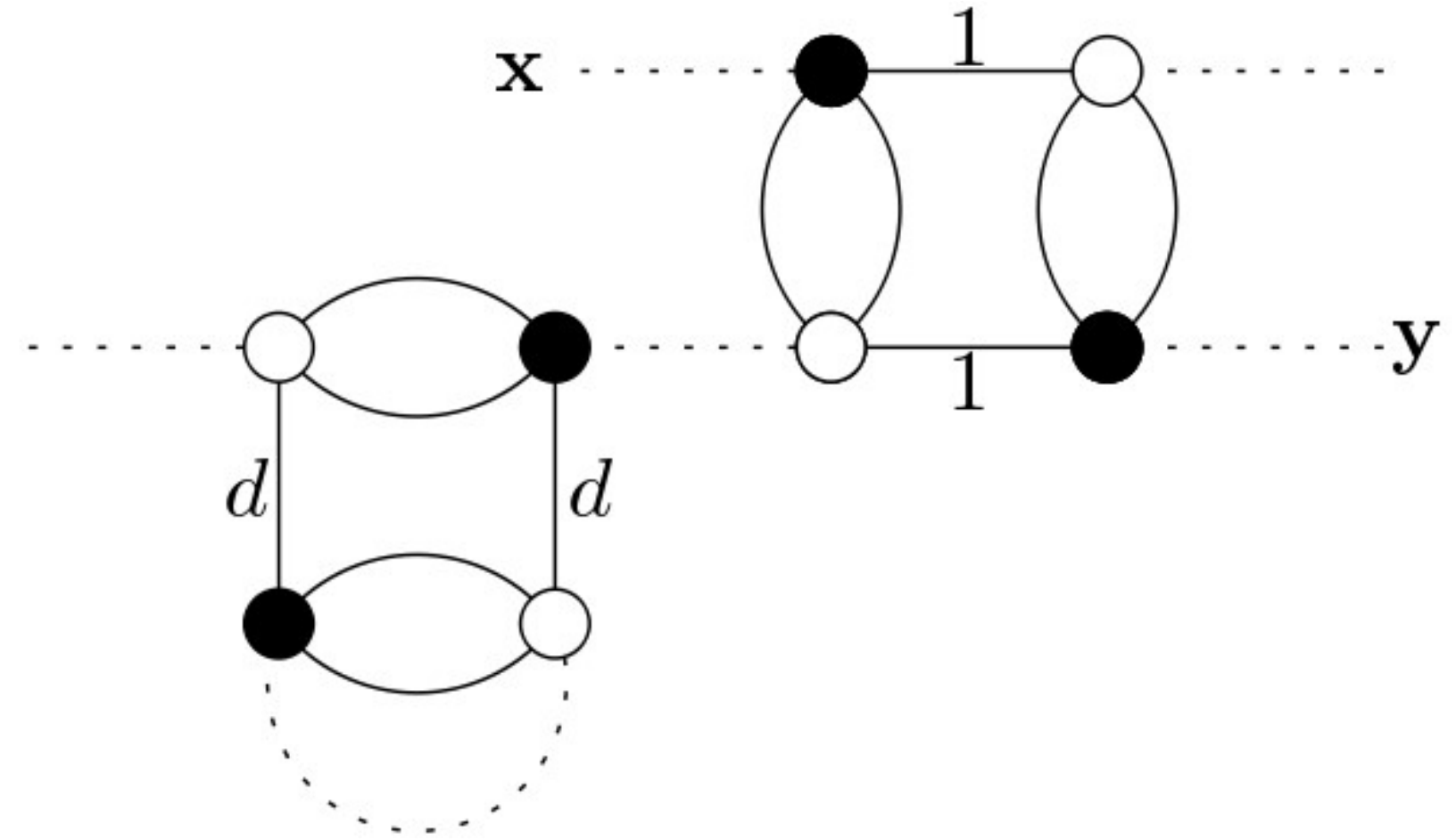} + \includegraphics[scale=0.25]{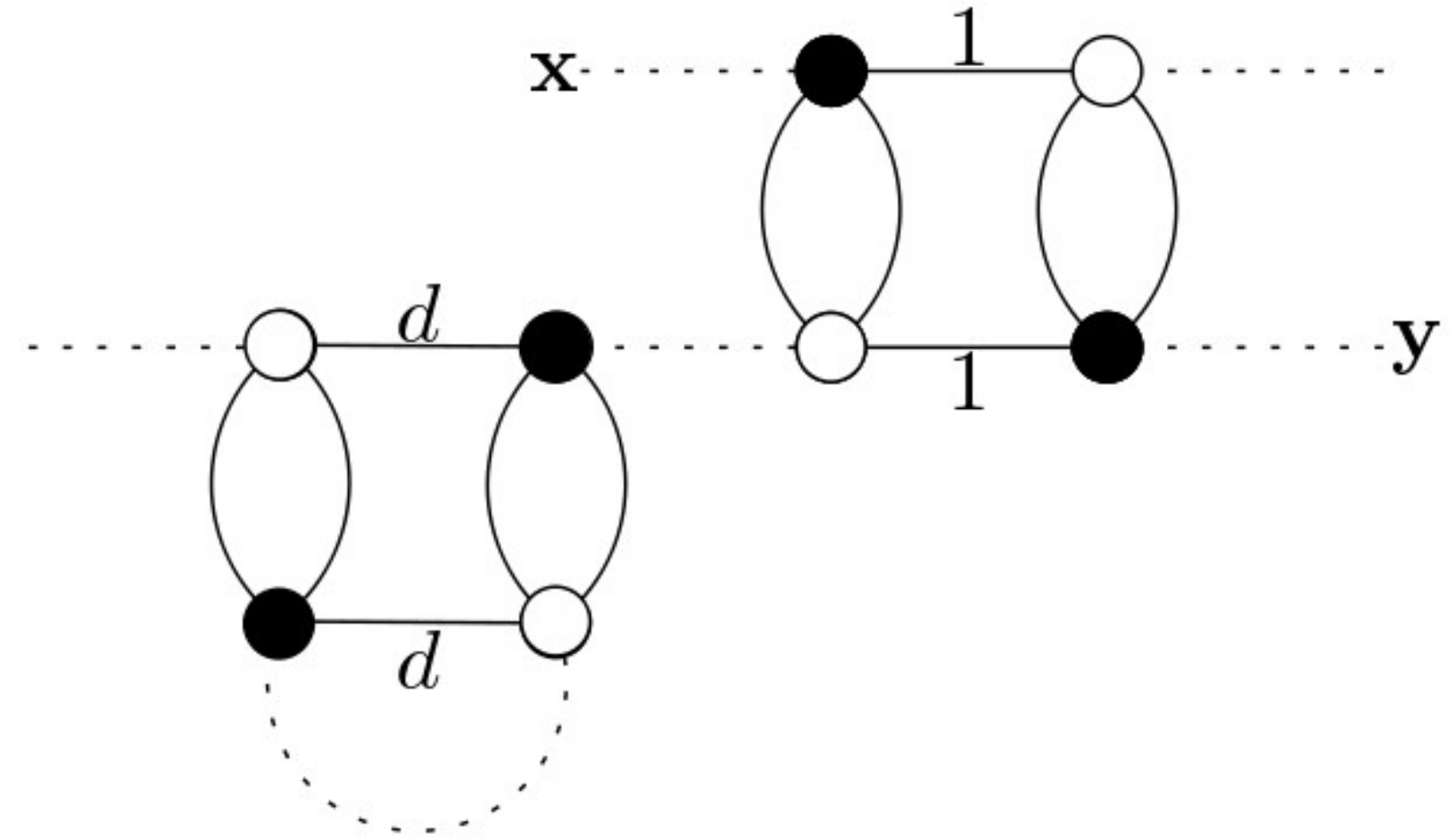} + \includegraphics[scale=0.25]{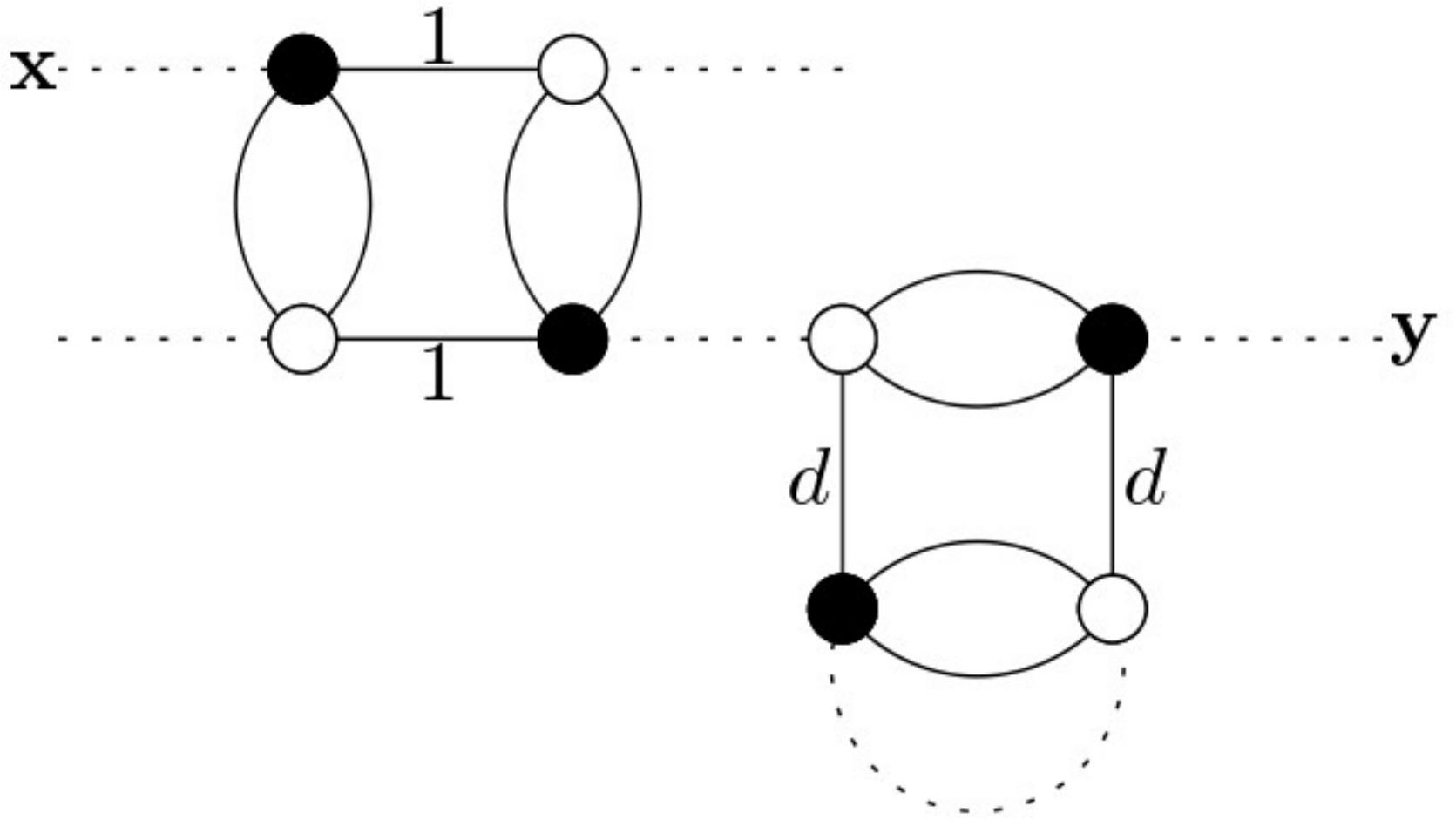} \nonumber \\
 &+ \includegraphics[scale=0.25]{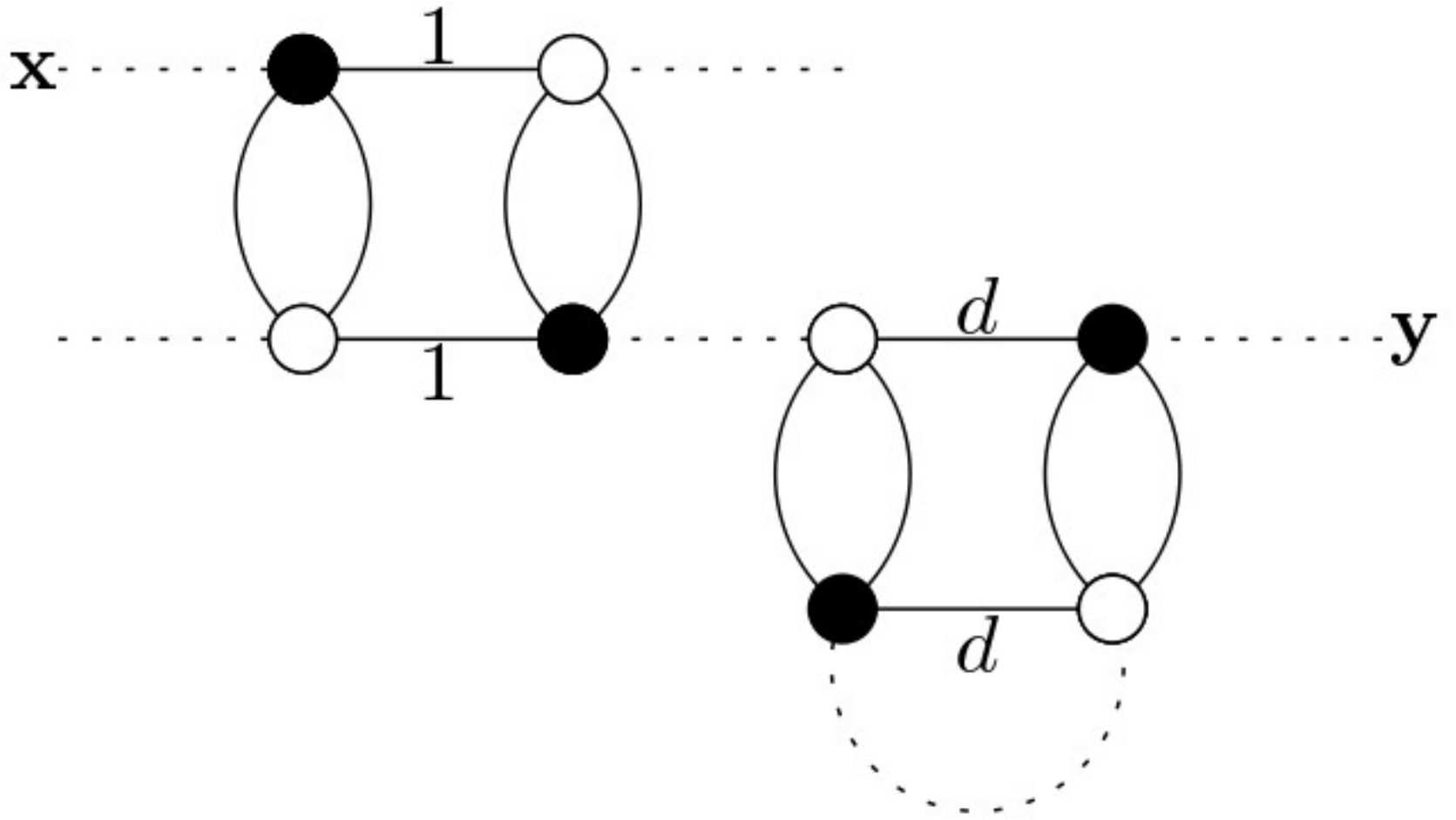} \Bigg) + O(\lambda^3),
\end{align}
\begin{align}
     &-\frac{2\lambda}{x_1^2+y_2^2+y_3^2}\sum_{b_1}
  \frac{\mathrm{G}^{(4)}_1(\mathbf{x},\mathbf{y}) -\mathrm{G}^{(4)}_1(b_1,x_2,x_3,\mathbf{y})}{b_1^2-x_1^2} \nonumber\\ 
  &= 4\includegraphics[scale=0.20]{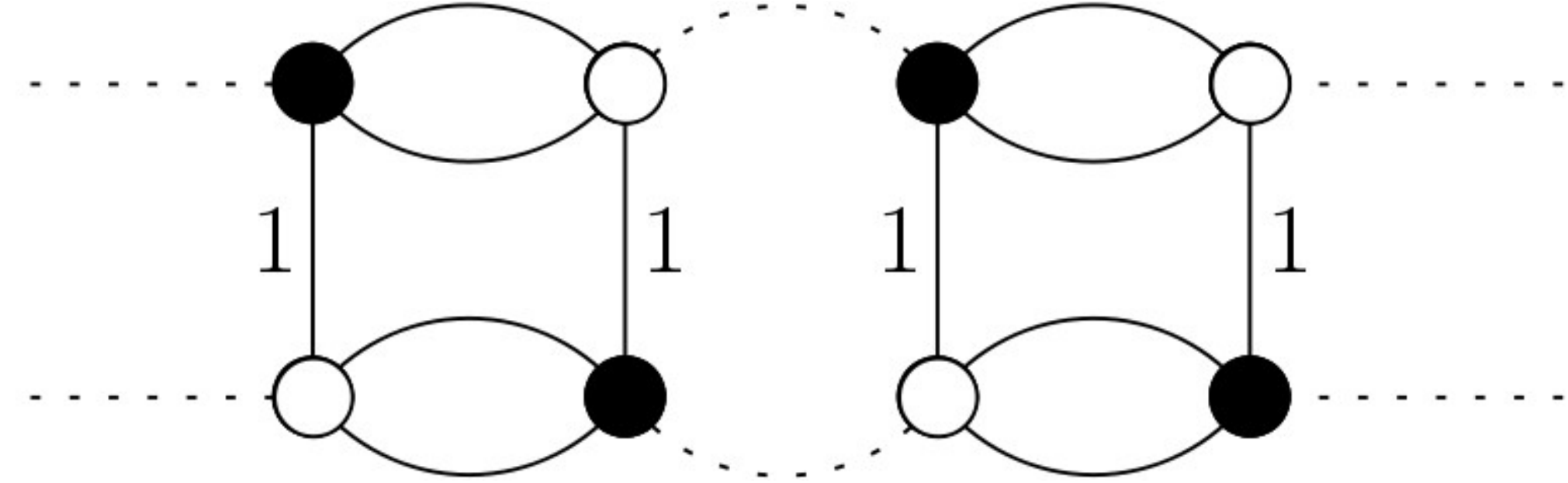} + 4\includegraphics[scale=0.25]{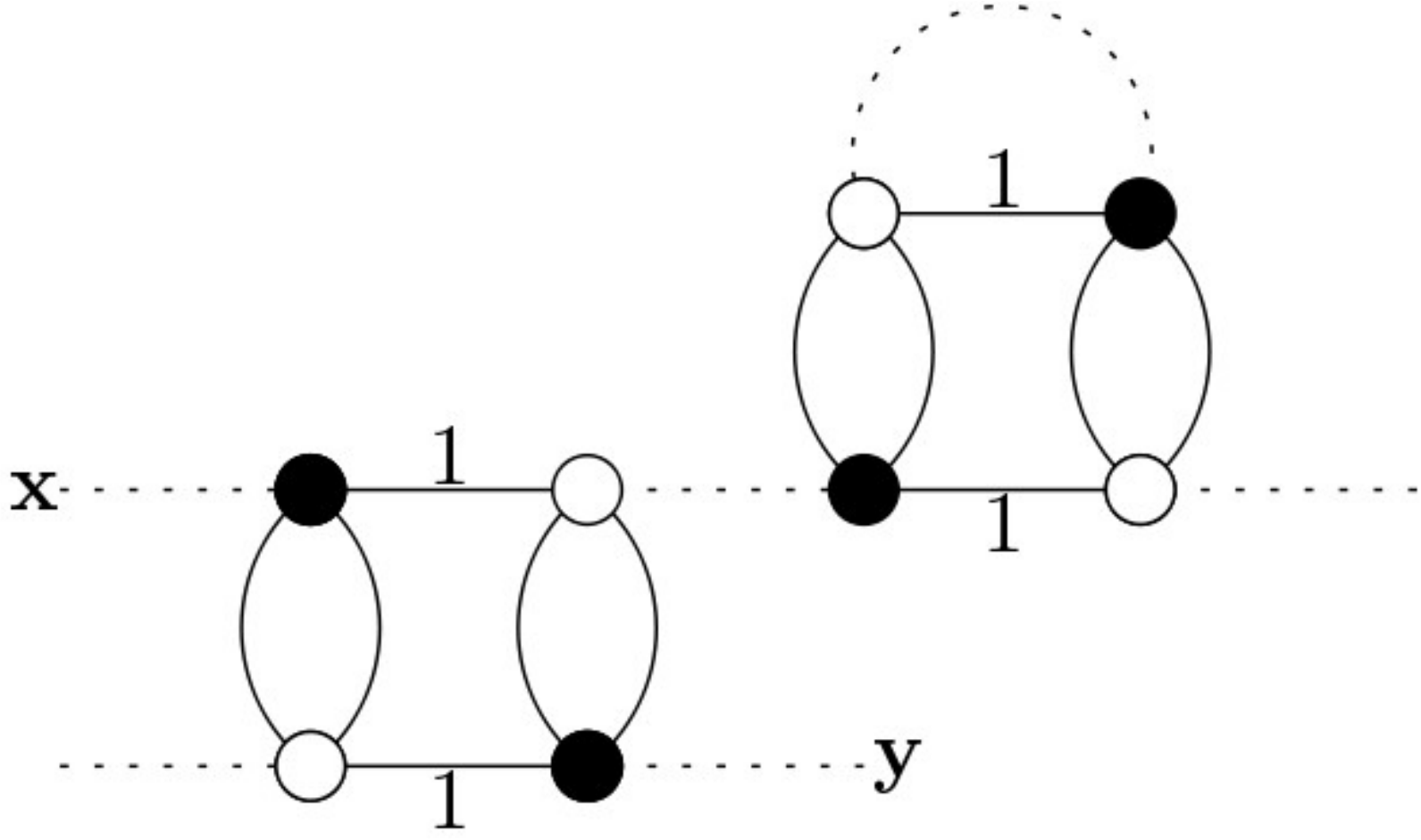}  + O(\lambda^3),  
\end{align}
\begin{align}
 &-\frac{2\lambda}{x_1^2+y_2^2+y_3^2}\Bigg(\sum_{b_2}
  \frac{\mathrm{G}^{(4)}_1(\mathbf{x},\mathbf{y}) -
  \mathrm{G}^{(4)}_1(\mathbf{x},y_1,b_2,y_3)}{b_2^2-y_2^2} + \sum_{b_3}
  \frac{\mathrm{G}^{(4)}_1(\mathbf{x},\mathbf{y}) -\mathrm{G}^{(4)}_1(\mathbf{x},y_1,y_2,b_3)}{b_3^2-y_3^2} \Bigg) \nonumber \\
  &= 4\includegraphics[scale=0.25]{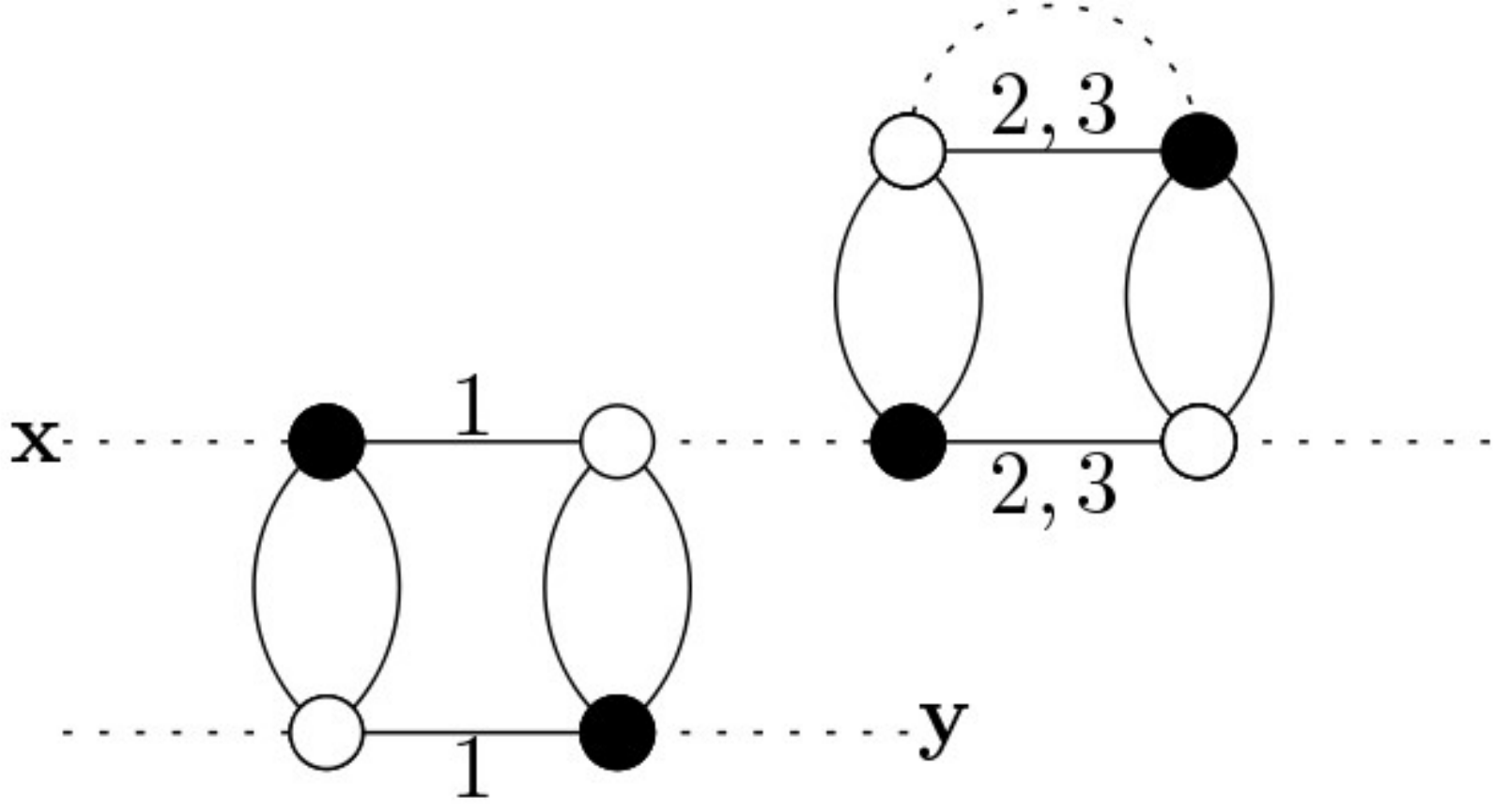} + 4\includegraphics[scale=0.25]{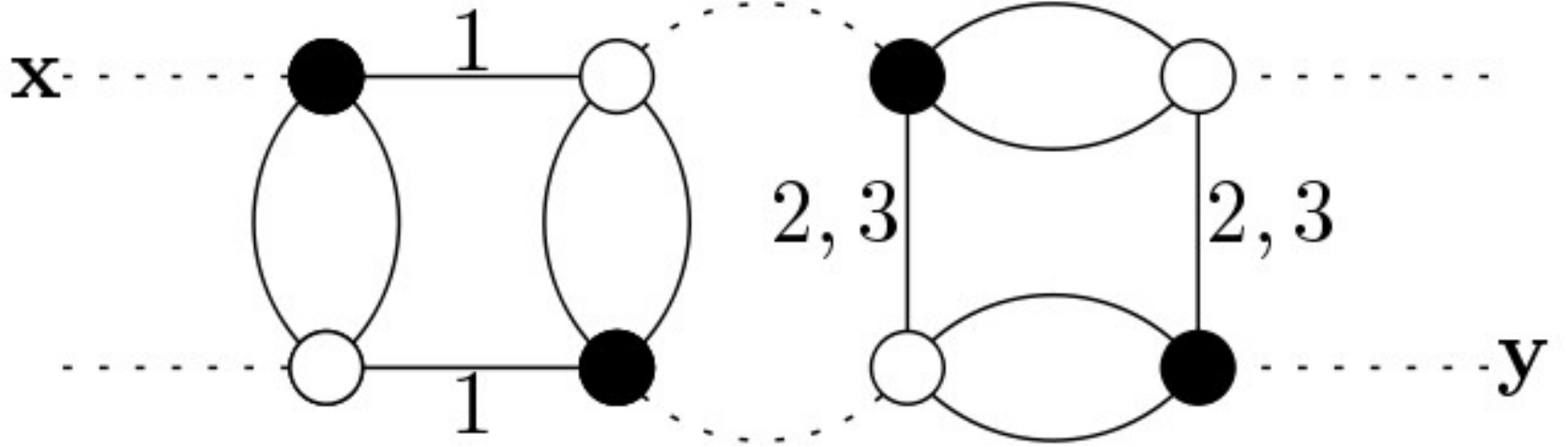}  + O(\lambda^3),
 \end{align}
 \begin{align}
     &-\frac{2\lambda}{x_1^2+y_2^2+y_3^2}\Bigg(\frac{\mathrm{G}^{(4)}_2(\mathbf{x},y_1,y_2,x_3)-
  \mathrm{G}^{(4)}_2(\mathbf{x},\mathbf{y})}{y_3^2-x_3^2} + \frac{\mathrm{G}^{(4)}_3(\mathbf{x},y_1,x_2,y_3)-
  \mathrm{G}^{(4)}_3(\mathbf{x},\mathbf{y})}{y_2^2-x_2^2}\Bigg)\nonumber \\
  &= 4\includegraphics[scale=0.25]{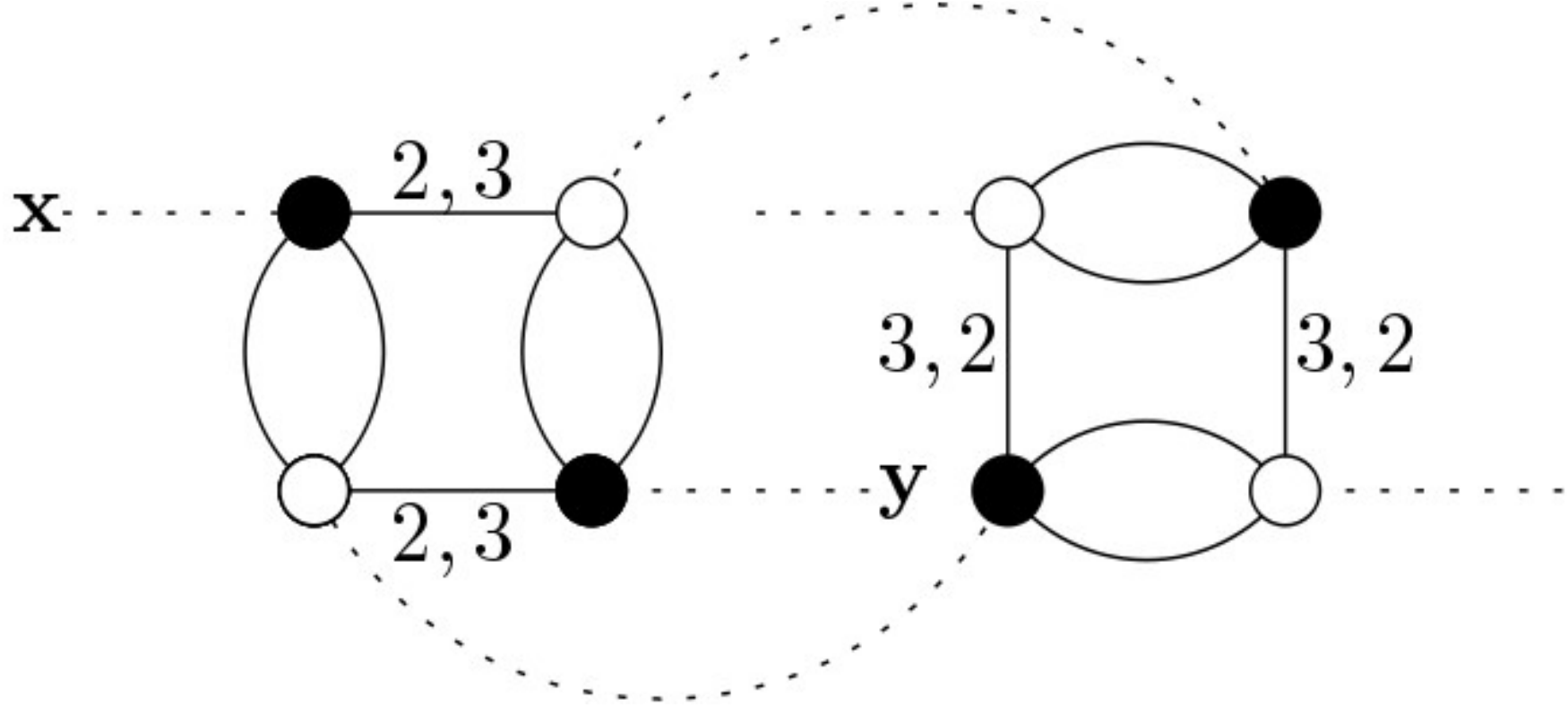} + 4\includegraphics[scale=0.25]{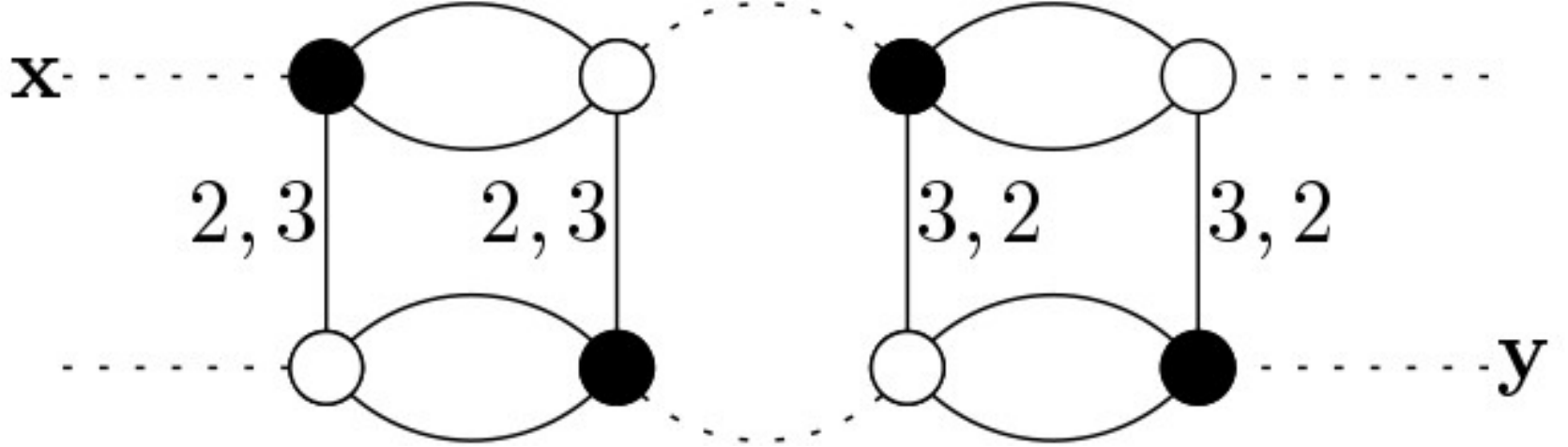}  + O(\lambda^3).
 \end{align}
Finally the expansion of the $4$-point function is
\begin{align}
    &\mathrm{G}^{(4)}_1(\mathbf{x},\mathbf{y}) = 2 \includegraphics[scale=0.25]{4ptpillow.pdf} + 4 \includegraphics[scale=0.25]{4pt2pillow.pdf} +  4\includegraphics[scale=0.2]{4pt2pillow2.pdf} \nonumber \\
    &+ 4 \includegraphics[scale=0.25]{4pt2pillow2c2,3.pdf} + 4 \includegraphics[scale=0.2]{4ptc23pillow1.pdf} + 4\includegraphics[scale=0.25]{4ptdiff4pt1.pdf} \nonumber \\
    &+4\includegraphics[scale=0.25]{4ptdiff4pt2.pdf} + 4 \sum \limits_{d=1}^3 \Bigg(\includegraphics[scale=0.25]{4ptPIpillow2x.pdf} + \includegraphics[scale=0.25]{4ptPIpillow1x.pdf} \nonumber\\
    &+ \includegraphics[scale=0.25]{4ptPIpillow2y1.pdf} + \includegraphics[scale=0.25]{4ptPIpillow1y1.pdf}+ \includegraphics[scale=0.25]{4ptPIpillow2x1.pdf} \nonumber \\
    &+ \includegraphics[scale=0.25]{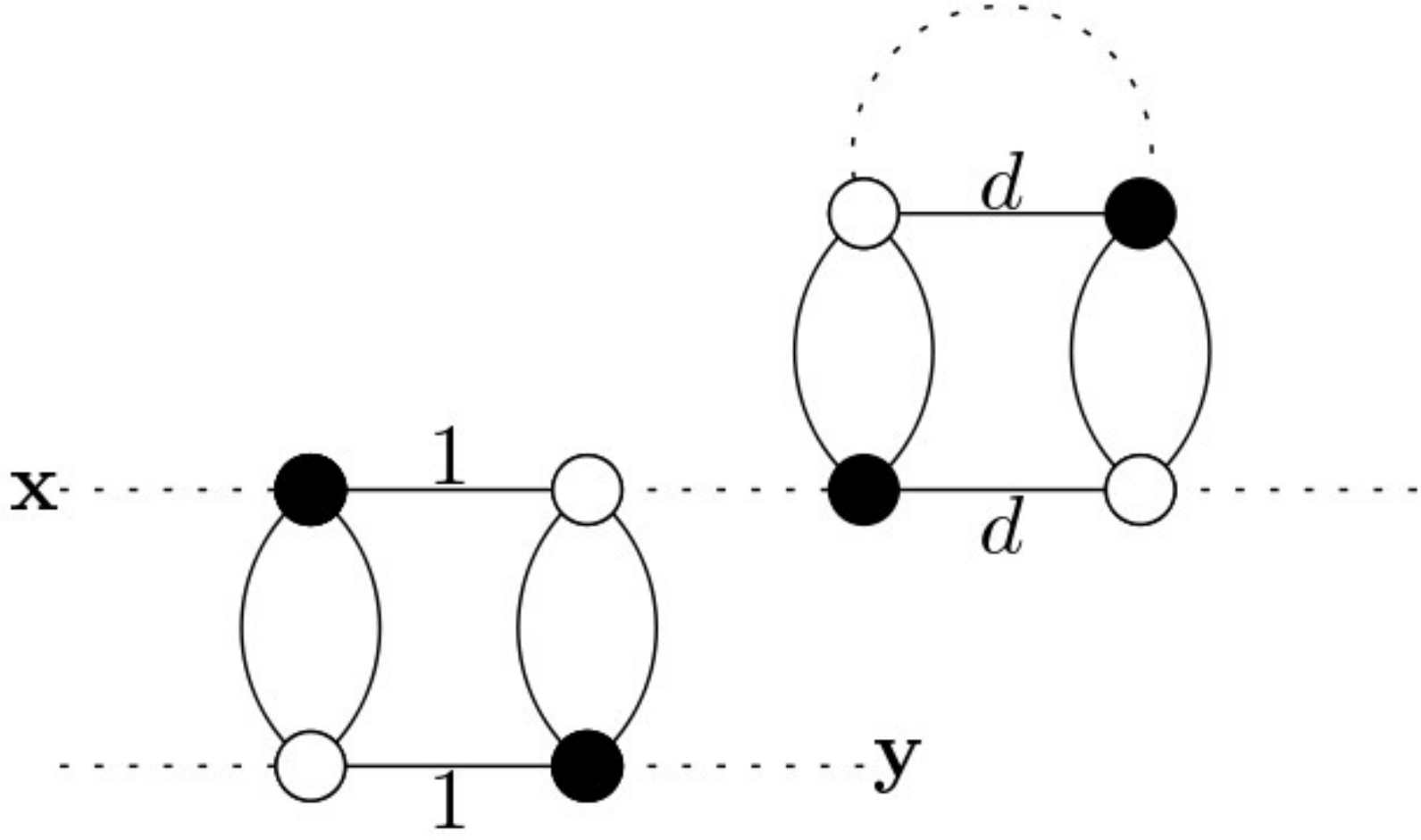} + \includegraphics[scale=0.25]{4ptPIpillow2y.pdf} + \includegraphics[scale=0.25]{4ptPIpillow1y.pdf}\Bigg)  + O(\lambda^3).
\end{align}
After taking the large $N$ limit, we get the following SDE and perturbative expansion
\begin{align}
    &\mathrm{G}^{(4)}_1(\mathbf{x},\mathbf{y}) = -\frac{2\lambda}{x_1^2+y_2^2+y_3^2}\Bigg(\sum_{c=1}^3\int\mathrm{d}\mathbf{q}_{\hat{c}} \mathrm{G}^{(2)}(\mathbf{q}_{\hat{c}}s_c)\mathrm{G}^{(4)}_1(\mathbf{x},\mathbf{y}) + \mathrm{G}^{(2)}(\mathbf{y})\frac{\mathrm{G}^{(2)}(\mathbf{x})-\mathrm{G}^{(2)}(y_1,x_2,x_3)}{y_1^2-x_1^2}\Bigg) \nonumber \\
    &=2 \includegraphics[scale=0.25]{4ptpillow.pdf} + 4 \includegraphics[scale=0.25]{4pt2pillow.pdf} + 4 \sum \limits_{d=1}^3 \Bigg(\includegraphics[scale=0.25]{4ptPIpillow2x.pdf} \nonumber \\
    &+ \includegraphics[scale=0.25]{4ptPIpillow2y1.pdf} + \includegraphics[scale=0.25]{4ptPIpillow2x1.pdf} + \includegraphics[scale=0.25]{4ptPIpillow2y.pdf} \Bigg)  + O(\lambda^3).
\end{align}

\subsection*{\texorpdfstring{$4$}{4}-point function with disconnected boundary}

The SDE for the $4$-point function with a disconnected boundary graph is
\begin{align}
    &\mathrm{G}^{(4)}_{\mathrm{m}}(\mathbf{x},\mathbf{y}) = -\frac{2\lambda}{|\mathbf{x}|^2}  \sum_{c=1}^3 \Bigg\{\sum_{\mathbf{q}_{\hat{c}}}\mathrm{G}^{(2)}(\mathbf{q}_{\hat{c}}x_c)\mathrm{G}^{(4)}_{\mathrm{m}}(\mathbf{x},\mathbf{y}) + \mathfrak{f}_{\mathrm{m}|\mathrm{m},
    x_c}^{(c)}\left(\mathbf{x},\mathbf{y}\right) + \mathfrak{f}_{\mathrm{m}|\mathrm{m},
    x_c}^{(c)}\left(\mathbf{y},\mathbf{x}\right) \nonumber\\
    &+\sum_{b_c}\frac{1}{b_c^2-x_c^2}\left(\mathrm{G}^{(4)}_{\mathrm{m}}(\mathbf{x},\mathbf{y})-\mathrm{G}^{(4)}_{\mathrm{m}}(\mathbf{x}_{\hat{c}}b_c,\mathbf{y})\right)+\frac{1}{y_c^2-x_c^2}\left(\mathrm{G}^{(4)}_c(\mathbf{x},\mathbf{y})-\mathrm{G}^{(4)}_{c}(\mathbf{x}_{\hat{c}}y_c,\mathbf{y})\right) \nonumber \\
    &+ \mathrm{G}^{(2)}(\mathbf{x})\bigg(\mathrm{G}^{(4)}_c(\mathbf{y}_{\hat{c}}x_c,\mathbf{y}) +\sum\limits_{d\neq c}\displaystyle\sum_{\substack{q_b \\ b \neq c,d}}\mathrm{G}^{(4)}_d(x_c,q_b,y_d,\mathbf{y}) + \sum\limits_{\mathbf{q}_{\hat{c}}}\mathrm{G}^{(4)}_{\mathrm{m}}(\mathbf{q}_{\hat{c}}x_c,\mathbf{y})\bigg)\Bigg\},
\end{align}
where $(x_c,q_b,y_d)$ with $\{b,c,d\} = \{1,2,3\} $ is implicitly reordered.
Let us again check the perturbative expansion up to $2^{\text{nd}}$ order in the coupling constant. We can note that the first graphs appearing in $\mathrm{G}^{(4)}_{\mathrm{m}}$ are of order $\lambda^2$, hence all terms in the SDE involving $\lambda\mathrm{G}^{(4)}_{\mathrm{m}}$ will start to contribute only at order $\lambda^3$, and the same goes for the terms $\lambda\mathfrak{f}_{\mathrm{m}|\mathrm{m}}$. The other terms give
\begin{align}
    &-\frac{2\lambda}{|\mathbf{x}|^2}  \mathrm{G}^{(2)}(\mathbf{x})\sum_{c=1}^3\mathrm{G}^{(4)}_c(\mathbf{y}_{\hat{c}}x_c,\mathbf{y}) = 4\sum_{c=1}^3\includegraphics[scale=0.25]{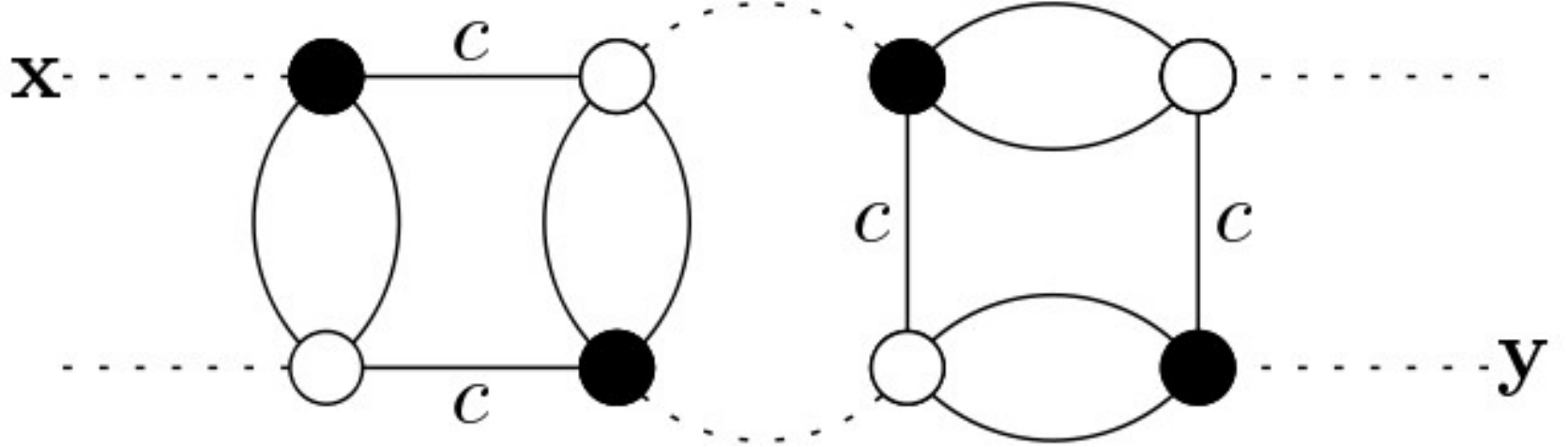} + O(\lambda^3), \\
    &-\frac{2\lambda}{|\mathbf{x}|^2} \mathrm{G}^{(2)}(\mathbf{x})\sum_{c=1}^3\sum\limits_{d\neq c}\displaystyle\sum_{\substack{q_b \\ b \neq c,d}}\mathrm{G}^{(4)}_d(x_c,q_b,y_d,\mathbf{y}) = 4\sum_{c=1}^3\sum\limits_{d\neq c} \includegraphics[scale=0.25]{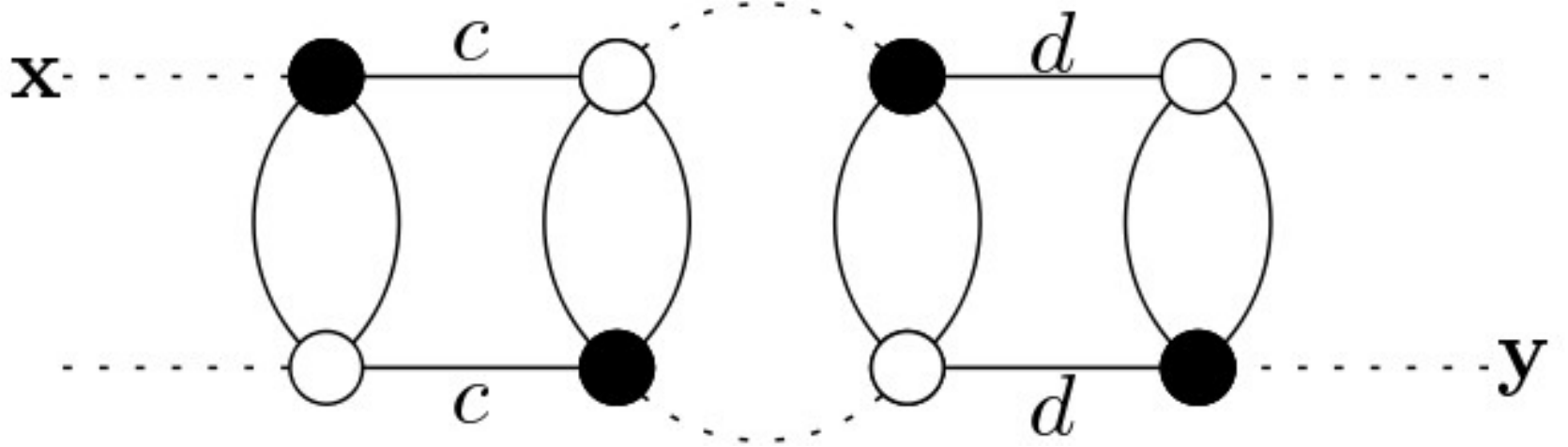} + O(\lambda^3), \\
    &-\frac{2\lambda}{|\mathbf{x}|^2}\sum_{c=1}^3\frac{\mathrm{G}^{(4)}_c(\mathbf{x},\mathbf{y})-\mathrm{G}^{(4)}_{c}(\mathbf{x}_{\hat{c}}y_c,\mathbf{y})}{y_c^2-x_c^2} \\
    &= 4\sum_{c=1}^3\Bigg(\includegraphics[scale=0.25]{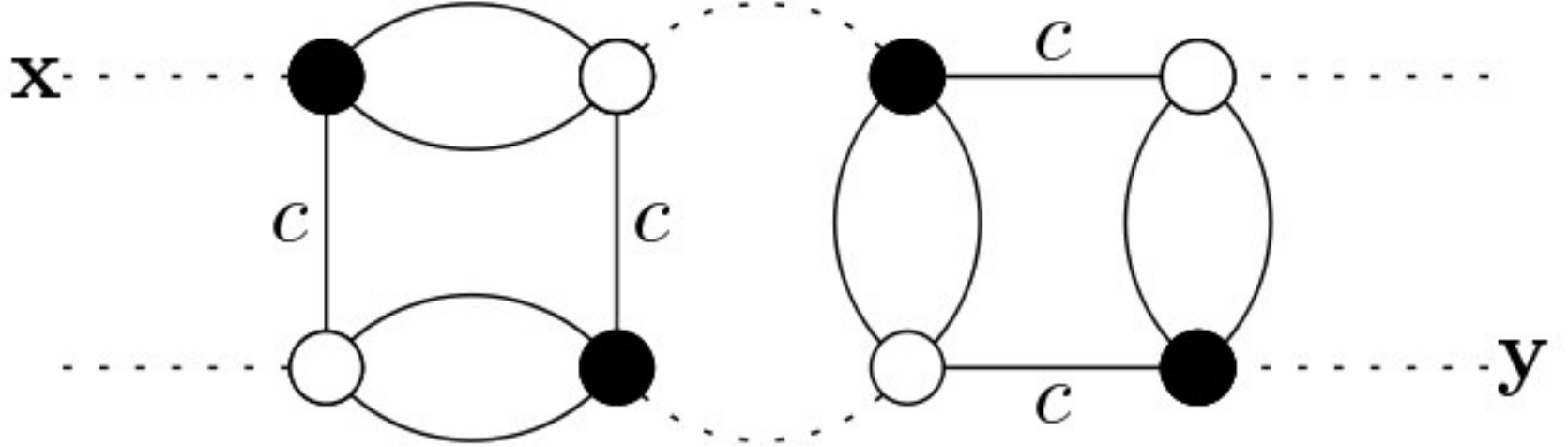}+\includegraphics[scale=0.25]{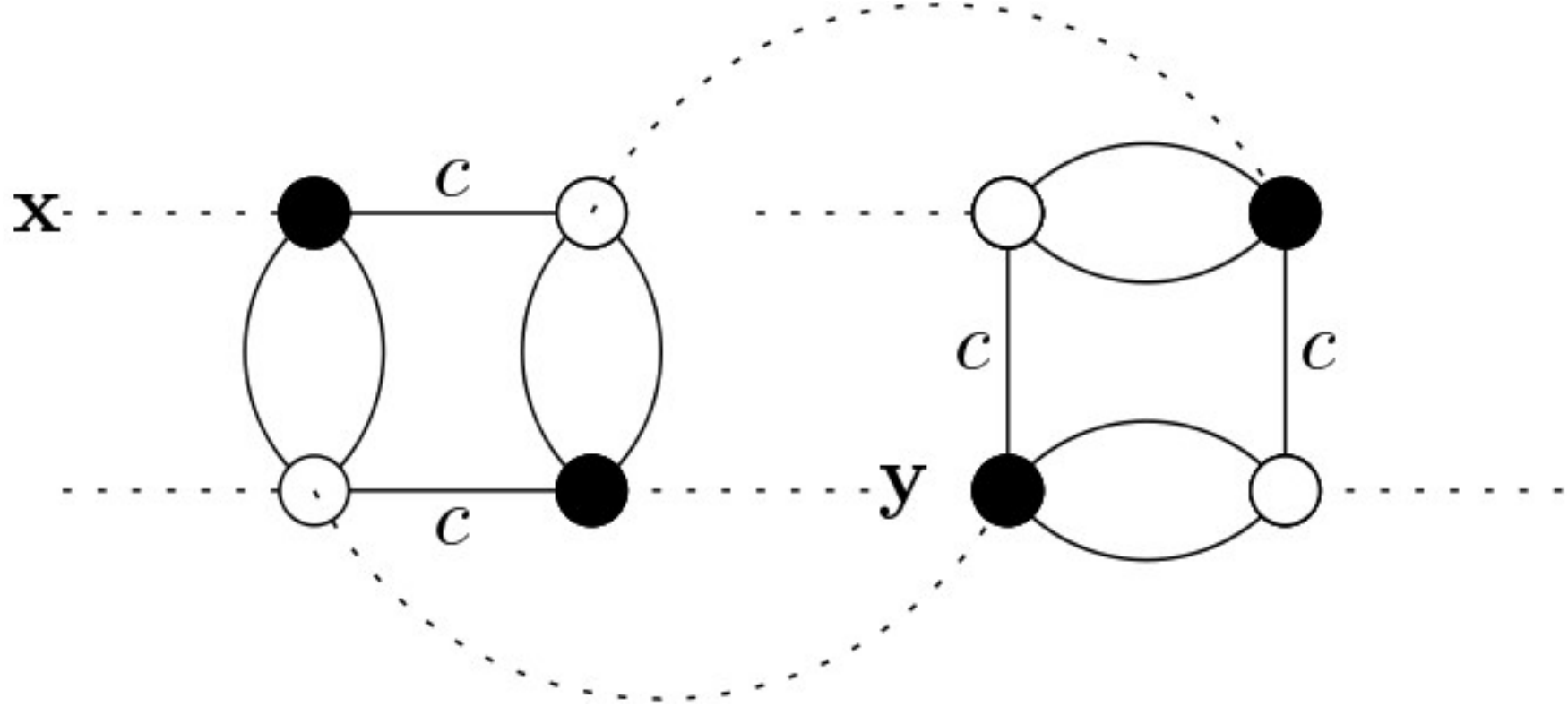}\Bigg) + O(\lambda^3). \nonumber
\end{align}
In the large $N$ limit, only one of these graphs survives and the SDE becomes
\begin{align}
    &\mathrm{G}^{(4)}_{\mathrm{m}}(\mathbf{x},\mathbf{y}) = 4\sum_{c=1}^3\sum\limits_{d\neq c} \includegraphics[scale=0.25]{4ptdiscLO.pdf} + O(\lambda^3) \\
    &= -\frac{2\lambda}{|\mathbf{x}|^2}  \sum_{c=1}^3\Bigg\{\int \mathrm{d}\mathbf{q}_{\hat{c}}\mathrm{G}^{(2)}(\mathbf{q}_{\hat{c}}x_c)\mathrm{G}^{(4)}_{\mathrm{m}}(\mathbf{x},\mathbf{y})+\mathrm{G}^{(2)}(\mathbf{x})\bigg(\sum\limits_{d\neq c}\int\mathrm{d}q_b\mathrm{G}^{(4)}_d(x_c,q_b,y_d,\mathbf{y}) + \int\mathrm{d}\mathbf{q}_{\hat{c}}\mathrm{G}^{(4)}_{\mathrm{m}}(\mathbf{q}_{\hat{c}}x_c,\mathbf{y})\bigg)\Bigg\}. \nonumber
\end{align}

\printbibliography

\end{document}